\documentclass[aps,prb,reprint,twocolumn,floatfix,superscriptaddress,amsmath,amssymb,amsfonts]{revtex4-2}
\AtBeginDocument{\usepackage{booktabs}}               
\makeatletter

\usepackage{graphicx}
\usepackage{dcolumn}
\usepackage{bm}
\usepackage{xcolor}
\usepackage{multirow}
\usepackage[percent]{overpic}
\usepackage[newcommand]{ragged2e}
\usepackage{rotating}
\usepackage{tikz}
\usepackage{gensymb}
\usepackage{mathtools}

\usepackage{caption}
\DeclareCaptionJustification{justified}{\justifying}
\usepackage{subcaption}

\usepackage[colorlinks=true, urlcolor=blue, linkcolor=blue, citecolor=blue, pdftex]{hyperref}
\usepackage[mathlines]{lineno}

 \g@addto@macro\bfseries{\boldmath}
\makeatother

\usepackage{float}

\begin{document}

\preprint{APS/123-QED}

\title{Classical and quantum phases of the pyrochlore $S=\frac{1}{2}$ magnet with Heisenberg and Dzyaloshinskii-Moriya interactions}

\author{Vincent Noculak}
\altaffiliation[These authors contributed equally to this work.] {}
\affiliation{Dahlem Center for Complex Quantum Systems and Fachbereich Physik, Freie Universit\"at Berlin, Arnimallee 14, 14195 Berlin, Germany}
\affiliation{Helmholtz-Zentrum Berlin f\"ur Materialien und Energie, Hahn-Meitner-Platz 1, 14109 Berlin, Germany}
\author{Daniel Lozano-G\'omez}
\altaffiliation[These authors contributed equally to the project.]{}
\affiliation{Department of Physics and Astronomy, University of Waterloo, Waterloo, Ontario N2L 3G1, Canada}
 \author{Jaan Oitmaa}
 \affiliation{School of Physics, The University of New South Wales, Sydney 2052, Australia}
 \author{Rajiv R. P. Singh}
 \affiliation{ Department of Physics, University of California Davis, California 95616, USA}
\author{Yasir Iqbal}
\affiliation{Department of Physics and Quantum Center for Diamond and Emergent Materials (QuCenDiEM), Indian Institute of Technology Madras, Chennai 600036, India}
\author{Michel J. P. Gingras}
\affiliation{Department of Physics and Astronomy, University of Waterloo, Waterloo, Ontario N2L 3G1, Canada}
\author{Johannes Reuther}
\affiliation{Dahlem Center for Complex Quantum Systems and Fachbereich Physik, Freie Universit\"at Berlin, Arnimallee 14, 14195 Berlin, Germany}
\affiliation{Helmholtz-Zentrum Berlin f\"ur Materialien und Energie, Hahn-Meitner-Platz 1, 14109 Berlin, Germany}
\affiliation{Department of Physics and Quantum Center for Diamond and Emergent Materials (QuCenDiEM), Indian Institute of Technology Madras, Chennai 600036, India}

\date{\today}

\begin{abstract}
We investigate the  ground state and critical temperature ($T_c$) phase diagrams of the classical and quantum $S=\frac{1}{2}$ pyrochlore lattice with nearest-neighbor Heisenberg and Dzyaloshinskii-Moriya interactions (DMI). We consider ferromagnetic and antiferromagnetic Heisenberg exchange interaction as well as direct and indirect DMI. At the classical level, three ground states are found: all-in/all-out, ferromagnetic and a locally ordered $XY$ phase, known as $\Gamma_5$, which displays an accidental classical U(1) degeneracy at the mean-field level. Quantum zero-point energy fluctuations computed to order $1/S$ are found to lift the classical ground state degeneracy and select the so-called $\psi_3$ state out of the degenerate manifold in most parts of the $\Gamma_5$ regime. Likewise, thermal fluctuations treated classically at the Gaussian level entropically select the $\psi_3$ state at $T=0^+$. In contrast to this low-temperature state-selection behavior, classical Monte Carlo simulations find that the system orders at $T_c$ in the non-coplanar $\psi_2$ state of $\Gamma_5$ for antiferromagnetic Heisenberg exchange and indirect DMI with a transition from $\psi_2$ to $\psi_3$ at a temperature $T_{\Gamma_5} <T_c$. The same method finds that the system orders via a single transition at $T_c$ directly into the $\psi_3$ state for most of the region with ferromagnetic Heisenberg exchange and indirect DMI. Such ordering behavior at $T_c$ for the $S=\frac{1}{2}$ quantum model is corroborated by high-temperature series expansion. To investigate the $T=0$ quantum ground state of the model, we apply the pseudo-fermion functional renormalization group (PFFRG). The quantum paramagnetic phase of the pure antiferromagnetic $S=\frac{1}{2}$ Heisenberg model is found to persist over a finite region in the phase diagram for both direct or indirect DMI. Interestingly, we find that a combined ferromagnetic Heisenberg and indirect DMI, near the boundary of ferromagnetism and $\Gamma_5$ antiferromagnetism, may potentially realize a $T=0$ quantum ground state lacking conventional magnetic order. Otherwise, for the largest portion of the phase diagram, PFFRG finds the same long-range ordered phases (all-in/all-out, ferromagnetic and $\Gamma_5$) as in the classical model.

\end{abstract}

\maketitle

\section{\label{sec:introduction}Introduction}

\subsection{Background \& motivation}

Spin fluctuations, being either of quantum or thermal nature, play a fundamental role in shaping the magnetic properties of frustrated magnetic systems~\cite{Springer_frust,balentsSpinLiquidsFrustrated2010b}. In the simplest scenario, when fluctuations act on a system with conventional long-range magnetic order, the static dipolar magnetic order parameter is reduced or even completely suppressed, resulting in a lesser ordered state. However, the opposite and, at first sight, counter intuitive situation also occurs where fluctuations are the very reason for the existence of long-range order, a phenomenon known as ``order-by-disorder'' (ObD)~\cite{Villain-1980,Shender-1982,Henley-1989,Prakash}. 
Order-by-disorder takes place when a classical ($\cramped{S\rightarrow\infty}$) spin system displays an accidental degeneracy of ground states that is not the result of global symmetries but, instead, of fine-tuned spin-spin interactions. Such a system then becomes highly susceptible to fluctuation-driven selection effects within the degenerate manifold of classical ground states. These effects can either arise through quantum fluctuations~\cite{Shender-1982}, most straightforwardly described by the harmonic zero-point energy of spin waves as described in a semiclassical $1/S$ approximation, or via an entropic selection coming from the thermal fluctuations about the classically degenerate states~\cite{Villain-1980,Henley-1989,Prakash}. Independently of the precise selection mechanism, the free-energy scale(s) associated with ObD are typically small compared to the bare spin-spin coupling terms in the Hamiltonian. As a result, the manifestation of ObD may be more prominently observed in parameter regimes where two different phases are strongly competing with their free-energy difference small. The smallness of the free-energy difference of competing phases combined with the possibility of ObD can give rise to rich phase diagrams driven by the complex interplay of subleading or perturbative energetic terms in the Hamiltonian and fluctuation effects, resulting in many cases in unconventionally ordered magnetic regimes. 
Additional complexity can arise when quantum and thermal fluctuations do not cooperate but select different states~\cite{Schick-2020}. Given all these subtleties, the simplest and most common, but also somewhat incomplete way to theoretically tackle ObD has often been to treat only one source or mechanism of fluctuations (thermal or quantum) while omitting the other. This perspective is manifestly inadequate, for example, when ObD  is either responsible for the finite temperature phase transition to long-range magnetic order itself or, perhaps less dramatic but nonetheless of importance, when it selects at the critical temperature a subset of all the states that are accidentally degenerate within a manifold of the critical states such as would be predicted in a mean-field theory treatment~\cite{Oitmaa-2013,Javanparast15,Javanparast-GTO,Sizyuk}.

The investigation of pyrochlore spin systems has over the past thirty years or so developed into a fruitful research enterprise that has provided a superb material-relevant context for exploring the rich physics of ObD~\cite{bramwell94,Champion-2003,Savary-2012,Zhitomirsky-2012,Oitmaa-2013,Zhitomirsky-2014,Javanparast15,Sarkis-2020}. On the one hand, the research has been fueled by an abundance of material realizations of the pyrochlores lattice~\cite{Gardner-RMP,Hallas-AnnRevCMP,Rau-2019}. From a more theoretical perspective, pyrochlore systems and, more specifically, spin ices~\cite{Bramwell-Science,balentsSpinLiquidsFrustrated2010b,Springer-spin-ice,Gardner-RMP}, 
harbor a variety of interesting physical phenomena, such as effective magnetic monopoles~\cite{Castelnovo-2012}, dipolar spin correlations~\cite{Stillinger-1973,Isakov04} and emergent gauge theories~\cite{Henley-2010}. Indeed, many of these phenomena result from the fact that the simplest possible pyrochlore spin models -- nearest-neighbor antiferromagnet Heisenberg and Ising models -- exhibit an exponentially large classical ground state degeneracy~\cite{Anderson-1956,Villain-1979,Reimers-1991a,Moessner98,Castelnovo-2012}, thus providing, at least naively, an ideal starting point for investigating ObD. It turns out, however, that the exponential degeneracy (i.e. extensive zero-temperature entropy) of these two models is too large and too robust for enabling dominant ObD effects. More precisely, these systems are known to be immune against thermal ObD effects~\cite{Reimers-1992,Zinkin-1996,Moessner-1998b} thus realizing a zero temperature “cooperative paramagnet”~\cite{Villain-1979} with finite residual entropy~\cite{Anderson-1956}, referred to as a “classical spin liquid”~\cite{Moessner98,Moessner-1998b,Reimers-1992,Zinkin-1996,Henley-2010}. Similarly, harmonic and higher order $1/S$ expansions do not select a unique quantum ground state in the pure nearest-neighbor pyrochlore Heisenberg antiferromagnet~\cite{Henley-2001,Sobral-1997,Tsunetsugu-2002,Henley-2006,Hizi-2006,Hizi-2007,Hizi-2009,Iqbal19}. In other words, pronounced manifestations of ObD in the nearest-neighbor Heisenberg pyrochlore antiferromagnet only occur when perturbing interactions already partially lift the extensive classical ground state degeneracy of that model~\cite{bramwell94,Champion-2003,Savary-2012,Zhitomirsky-2012,Oitmaa-2013,McClarty2014,Javanparast-GTO,Javanparast15,wong13,Yan-2017,chern10}.

\begin{figure}[ht!]
    \centering
     \begin{overpic}[width = 0.7\columnwidth]{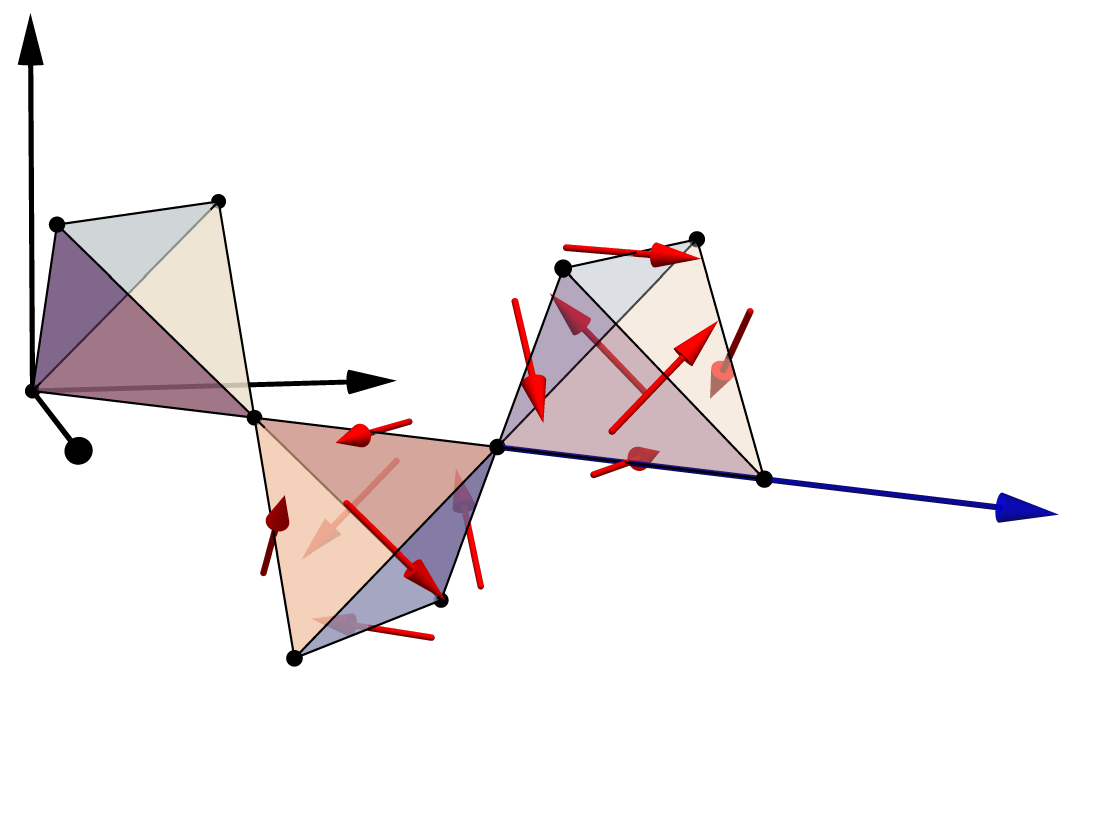}
    \put(100,27){\textcolor{blue}{$[110]$}}
    \put(42,37){$0$}
    \put(65,55){$1$}
    \put(45,52){$2$}
    \put(70,35){$3$}
    \put(1,75){$z$}
    \put(30,45){$y$}
    \put(1,30){$x$}
    \end{overpic}
\caption{Illustration of the DM vectors and sublattice labels. The DM vectors shown in the ``up'' tetrahedron (rightmost tetrahedron) correspond to $\bm D_{0\mu}$, $\bm D_{\nu 1}$ and $\bm D_{32}$ where $\mu\in\{1,2,3\}$ and $\nu\in\{3,2\}$ are sublattice indices, whereas the DM vectors shown in the ``down'' tetrahedron (middle tetrahedron) correspond to $\bm D_{\mu0}$, $\bm D_{1\nu}$ and $\bm D_{23}$ where $\mu\in\{1,2,3\}$ and $\nu\in\{3,2\}$. The DM vectors shown correspond to the direct DM interaction where $D>0$.
}
        \label{fig:pcLattice}
\end{figure}

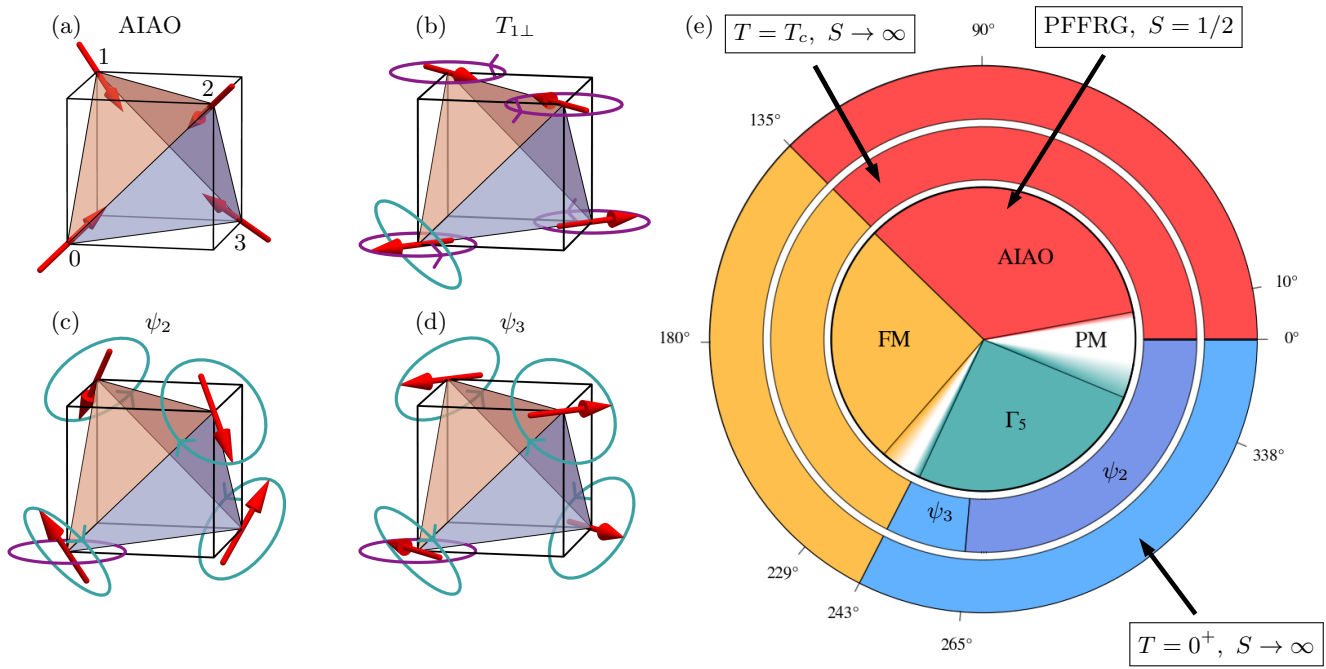
\begin{figure*}[ht!]
\centering
    \begin{overpic}[width=\textwidth]{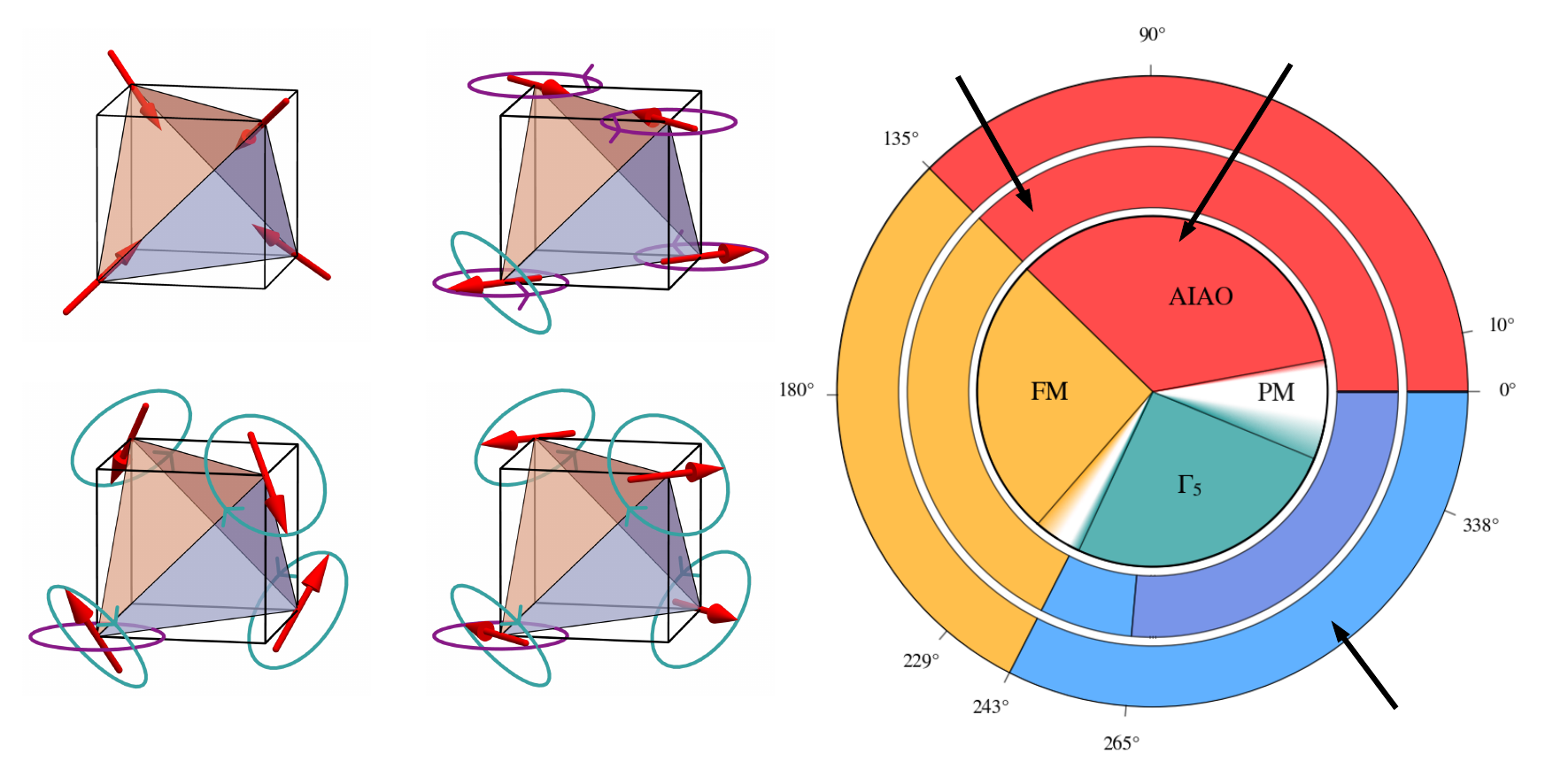}
     \put(12,26){$\psi_2$}
     \put(10,48){AIAO}
     \put(38.3,26){$\psi_3$}
     \put(38,48){$T_{1\perp}$}
     \put(55,47.5){$\boxed{T=T_c,\ S\to \infty }$}
     \put(78,48){$\boxed{\mathrm{PFFRG},\ S=1/2}$}
     \put(85,2){$\boxed{T=0^+,\ S\to \infty}$}
     \put(83,15){$\psi_2$}
     \put(70,12){$\psi_3$}

     \put(8.5,45.7){1}
     \put(16.2,43.3){2}
     \put(6.35,30.7){0}
     \put(18.6,31.9){3}

     \put(5,48){(a)}
     \put(32,48){(b)}
     \put(5,26){(c)}
     \put(32,26){(d)}
     \put(52,48){(e)}

    \end{overpic}
    \caption{
    (a)-(d) Magnetic ${\bm q}={\bm 0}$ orders for the Hamiltonian in Eq. \eqref{eq:hamiltonian}, illustrated for one ``up'' tetrahedron, where the turquoise (purple) rings in panels (b)-(d) indicate spin rotations about axes in the local (global) coordinate frame to construct the spin states defined by the $\Gamma_5$ manifold (coplanar manifold). Note that only one out of three coplanar manifolds is shown by the purple rings which is, in the present case, obtained by rotating an $x$-$y$-coplanar $\psi_3$ spin state [red arrows in panel (d)] or a $T_{1\perp}$ spin state [red arrows in panel (b)] about the perpendicular global $\hat z$ (cubic $[001]$) axis. In a similar way, the other two coplanar manifolds are defined in the global $x$-$z$ and $y$-$z$ planes, respectively. In panels (b)-(d), the two rings corresponding to the $\Gamma_5$ (turquoise) and coplanar (purple) manifolds illustrate how these two manifolds only intersect at a $\psi_3$ spin configuration. This already suggests the importance that $\psi_3$ states will play, through thermal and quantum fluctuations, in systems defined by interaction parameters that put them in a regime of competing $\Gamma_5$ and $T_{1\perp}$ orders. (e) Phase diagram of the $S=\frac{1}{2}$  Hamiltonian in Eq. \eqref{eq:hamiltonian} parameterized by $\theta$ with $J = \text{cos}(\theta)$ and $D = \text{sin}(\theta)$.
    Here, the outermost ring illustrates the $T=0^+$ order as observed in classical Monte Carlo ($S\to \infty$), the middle ring represents the selected state at $T=T_c$ for the classical model ($S\to \infty$), and the inner circle represents the $T=0$ quantum phase diagram as obtained by the pseudo-fermion functional renormalization group (PFFRG, $S= 1/2$). The white regions in the inner (PFFRG $S=1/2$) circle correspond to regions where the PFFRG method identifies an absence of conventional long-range magnetic order at $T=0$.
    This phase diagram constitutes the main result of our paper, whose details are elaborated upon in Secs. \ref{sec:model}-\ref{sec:results}.
    }
    \label{fig:phasediagram}
\end{figure*}

In this paper, we study in detail the physics of the above partial degeneracy-lifting effects by considering as perturbation to the nearest-neighbor Heisenberg exchange interactions
the Dzyaloshinskii-Moriya (DM) interaction~\cite{Dzyaloshinsky58,Moriya60,Moriya60b} -- a symmetry-allowed coupling that is inevitably present in real pyrochlore materials~\cite{Elhajal05,chern10}.
In particular, we investigate the classical and quantum ($S=\frac{1}{2}$) phase diagrams of a pyrochlore model with nearest-neighbor Heisenberg and DM interactions and resolve the subtle interplay between thermal and quantum ObD in the cases where it arises. 
We discuss in Sec.~\ref{sec:material_persp} a perspective as per the pertinence and relationship of this model with real pyrochlore materials.

The Hamiltonian for the Heisenberg antiferromagnet plus DM interactions is
\begin{equation}
\label{eq:hamiltonian}
    \mathcal{H}= \sum_{\langle ij \rangle} J {\bm S}_{i}\cdot {\bm S}_{j} + \sum_{\langle ij \rangle}{\bm D}_{ij} \cdot ( {\bm S}_{i} \times {\bm S}_{j} ) .
\end{equation}
Here, $\langle ij\rangle$ are pairs of nearest-neighbor sites on the pyrochlore lattice and ${\bm D}_{ij}$ are the so-called Moriya vectors. According to the system's symmetries and following Moriya's rules~\cite{Moriya60,Moriya60b,Elhajal05}, these vectors are defined as shown in Fig.~\ref{fig:pcLattice} where ${\bm D}_{ij}=\pm D(-1,1,0)$ for a bond $\langle ij \rangle$ along the $[110]$ cubic direction while the direction of all other vectors ${\bm D}_{ij}$ follow by applying lattice symmetries. As we will discuss below, positive and negative $D$ lead to distinctly different models realizing different magnetic phases. We adopt the previous convention of denoting $D>0$ as {\it direct} and $D<0$ as {\it indirect} DM interaction (DMI)~\cite{Elhajal05}. Furthermore, we parametrize the $J$ and $D$ couplings via
\begin{equation}
    J \equiv \cos(\theta)
    \;\;\textrm{and} \;\; 
    D \equiv  \sin(\theta) ,
    \label{eq:parametrization}   
\end{equation}
which implies that $J^2+D^2=1$ such that from now on all energy scales are given in units of $\sqrt{J^2+D^2}$ and the temperature scale is in units of $\sqrt{J^2+D^2}/k_{\textrm B}$. We investigate the full range of angles $\theta\in [0,2\pi)$ where $J$ and $D$ can both be either positive or negative. In the following two subsections, we give a brief and general overview of our paper and key results.

\subsection{Classical ground states and ordered phases}
\label{sec:classgs}

The ground state phase diagram of the Heisenberg-DM model of Eq.~(\ref{eq:hamiltonian}) shows an interesting interplay of phases already at the classical level. Besides the more conventional all-in-all-out (AIAO)~\cite{Bramwell-1998,Yan-2017,Elhajal05} and globally uniform ferromagnetic (FM) orders~\footnote{Despite the presence of anisotropic terms in the Hamiltonian~(\ref{eq:hamiltonian}), the uniform ferromagnetic state with an arbitrary direction of the magnetization is an exact eigenstate of the Hamiltonian and therefore experiences no macroscopic quantum spin fluctuations.} for $\theta\in(0\degree,135\degree)$ and $\theta\in[135\degree,243\degree)$, respectively. 
A large part of the classical ground state phase diagram for indirect DM interaction, i.e., for $\theta \in (243\degree,360\degree)$, is occupied by a phase with an unusual degenerate ground state manifold that consists of two subsets of states which we now describe.

First, a U(1) subset of the degenerate ground states is given by the so-called $\Gamma_5$ manifold characterized by a ${\bm q}=\bm 0$ propagation vector. That is, a  spin configuration that is identical for every primitive tetrahedron basis cell and where the spins order ``uniformly'' in such a manner that each lies in its {\it local} $x$-$y$ plane that is normal to the axis piercing through that site and connecting the centers of the two adjoining tetrahedra (see Appendix~\ref{appendix:configurations} for the definition of the local coordinate frame). Incidentally, these axes correspond to the directions in which the spins point in the AIAO phase 
[see Fig.~\ref{fig:phasediagram}(a)], and which correspond to the ordered phases of FeF$_3$~\cite{Ferey-1986,Sadeghi-2015}, Cd$_2$Os$_2$O$_7$~\cite{Yamaura-2012}, Na$_3$Co(CO$_3$)$_2$Cl~\cite{ Fu-2013}, Nd$_2$Zr$_2$O$_7$~\cite{Lhotel-2015} and the pyrochlore iridates~\cite{Tomiyasu-2012,Lefran-2015}.
Remarkably, the two discrete sets of so-called $\psi_2$ and $\psi_3$ states that span the $\Gamma_5$ manifold 
[see Fig.~\ref{fig:phasediagram}(c), (d)] remain energetically degenerate at the classical level for all symmetry-allowed nearest-neighbor couplings~\cite{Savary-2012}.
These  have been extensively studied as possible ground states of pyrochlore materials~\cite{Savary-2012,Zhitomirsky-2012,wong13,Champion-2003,Oitmaa-2013,Rau-Petit,Sarkis-2020} and are also the key competitors in our investigation of ObD in the Heisenberg-DM model. Second, the remaining subset of degenerate states within the 
ground state manifold for the $\theta\in(243\degree,360\degree)$ indirect DM range is given by three $\bm q=\bm 0$ coplanar U(1) manifolds with spins lying in the {\it global} $x$-$y$, $x$-$z$, or $y$-$z$ planes~\cite{Elhajal05,Canals08} which traces its origin to the previously noted degeneracy between the $\Gamma_{5}$ and a splayed ferromagnetic state~\cite{wong13}. The aforementioned $\psi_3$ states can be thought of as lying at the intersections of the $\Gamma_5$ and coplanar manifolds, see Fig.~\ref{fig:phasediagram} for details. This property already suggests that the $\psi_3$ state will play an important role in the selection effects from thermal and quantum fluctuations. The fact that the full ground state manifold is a composition of the $\Gamma_5$ and coplanar states is, indeed, a characteristic property of our model that occurs when \emph{only} indirect DM interactions, and no other nearest-neighbor anisotropic bilinear exchange interactions, perturb the Heisenberg model~\cite{wong13,Yan-2017}. In the following, we shall denote this composition of $\Gamma_5$ and coplanar manifolds as ``$\Gamma_5$/copl'' and the corresponding regime $\theta\in(243\degree,360\degree)$ in the classical ground state phase diagram where it is realized as the ``$\Gamma_5$/copl regime''. 
Due to this unusual classical ground state degeneracy, the relative simplicity of the model, but also the relevance for real materials~\cite{Onose,Mena,Riedl-2016,Sadeghi-2015,Riedl-2016,Shu-2019}, the Heisenberg-DM model on the pyrochlore lattice represents an intriguing system for studying ObD from thermal and quantum fluctuations as well as their combined, cooperating or competing, effects. 
We return to the materials perspective of the model considered in this paper, Eq.~\eqref{eq:hamiltonian}, in Sec.~\ref{sec:material_persp}.

For clarity sake, it is worth emphasizing that ObD in the model \eqref{eq:hamiltonian} may be discussed from different perspectives. First, when restricting to the classical version $(S\to \infty)$ of the model~(\ref{eq:hamiltonian}), the entropic selection that takes place at infinitely small temperatures $T=0^+$ may differ from the selection at the nonzero transition temperature, $T_c$. We indeed find this to be the case in parts of the phase diagram [compare the two outermost rings in the phase diagram of Fig.~\ref{fig:phasediagram}(e) in the $\theta\in (265\degree,360\degree)$ indirect DM range].
Second, at $T=0$, the selection via quantum fluctuations in the semiclassical regime $(1/S\ll 1)$ may differ with the extreme quantum case of $S=\frac{1}{2}$, which may likewise differ from the ObD effects operating at $T=T_c$. In particular, the question of the nature of the ordering at $T_c$ for the $S=\frac{1}{2}$ system represents a rather challenging case. Our paper aims at disentangling all these perspectives on ObD and maps out detailed classical and quantum phase diagrams of the Heisenberg-DM model. Particularly, by considering both signs of the two ($J$ and $D$) couplings, our study includes the previously unexplored parameter regime of ferromagnetic Heisenberg interactions. In order to tackle all the aforementioned tasks, we apply an effective combination of several analytical and numerical approaches including classical Monte Carlo, classical low-temperature expansion, $1/S$ spin-wave theory, pseudo-fermion functional renormalization group (PFFRG) and high-temperature series expansion (HTSE). This paper is also one of only very few studies, which treat an anisotropic three-dimensional spin system via PFFRG~\cite{Revelli-2019}. More specifically, pyrochlore models with anisotropic bilinear spin-spin interactions have, to the best of our knowledge, not previously been tackled using PFFRG.

\subsection{Summary of results}

The main results and structure of the paper are as follows: After providing a description of the model in Sec.~\ref{sec:model} and a brief discussion in Sec.~\ref{sec:method}
of the methods employed, we present our results in Sec.~\ref{sec:results}. We start with the classical model at $T=0^+$ where thermal ObD selects the $\psi_3$ state in the entire $\Gamma_5$/copl region
($243\degree \lesssim \theta < 360\degree$). A complete picture of ObD in the classical model is then provided by classical Monte Carlo simulations (Sec.~\ref{subsec:mc_results}) which indicates that in the fourth quadrant and in a small part of the third quadrant of the phase diagram, i.e., for $\theta\in[265\degree,360\degree)$, the selection at $T=T_c$ differs from the one at $T=0^+$ in that the $\psi_2$ states are chosen at $T_c$~\cite{Javanparast15,Oitmaa-2013}
[c.f. middle ring in Fig.~\ref{fig:phasediagram}(e)]. We construct a full phase diagram of the classical model in $\theta$-$T$ space, specifying the regions where either $\psi_2$ or $\psi_3$ states are selected by thermal fluctuations. 
We continue discussing the quantum model in Secs.~\ref{subsec:lsw},~\ref{subsec:pffrg}, and~\ref{sec:high-T_results}. Within the semiclassical spin-wave theory, we find that in most of the $\Gamma_5$/copl region, the $\psi_3$ order undergoes the largest reduction of the ground state energy from zero-point contributions, indicating the same $\psi_3$ selection as for the classical model at $T=0^+$. We then proceed with the study of the spin-$1/2$ quantum model which is first treated within PFFRG. This method exposes an absence of any type of magnetic instability in a regime $-9\degree \lesssim \theta \lesssim 8\degree$ around the pure Heisenberg antiferromagnet ($\theta=0$). Similarly, we find indications for a second, narrow nonmagnetic phase at $237\degree \lesssim \theta \lesssim 241.5\degree$, driven by the competition between the $\Gamma_5$, coplanar, and ferromagnetic states. Concerning ObD in the $\Gamma_5$/copl region, the PFFRG method finds that fluctuations in the $\Gamma_5$ manifold clearly dominate over fluctuations in the aforementioned coplanar manifold (which is also found in the classical model via Monte Carlo). Due to intrinsic methodological limitations in its current implementation, PFFRG is  incapable of differentiating whether $\psi_2$ or $\psi_3$ ground state order is selected. Towards addressing this limitation of PFFRG, we employ HTSE which provides indications that the $S=\frac{1}{2}$ quantum model exhibits a very similar selection at $T_c$ as in the classical model, i.e., with $\psi_2$ states chosen for $\theta\in[265\degree,360\degree)$ and $\psi_3$ states chosen for $\theta\in(243\degree,265\degree)$, see Fig.~\ref{fig:phasediagram}(e). Overall, one may conclude that in both the classical and quantum models, the ObD effects at $T=0^{+}$ and $T=T_c$ are distinctly different (at least in the fourth quadrant with antiferromagnetic Heisenberg exchange, $J>0$, and indirect DMI, $D<0$), selecting $\psi_3$ and $\psi_2$ states, respectively. On the other hand, for most of the phase diagram, selection effects of the classical and quantum models resemble each other both at $T=0^{+}$ and $T=T_c$.

\section{\label{sec:model}
Materials perspective, Model \& irreducible representations}

\subsection
{\label{sec:material_persp}
{Materials perspective}
}

As mentioned in the Introduction, the anisotropic DM spin-spin interaction of magnitude $D$~\cite{Dzyaloshinsky58,Moriya60,Moriya60b} is symmetry-allowed and expected to be nonzero in real pyrochlore materials~\cite{Elhajal05,chern10}.
The DM interaction originates from spin-orbit interactions and can be obtained by extending the Anderson theory of superexchange to include spin-orbit coupling~\cite{Moriya60}. It can also be obtained starting from a Hubbard model~\cite{Coffey,Coffey_err}.
More generally, spin-orbit interaction will also lead to anisotropic exchange beyond the DM interactions, such as symmetric pseudo-dipolar couplings, but weaker than the DM interaction, as originally derived by Moriya~\cite{Moriya60}. For the pyrochlore lattice, there are four symmetry allowed nearest-neighbor bilinear anisotropic spin-spin couplings of the form $J_{ij;\alpha\beta} S_{i}^{\alpha} S_{j}^{\beta}$~\cite{McClarty-2009,thompson_2011,ross11,Yan-2017}.
Most commonly, these $J_{ij;\alpha\beta}$ are expressed 
 either using a global Cartesian frame~\cite{Yan-2017} or a local quantization frame~\cite{McClarty-2009,ross11}. However, they can also be written in a form that exposes  
 isotropic Heisenberg exchange $J$, DM coupling $D$,
a symmetric pseudo-dipolar coupling, and an Ising coupling between the local Ising (cubic $\langle 111\rangle$) directions~\cite{McClarty-2009,thompson_2011}.

In this paper, we focus on the Heisenberg and DM interactions, and ignore the other two anisotropic bilinear couplings allowed by symmetry.
From a real materials perspective, one would then contend that we are {\it de facto} focusing on $3d$ transition metal ions where the spin-orbit interaction can be treated perturbatively with the DM interaction being the leading anisotropic bilinear coupling~\cite{Moriya60}, with $D/J\lesssim O(10^{-1})$.
This situation may be relevant to compounds such as 
FeF$_3$, Lu$_2$V$_2$O$_7$, and NaCaNi$_2$F$_7$, each with $3d$ magnetic ions (Fe$^{3+}$, Ni$^{2+}$, V$^{4+}$, respectively) residing on a pyrochlore lattice. 
This expectation is corroborated by density-functional theory  (DFT) calculations for FeF$_3$~\cite{Sadeghi-2015} and Lu$_2$V$_2$O$_7$~\cite{Riedl-2016}, as well as inelastic neutron scattering measurements for the latter~\cite{Mena}, and fits to diffuse neutron scattering for NaCaNi$_2$F$_7$~\cite{Shu-2019}.

Let us briefly comment on the pertinence of our paper for each of the above three materials.
First, note that $S=\frac{5}{2}$ for Fe$^{3+}$ in FeF$_3$, $S=1$ for Ni$^{2+}$ in NaCaNi$_2$F$_7$, both with antiferromagnetic nearest-neighbor Heisenberg exchange ($J>0$), while  $S=\frac{1}{2}$ for the Lu$_2$V$_2$O$_7$ ferromagnet ($J<0$).
For such $3d$ systems, one expects, very roughly, $\cramped{| \theta|  \lesssim 10 ^{\degree}}$ for antiferromagnets and 
$\cramped{\theta \lesssim 180^{\degree}\!\pm 10^{\degree}}$ for ferromagnets.
In this case, the $S=\frac{1}{2}$ Lu$_2$V$_2$O$_7$ compound is predicted to sit robustly in the ferromagnetic phase of the inner $S=\frac{1}{2}$ circle of the phase diagram in Fig.~\ref{fig:phasediagram}(e) and far removed from the strongly frustrated regime at  $237\degree \lesssim \theta \lesssim 241.5\degree$.
 In this paper, we consider only the classical $S=\infty$ or the $S=\frac{1}{2}$ cases. As such, we shall not be able to comment on the specific location of the 
 $S=\frac{5}{2}$ FeF$_3$ and $S=1$ NaCaNi$_2$F$_7$ 
 antiferromagnets in the phase diagram of
 Fig.~\ref{fig:phasediagram}(e). However, one may speculate that with $S=5/2$, FeF$_3$ would be well approximated by a classical model.  Within such a description, the DM interaction  for FeF$_3$ is predicted to be direct ($D>0$)~\cite{Sadeghi-2015} and responsible for the AIAO ordered phase found experimentally~\cite{Ferey-1986}. However, the situation is presumably more complicated due to the predicted non-negligible bi-quadratic exchange of the form $B({\bm S}_i\cdot {\bm S}_j)^2$ in this compound~\cite{Sadeghi-2015}.
 For the $S=1$ NaCaNi$_2$F$_7$ antiferromagnet, one may anticipate  that whether it falls within a regime lacking long-range order or not would depend rather precisely on the magnitude of the various nearest-neighbor anisotropic spin-spin couplings, including the DM interaction~\cite{Shu-2019}, as well as exchange beyond nearest neighbors~\cite{Iqbal19}. 
It would be interesting to incorporate all these aspects of the spin model of NaCaNi$_2$F$_7$ in a future study and investigate how close the resulting model is to a quantum disordered (spin liquid) phase. However, we note that the Na/Ca site disorder in NaCaNi$_2$F$_7$ will generate random exchange, which is presumably at the origin of the spin-freezing observed in this material~\cite{Shu-2019}, hence complicating comparison between experiments and theory.

Finally, we note that there is a paucity of $S=\frac{1}{2}$ transition metal ion pyrochlore antiferromagnets.
One notable example is Lu$_2$Mo$_2$O$_5$N$_2$ in which the magnetic Mo$^{5+}$ $S=\frac{1}{2}$ ions reside on a pyrochlore lattice~\cite{Clark-Lu2Mo2O5N2}. 
Density functional theory calculations provide an estimate of the Heisenberg exchange couplings up to third nearest neighbors for this material as well as point to the presence of an indirect DM ($D<0)$ coupling with an estimated magnitude $|D|/J\sim 10\%$~\cite{Iqbal-2017}.
Considering only this estimate of $|D|/J$ would, on the basis of the PFFRG calculations reported later in the paper, put Lu$_2$Mo$_2$O$_5$N$_2$ in the nonmagnetic (white) region of the phase diagram in  Fig.~\ref{fig:phasediagram}(e).
In contrast, previous PFFRG calculations ~\cite{Iqbal-2017} that did not incorporate DM interactions found that, while exchange beyond nearest neighbors was not sufficiently strong to drive long-range order, those induced significant highly-structured ``gearwheel'' like correlations in the paramagnetic phase. This suggests either a proximity of the system to an instability towards a long-range ordered phase or could be reflective of a valence-bond-crystal stabilized by quantum melting of the parent classical spiral order, an observation made on other lattices~\cite{Ferrari-2017,Kiese-2023}.
How DM interactions of order $|D|/J\sim 10\%$ compete or cooperate with exchanges beyond nearest neighbor to still allow for a disordered phase or, instead, lead to a long-range ordered phase is an interesting question that goes beyond the scope of the present paper but would be worthwhile to investigate. 
We note, however, that single crystals of Lu$_2$Mo$_2$O$_5$N$_2$ remain to this day unavailable. Definite progress in understanding the properties of Lu$_2$Mo$_2$O$_5$N$_2$ and quantitative comparison with theory must
await the synthesis of high-quality single crystals of that compound.




To summarize, consideration of the model defined by Eq.~\eqref{eq:hamiltonian} is, naively, more physically pertinent for systems with $3d$ transition metal ions where the spin-orbit interactions give a leading perturbative DM interaction~\cite{Moriya60}. In such a situation, one might naively expect $\theta$ to be within $\pm 10\degree$ 
about $\theta=0\degree$ for antiferromagnets ($J>0$), and 
about $\theta=180\degree$ for ferromagnets ($J<0$), and 
thus far away from the interesting (seemingly) magnetically disordered white sector separating the FM and $\Gamma_5$ phase in the phase diagram of  Fig.~\ref{fig:phasediagram}(e). 
On the other hand, from an entirely different perspective, 
rare-earth pyrochlore systems such as $R$$_2$(Ti, Sn, Hf, Zr, Ge)$_2$O$_7$, with $R$ a $4f$ transition metal ion ($R=$Ce, Sm, Er, Yb, etc), are described by \emph{effective} pseudospin-1/2 degrees of freedom interacting via the aforementioned four symmetry-allowed bilinear couplings 
$J_{ij;\alpha\beta}$~\cite{ross11,Yan-2017,Rau-2019}, which are generically of similar magnitude~\cite{Savary-2012,Sarkis-2020}, and not solely the $J$-$D$ couplings of Eq.~\eqref{eq:hamiltonian}.
However, rather interestingly, effective spin-1/2 models with dominant Heisenberg exchange and DM interactions in rare-earth ytterbium-based (breathing) pyrochlore magnets are also possible and quantitatively relevant~\cite{Rau-octa,Rau-BYZO,Dissanayake-2022}.
Considering $S=\frac{1}{2}$ systems, the present study is an important bridge between the strictly isotropic pyrochlore  systems studied before~\cite{Iqbal-2017,Iqbal19} and the more general models with effective $S=\frac{1}{2}$ pseudospins with general anisotropic nearest-neighbor $J_{ij;\alpha\beta}$ couplings~\cite{ross11,Yan-2017,Rau-2019}.
More pragmatically, our paper constitutes a detailed study of systems that reside on or near a ``hyperplane cut'' of the phase diagram in this four-dimensional $J_{ij;\alpha\beta}$ parameter space, one which varies the Heisenberg exchange $J$, and the DM interaction $D$, but sets to zero both the bilinear Ising coupling and the pseudo-dipolar coupling. From previous works~\cite{wong13,Yan-2017}, we already know that there exists interesting phase competition and phase boundaries in that hyperplane, but the details of ObD and of the competing phases, the latter in particular for the quantum $S=\frac{1}{2}$ case, have not been investigated in much detail.
This is the key motivation for our paper which we report below, and henceforth not claiming any (current) direct relevance to any specific real materials.

\subsection{Model}

Having discussed the relevance of our paper for real materials, from now on we consider the pyrochlore lattice with nearest-neighbor Heisenberg and DM interaction as given by the Hamiltonian in Eq.~(\ref{eq:hamiltonian}). The pyrochlore lattice, shown in Fig.~\ref{fig:pcLattice}, is a face-centered cubic (fcc) space lattice with a four-site basis. Basis sites are positioned at the origin of the unit cell and the midpoints of the fcc lattice primitive vectors which we specify as ${\bm {a}}_1=\frac{1}{2}(0,1,1)$, ${\bm {a}}_2=\frac{1}{2}(1,0,1)$ and ${\bm {a}}_3=\frac{1}{2}(1,1,0)$. 
Nearest-neighbor bonds build up the lattice from corner-sharing tetrahedra. This structure leads to high geometrical frustration for nearest-neighbor antiferromagnetic Heisenberg interactions~\cite{Villain-1979,Reimers-1991a,Moessner-1998b}.
Enforcing the Hamiltonian to be invariant under lattice symmetries specifies the orientation of $\bm{D}_{ij}$ for each nearest-neighbor bond. This allows us to adopt the parametrization given by Eq.~(\ref{eq:parametrization}) with the DM vector ${\bm D}$ for the pair of sublattices $0$ and $1$ illustrated in Fig.~\ref{fig:pcLattice} having the orientation ${\bm{D}}_{01} = D(0,-1,1)$. The DM vectors for all other nearest-neighbor bonds within one tetrahedron are shown in Fig.~\ref{fig:pcLattice} as well. The restricted DM vector orientation stems from its microscopic origin. The DMI is a bilinear spin interaction that is antisymmetric under the exchange of site arguments and originates from spin-orbit coupling~\cite{Moriya60} and, as a result,  the interaction is anisotropic in spin space. We now briefly discuss in the next three paragraphs the symmetry properties of ${\bm D}_{ij}$. 
 
Depending on the underlying states of a magnetic ion on which an effective spin is defined, the effective spin can transform differently under the application of lattice symmetries~\cite{Huang14}. One can consider the general case that the spins transform under the application of a lattice symmetry as  $\bm{S}_{i} \rightarrow \bm{U} \bm{S}_{i}$. In a lattice Hamiltonian consisting of general bilinear spin interaction terms $\bm{S}_{i}^{T} \bm{M}_{ij} \bm{S}_{j}$, with $3\times3$ interaction matrices $\bm{M}_{ij}$, the spin transformation associated with a lattice symmetry can also be recast as a transformation of the interaction matrices. A term with an antisymmetric $\bm{M}_{ij}$ can be written in the form of the DM interaction term in Eq.~\eqref{eq:hamiltonian}. Lattice symmetries then transform the DM vector $\bm{D}_{ij}$, restricting the allowed DMI. That is, for a DMI to be non-vanishing, the antisymmetric form of the DM interaction requires that the center of the corresponding bond is not a center of lattice inversion symmetry~\cite{Moriya60}.

In the pyrochlore lattice, we can transform each of the twelve nearest-neighbor bonds, inequivalent by pure lattice translation symmetries, onto each other by applying lattice symmetries of a C3 rotation about the [111] axis, C2 symmetry about the [001] axis, and inversion symmetry. Consequently, and as stated above, the DMI of each nearest-neighbor bond is uniquely identified after specifying it for one single bond. As we assume the spin to transform as a magnetic dipole, it is invariant under inversion whereas the $C_3$ rotation permutes spin components cyclically.
 
By employing the inversion and $C_2$ rotation symmetry about the $[0 \bar 1 1]$ axis passing through the origin in Fig.~\ref{fig:pcLattice}, one finds that the $01$-bond maps onto itself. The spin components $S_y$ and $S_z$ permute under this operation. Restricting to antisymmetric bilinear interactions that are invariant under such a permutation, the DM vector orientation on this bond is fixed as $\bm{D}_{01} = D(0,-1,1)$ with the magnitude and sign of $D$ unconstrained.

\subsection{States classification} 
\label{subsection:state_class}

The possible $\bm q=\bm 0$ orders allowed by Kramers ions for a nearest-neighbor pyrochlore quadratic Hamiltonian have been widely studied and classified according to their transformation under the tetrahedral group $T_d$~\cite{Yan-2017,McClarty-2009,Curnoe-2007}. This classification is most readily manifested by rewriting the quadratic Hamiltonian as a sum of separate
interaction terms contributed by each 
(``up'' and ``down'') tetrahedron 
\begin{equation}
    \mathcal{H}=\sum_{\boxtimes} \mathcal{H}^{\boxtimes}=\sum_{\boxtimes}\sum_{i,j\in\boxtimes }\bm S_i^T \bm M_{ij} \bm S_j .
    \label{eq:single-tet-Hamiltonian}
\end{equation}
Here, $\mathcal{H}^{\boxtimes}$ is referred to as the single tetrahedron Hamiltonian and ${\bm M}_{ij}$ corresponds to the spin-spin interaction matrix. This matrix is invariant under the action of any transformation in $T_d$ and can therefore be decomposed according to the irreducible representations (irreps) of this group. For the Hamiltonian~\eqref{eq:hamiltonian}, the irreps for $T_d$ are labeled as $A_2$, $E$, $T_{1\parallel}$, $T_{1\perp}$ and $T_2$, 
describing subspaces of dimension $ 1,\ 2,\ 3,\ 3$ and $3$,  respectively. Using the irreps decomposition, the single tetrahedron Hamiltonian in Eq.~\eqref{eq:single-tet-Hamiltonian} can be expressed in the following way:
\begin{eqnarray}
\mathcal{H}^{\boxtimes}=\frac{1}{2}&& \Big [
a_{A_2}\left(m^\boxtimes_{A_2}\right)^2+a_{E}\left(\bm{m}^\boxtimes_{E}\right)^2+a_{T_2}\left(\bm m^\boxtimes_{T_2}\right)^2\nonumber\\
    &&+a_{T_{1\parallel}}\left(\bm m^\boxtimes_{T_{1\parallel}}\right)^2+a_{T_{1\perp}}\left(\bm m^\boxtimes_{T_{1\perp}}\right)^2\Big],
\end{eqnarray}
where the $\{a_I\}$ interaction parameters are linear functions of $J$ and $D$, and are defined by the $J$ and $D$ dependence of the elements of  ${\bm M}_{ij}$ 
(see Appendix~\ref{appendix:Irrep_Energy_Parameters} for more details). These parameters correspond to the energy of a spin configuration defined by the $I$th irrep, and $\bm  m^\boxtimes_{I}$ is the local (irrep) order parameter for tetrahedron $\boxtimes$ transforming under the $I$th irrep. Within this formalism, the $A_2$ irrep corresponds to an AIAO state, the $E$ irrep corresponds to $\Gamma_5$ 
(separated into the $\psi_2$ and $\psi_3$ states, or local $x$ and $y$ basis vectors \cite{Poole_2007}), the $T_2$ irrep describes the Palmer-Chalker (PC) state~\cite{Palmer-Chalker}, the $T_{1\parallel}$ corresponds to a simple colinear ferromagnet, and the $T_{1\perp}$ to certain types of coplanar states, which we denote as $T_{1\perp}$ states, and discuss them further below.  See Figs.~\ref{fig:phasediagram}(a-d) for an illustration of the spin configurations for the AIAO, $T_{1\perp}$, $\psi_2$ and $\psi_3$ states. A PC state can be visualized by flipping one of the two pairs of anti-aligned spins of $\psi_3$. 
The $E$ irrep, of particular interest in our paper, defines a two dimensional manifold parameterized by the rotation of each spin about its \textit{local} $z$ axis resulting in states which can be decomposed as a normalized linear combination of the $\psi_2$ and $\psi_3$ states, see the turquoise circles in Fig.~\ref{fig:phasediagram}. For a further description of the states in terms of the individual spin components, we refer the reader to Appendix~\ref{appendix:configurations}.
 
\begin{figure}[ht]
\centering
\begin{overpic}[width=1.\columnwidth]{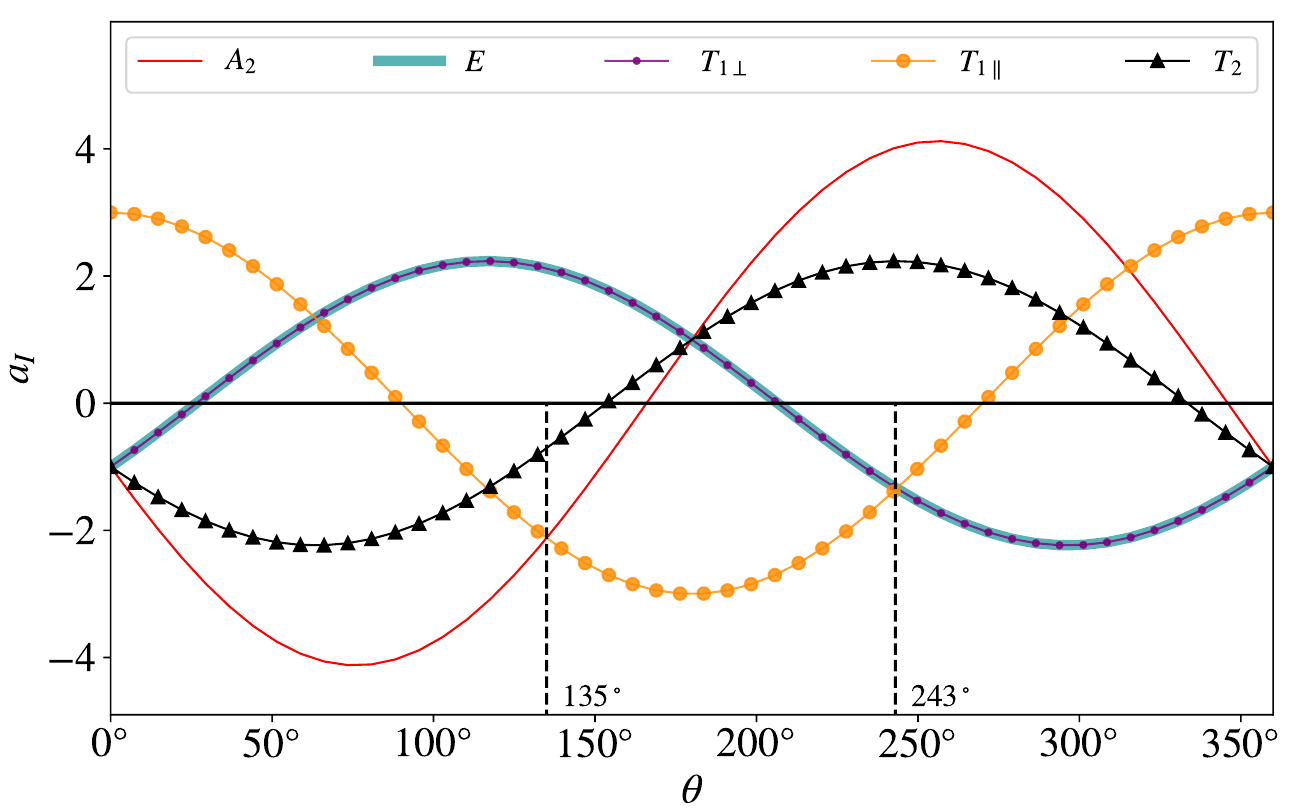}
\end{overpic}
\caption{Irreducible representation's energy parameter (IEP) $a_{I}$ as a function of $\theta$. Note that the critical angles $\theta=135\degree,243\degree$ for the phase boundaries between the all-in/all-out and the colinear ferromagnet (FM), and the latter with the $\Gamma_5$ phase, correspond to the ratios $D/J=-1$ and $2$, respectively.
Note also that the $a_I$ for the $E$ and $T_{1\perp}$ irreps are degenerate for all values of $\theta$ (purple dots and blue line overlapping).
}
\label{fig:Irreps}
\end{figure}

Following the irrep decomposition, the irrep with the lowest energy (i.e. the smallest $a_I$) is expected to describe the classical ground state of the system~\cite{Yan-2017}. Moreover, within a Ginzburg-Landau theory, the coefficient of the quadratic term in the expansion of the free energy in terms of the irreps as (competing) order parameters is proportional to $(3T+a_I)$, where $T$ is the temperature~\cite{Reimers-1991a,Enjalran}. As such, a smaller (more negative) $a_I$ results in a higher mean-field second order transition temperature (i.e. when $3T+a_I=0$) into the corresponding long-range ordered irrep state. We henceforth refer to a given $a_I$ as an irrep energy parameter, or IEP. Using the parametrization in Eq.~\eqref{eq:parametrization}, the five IEPs, and hence the corresponding spin configuration energy per tetrahedron, are plotted as a function of $\theta$ in Fig.~\ref{fig:Irreps}; for further details, we refer the reader to Appendix~\ref{appendix:Irrep_Energy_Parameters}.

The identification of the minimum IEP reveals three distinct regions in parameter space: two regions defined by $\theta \in (0\degree,135\degree)$ and $\theta \in (135\degree,243\degree)$ where the lowest energy configuration is specified by a single irrep (the $A_2$ and $T_{1\parallel}$, respectively). A third region defined by $\theta \in (243\degree,360\degree)$, where \emph{two irreps} ($E$ and $T_{1\perp}$) have identical $a_I$ (see overlapping $a_E$ and $a_{T_{1\perp}}$ IEPs in Fig.~\ref{fig:Irreps}), corresponds to a previous observation~\cite{wong13} that the Hamiltonian in Eq.~\eqref{eq:hamiltonian} resides on the phase boundary
between the $\Gamma_5$ phase and a splayed FM phase.
Furthermore, from an analysis of the resulting quadratic Ginzburg-Landau theory expressed in terms of the irreps, identical order parameter susceptibilities for the $E$ and $T_{1\perp}$ irreps are expected at high temperatures, i.e. $\lim_{T\rightarrow \infty} \chi_{E}= \lim_{T\rightarrow \infty} \chi_{T_{1\perp}}$. This equivalence is explored in Secs.~\ref{subsec:mc_results} and \ref{subsec:pffrg} for the classical and quantum cases, respectively, and further illustrated in Appendix~\ref{appendix:HTSE_proof} with a HTSE to order $\beta^2$. 
Additionally, as discussed in Sec.~\ref{sec:classgs}, the 
degenerate $E$ and $T_{1\perp}$ irreps combine to form the so-called coplanar \textit{manifold}~\cite{Canals08}. The spin configurations in this manifold, illustrated in Fig.~\ref{fig:phasediagram} by the purple circles, are a superposition of $\psi_3$ and $T_{1\perp}$ states. More specifically, for $\psi_3$ and $T_{1\perp}$ spin configurations  residing within the \textit{same} crystallographic plane of the conventional cubic unit cell, the $\mu$th spin of a state in the coplanar manifold is defined by 
\begin{equation}
\bm S^{\rm coplanar}_\mu(\gamma)=\cos(\gamma)\bm S^{\psi_{3}}_\mu+\sin(\gamma)\bm S^{T_{1\perp}}_\mu .
\label{eq:psi3-t1p}
\end{equation}
Here, $\bm{S}^{\psi_{3}}_\mu$ and $\bm{S}^{T_{1\perp}}_\mu$ denote the spin orientations on sublattice $\mu$ in each of the respective two types of irrep order. The angle $\gamma$ parametrizes the coplanar manifold and denotes a rotation about the global cubic axis perpendicular to the corresponding cubic plane where the $\psi_3$ and $T_{1\perp}$ orders are defined. 
In Fig.~\ref{fig:phasediagram}(b), this corresponds to the displayed purple circles perpendicular to the global cubic $[001]$ direction (not shown). 
The realization of this manifold is only possible because a $\psi_3$ state can be continuously converted into a $T_{1\perp}$ state in a way that preserves the energy of the resulting states, as described by Eq.~\eqref{eq:psi3-t1p}, with no other irreps involved.


\section{\label{sec:method}Methods}

In this section, we discuss the various methods we used to explore the classical and quantum properties of the Heisenberg-DM model of Eq.~\eqref{eq:hamiltonian} at zero and finite temperature. The reader mostly interested in the results can jump directly to Sec.~\ref{sec:results}.

\subsection{Classical Monte Carlo}
\label{subsec:class_mc}

To investigate the finite temperature behavior of the classical version of the model Hamiltonian in Eq.~(\ref{eq:hamiltonian}), we carried out classical Monte Carlo simulations with three-component (Heisenberg) spins of unit length $\vert {\bm S}_i\vert =1$. Calculations were performed on a periodic system of $N=4L^3$ spins with a simple Metropolis single-spin-flip algorithm supplemented with an over-relaxation update \cite{Creutz,Zhitomirsky-2012} where we used about $10^5$ sweeps to thermalize the system and $2\times10^5$ sweeps to measure the thermodynamic quantities discussed in this subsection. Furthermore, to ensure good statistics leading to a smooth behavior in all the thermodynamic quantities measured~\cite{binder1993monte}, the various sampled thermodynamic data were averaged over $20$ to $100$ independent Monte Carlo simulations (runs).
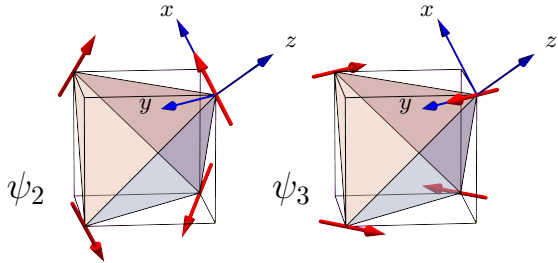
\begin{figure}[t!]
    \centering
    \begin{overpic}[width=0.8\columnwidth]{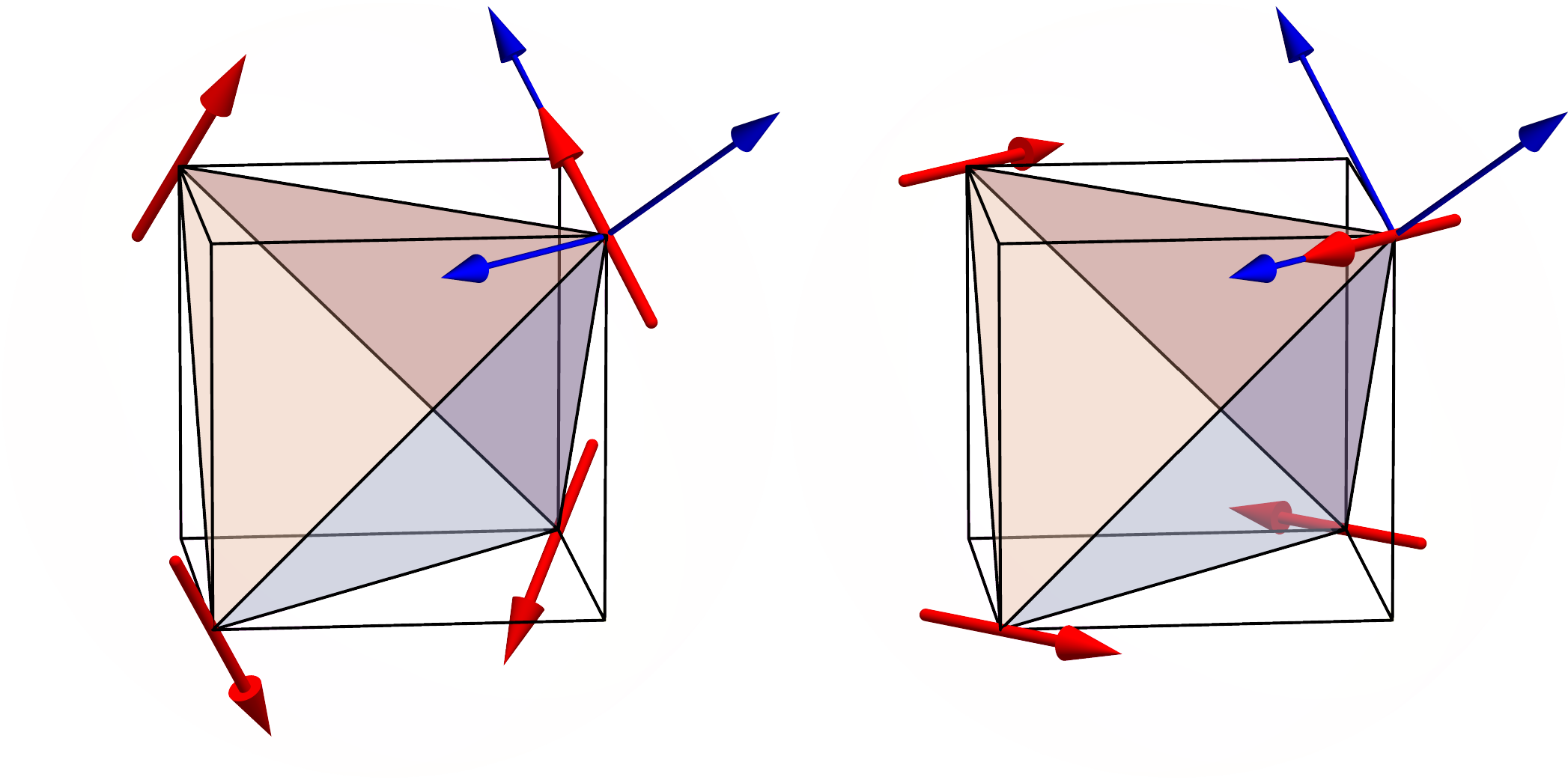}
     \put(48,15){\fontsize{14}{16} $\psi_3$}
     \put(-3,15){\fontsize{14}{16} $\psi_2$}
     \put(28,50){$x$}
     \put(24,32){$y$}
     \put(52,44){$z$}
      \put(78,50){$x$}
     \put(74,32){$y$}
     \put(102,44){$z$}
    \end{overpic}
    \caption{$\psi_2$ (left) and $\psi_3$ (right) where the local moments point along the local $x$ and $y$ axes for $\psi_2$ and $\psi_3$, respectively. Consider this figure along with Fig.~\ref{fig:phasediagram}(c) and (d).}
    \label{fig:gamma_5_local_basis}
\end{figure}

To expose the onset of the anticipated  $\bm{q}=\bm{0}$ orders, we compute global instantaneous system-averaged  $\bm m_{I,{\textrm{av}}}$ for different irreps $I$. To do so, we define an average over local irrep modes $\bm{m}_I^\boxtimes$ on all tetrahedra labeled by $\boxtimes$,
\begin{equation}
\bm m_{I,{\textrm {\textrm{av}}}}\equiv \frac{4}{N}\sum_\boxtimes \bm m_{I}^\boxtimes[\{\bm S\}],
\end{equation}
where $N$ is the number of pyrochlore sites, and $m_{I}^\boxtimes[\{\bm S\}]$ represents the numerical value of the $\bm m_I^\boxtimes$ irrep mode on a tetrahedron associated with a Monte Carlo measured configuration $\{\bm S\}$. We refer the reader to Appendix~\ref{appendix:configurations} for further details regarding the definition of $\bm m_I^\boxtimes$. Particularly relevant for the $\theta \in (243\degree,360\degree)$ range is the investigation of possible state selection (ordering) within the $\Gamma_5$ manifold, for which the corresponding global two-dimensional irrep mode has the form $\bm m_{E,{\textrm {av}}}= (m_{E,\rm av}^x,m_{E,{\textrm {av}}}^y)$ with
\begin{eqnarray}
    m_{E,{\textrm {av}}}^x&=& \frac{1}{N}\sum_{i} S^x_i,\label{eq:Mex}\\
    m_{E,{\textrm {av}}}^y&=&  \frac{1}{N}\sum_{i}S^y_i.\label{eq:Mey}
\end{eqnarray}
Here, $S^x_i$ and $S^y_i$ are the \textit{local} $x$ and $y$ components of a spin ${\bm S}_i$ at site $i$ (see Fig.~\ref{fig:gamma_5_local_basis}). Upon averaging over different independent system configurations (i.e. Monte Carlo runs), henceforth indicated by $\langle\cdots\rangle$, we obtain the Monte Carlo order parameter $m_E$ defined by
\begin{equation}
    m_E=\langle |\bm{m}_{E,{\textrm {av}}}|\rangle .
\end{equation}
Below, $m_E$ will serve as a diagnostic quantity to probe whether the system orders in the $\Gamma_5$ manifold in the first place. However, since $m_E$ cannot distinguish between the $\psi_2$ and $\psi_3$ states, we also compute the average angles of the spin orientations in the local $x$-$y$ plane \cite{Zhitomirsky-2012,wong13,AndradePRL2018} for individual Monte Carlo configurations
\begin{equation}
    \phi= \arctan\left(m_{E,{\textrm {av}}}^y/m_{E,{\textrm {av}}}^x\right) ,
    \label{eq:phidef}
\end{equation}
along with the order parameter
\begin{eqnarray}\label{eq:me6}
m_{E6}=\langle |\bm{m}_{E,{\textrm {av}}}|\cos(6\phi)\rangle.
\end{eqnarray}
With this parameter, we use a simple approach for labeling the $\psi_2$ and $\psi_3$ orders in the phase diagram of Fig.~\ref{fig:Phase_diagram_Gamma_5_T} below, by assigning $\psi_2$ order ($\psi_3$ order) to those regions in the phase diagram where $m_{E6}>0$ ($m_{E6}<0$). This approach, which is in line with the one employed in Refs.~\cite{AndradePRL2018,chern10}, has a potential
caveat as a
full characterization of the ordering within the $\Gamma_5$ manifold requires the probability distribution function $P(\phi)$ of angle $\phi$ for the sampled Monte Carlo configurations. The $P(\phi)$ distribution signals a $\psi_2$ order ($\psi_3$ order) if it exhibits \emph{sharp} maxima at $\phi=n\pi/3$ ($\phi=n\pi/3+\pi/6$), where $n\in\{0,1,2,3,4,5\}$~\cite{AndradePRL2018,Yan-2017}.
Previous work~\cite{AndradePRL2018} noted that the first term in a free energy theory which differentiates the $\Gamma_5$ states  (i.e. the $m_{E6}$ term) is dangerously irrelevant, resulting in an emergent length scale $\lambda(T)$ for which a robust 
(rigid) ObD selection only takes place when the system size $L \gg \lambda$. Consequently, because of computing limitations in terms of accessible system size and adequate statistics, this type of analysis becomes increasingly difficult near phase boundaries where the probability distribution $P(\phi)$ is rough and less well-peaked due to enhanced thermal fluctuations. 
As an example, we illustrated a histogram representation for $P(\phi)$ for $\theta=300\degree$ ($D/J=-{\sqrt 3}$) in Appendix~\ref{appendix:distribution}, where we also discuss how $P(\phi)$ evolves as a function of temperature and how this function is affected by the system size considered in our simulations.


Furthermore, to elucidate the aforementioned competition between the $E$ and the $T_{1\perp}$ irreps ($\Gamma_5$ and $T_{1\perp}$ ordering, respectively), we calculate the order parameter susceptibilities associated with each of these two states, employing the standard form used in Monte Carlo simulations of finite-size classical three-component spin systems~\cite{Landau-Binder}:
\begin{equation}
    \label{eq:IrrepChi}
    \chi_{ I}=\frac{\langle \bm m_{I,{\textrm {av}}}^2\rangle-\langle |  \bm  m_{I,{\textrm {av}}}|\rangle^2 }{T},
\end{equation}
where, $I$ is the irrep label of interest, $I= E$ or $T_{1\perp}$.

\subsection{Classical low temperature expansion}

To study the ObD selection within the $\Gamma_5$ manifold arising from thermal fluctuations at temperatures much lower than the critical temperature $T_c$, we study the entropic weight of the spin fluctuations about the $\psi_2$ and $\psi_3$ configurations \cite{McClarty2014,Yan-2017}. To this end, we consider the Hamiltonian in Eq.~\eqref{eq:hamiltonian} and express the spin orientation in terms of quadratic fluctuations about the reference  spin configuration as 
\begin{equation}
    \bm S_\mu\simeq \left(\delta S^{\tilde x}_\mu,\, \delta S^{\tilde y}_\mu, \, S-\frac{(\delta S^
{\tilde x}_\mu)^2}{2S}-\frac{(\delta S^{\tilde y}_\mu)^2}{2S}\right),\label{eq:CLTE_spins}
\end{equation}
where the subindex $\mu=0,1,2,3$ labels the different sublattices (see Fig.~\ref{fig:pcLattice}), the local ${\tilde z}$ component is defined along the ordered spin direction, and $\delta S^\alpha$ with $\alpha\in\{{\tilde x},{\tilde y}\}$ are the (local) transverse spin fluctuations. Inserting Eq.~\eqref{eq:CLTE_spins} into the Hamiltonian~\eqref{eq:hamiltonian} leads to an effective Hamiltonian for the spin fluctuations 
\begin{equation}
    \mathcal{H}_{\rm eff}=\mathcal{E}_0+\frac{1}{2}\sum_{\mu,\nu}\sum_{\alpha,\beta}\sum_{\bm q} \delta S_\mu^\alpha(-\bm q)H_{\mu\nu}^{\alpha\beta}(\bm q)\delta S_\nu^\beta(\bm q),
\end{equation}
where $\mathcal{E}_0$ is the classical energy associated with the ordered configuration, and $H_{\mu\nu}^{\alpha\beta}$ is the Hessian matrix that results from the second derivatives of the classical energies of the $\psi_2$ and $\psi_3$ states with respect to the transverse spin fluctuations. With the above effective Hamiltonian, the entropic weight associated with the fluctuations about the chosen configuration is 
\begin{equation}
\label{eq:entropy}
S(\psi)=\mathrm{constant} - \frac{1}{2}\sum_{\bm q}\ln\left(\det H(\bm q )\right),
\end{equation}
where $\psi$ labels the ordered configuration about which the fluctuations are considered. The entropic selection within the $\Gamma_5$ manifold between the $\psi_2$ and $\psi_3$ states can be exposed by computing the entropy difference, $\Delta S \equiv S(\psi_3)-S(\psi_2)$, where $\Delta S>0$ signals a $\psi_3$ selection while $\Delta S<0$ signals a  $\psi_2$ selection.

\subsection{Quantum $1/S$ spin waves}

To resolve the ground state configuration selected by quantum ObD at $T=0$ within the $\Gamma_5$ manifold, we perform a linear spin-wave calculation to determine the zero-point energy correction about both the $\psi_2$ and $\psi_3$ classical spin configurations~\cite{Savary-2012}. In this formalism, the spin operators are rotated into the local ordering directions and expressed in terms of Holstein-Primakoff bosons \cite{Maestro_2004,wong13,ross11,Yan-2017}

\begin{eqnarray}
\hat{S}_{\bm R\mu}^{z}&=&S-\hat{c}_{\bm R\mu}^\dagger\hat{c}_{\bm R\mu},\\
\hat{S}_{\bm R\mu}^{+}&=&(2S-\hat{c}_{\bm R\mu}^\dagger\hat{c}_{\bm R\mu})^{1/2}\hat{c}_{\bm R\mu}\simeq \sqrt{2S}\hat{c}_{\bm R\mu} ,\\
\hat{S}_{\bm R\mu}^{-}&=&\hat{c}_{\bm R\mu}^\dagger(2S-\hat{c}_{\bm R\mu}^\dagger\hat{c}_{\bm R\mu})^{1/2}\simeq \sqrt{2S}\hat{c}_{\bm R\mu}^\dagger,\label{eq:HP_transform}
\end{eqnarray}
where $\hat{S}^z_{\bm R\mu}$ is the rotated spin operator into the local ordering direction of the spin on sublattice $\mu$ in the tetrahedron centered at the fcc lattice vector $\bm R$. Introducing this transformation in Eq.~\eqref{eq:hamiltonian} results in the linear spin-wave Hamiltonian 
\begin{equation}
\mathcal{H}_{\mathrm{LSW}}=\mathcal{E}_0\left(1+\frac{1}{S}\right)+S\sum_{\bm q}\sum_{\mu}\omega_\mu(\bm q)\left[\hat{n}_\mu(\bm q)+\frac{1}{2}\right],
\label{eq:LSW_Hamiltonian}
\end{equation}
where $\{\omega_\mu(\textbf{q})\}$ are magnon frequencies with $\hbar$ set to one, and $\hat{n}_\mu(\textbf{q})$ is the boson occupation number of the corresponding spin wave mode. Based on these magnon frequencies, we compute the zero-point energies for the $\psi_2$ and $\psi_3$ configurations via
\begin{eqnarray}
E_0(\psi) &=& \frac{S}{2}\sum_{\bm q}\sum_{\mu}\omega_\mu(\bm q)\label{eq:Zero_point_energy}.
\end{eqnarray}
To determine the ground state configuration chosen by quantum ObD, the zero-point energy difference $\Delta E_0=E_0(\psi_3)-E_0(\psi_2)$ is computed. Consequently, $\Delta E_0<0$ signals $\psi_3$ order while $\Delta E_0>0$ signals $\psi_2$ order. For non-vanishing temperatures, consideration of the ObD selection demands the inclusion of the magnon occupation via a Bose-Einstein distribution factor in addition to the above zero-point energy in Eq.~(\ref{eq:Zero_point_energy}). 
Both of these effects are contained in the free energy
\begin{eqnarray}
F(\psi) &=& S\sum_{\bm q}\sum_{\mu}
\left[
\frac{\omega_\mu(\bm q)}{2}+ T  \ln \left( 1-e^{-\frac{\omega_{\mu}(\bm q)}{T}}\right)\right],
\label{eq:LSW_free_energy}
\end{eqnarray}
where the first term corresponds to the zero-point energy, Eq.~\eqref{eq:Zero_point_energy}, while the second term incorporates the magnon  thermal population  effects~\cite{Schick-2020}.

\subsection{PFFRG}\label{sec:PFFRG}

To investigate the properties of the quantum $S=\frac{1}{2}$ model, we apply the standard one-loop (plus Katanin) PFFRG approach~\cite{Reuther10}. This method yields static $T=0$ spin-spin correlations and, upon Fourier transform, static spin susceptibilities as a function of a renormalization group frequency cutoff parameter $\Lambda$. We implement $\Lambda$ as a sharp frequency cutoff with $\Lambda=0$ corresponding to the physical, cutoff-free limit. The central PFFRG flow equations are solved using Euler's method, where one integration step corresponds to a reduction of $\Lambda$ by a factor of $0.98$. To capture the system's dynamics, the vertex functions are described on a frequency grid with $N_{\omega}=64$ mesh points for each frequency variable. Furthermore, as a real-space approximation, spin-spin correlations are only computed up to distances of five nearest-neighbor lattice spacings, such that each site is correlated with 380 lattice sites around it, and longer correlations are treated as zero.

The numerical solution procedure of the flow equations begins in the limit of infinite cutoff $\Lambda\rightarrow\infty$ where only the bare couplings enter as initial conditions~\cite{Reuther10}. Continuing the solutions towards smaller $\Lambda$ can then be understood as gradually taking into account more of the system's low energy dynamics. A more intuitive picture of the renormalization group flow is obtained when associating $\Lambda$ with the temperature $T$~\cite{Iqbal16}. Even though there is no exact correspondence between both quantities, they share various conceptual similarities. Particularly, like $\Lambda$, the temperature $T$ can be understood as a frequency cutoff (this becomes most obvious in a Matsubara framework where a finite temperature implies the existence of a minimal non-vanishing discrete fermionic Matsubara frequency). This means that even though the PFFRG approach is implemented at $T=0$, the flow towards small $\Lambda$ bears similarities with a cooling protocol.

For example, in a magnetically ordered system, the flow of the maximum susceptibility in momentum space $\chi^{\Lambda}_{\text{max}}$ is expected to diverge at a finite critical cutoff scale, $\Lambda=\Lambda_c$, as we lower $\Lambda$, in similarity to divergencies at a critical temperature $T_c$~\cite{Baez17}. However, due to the aforementioned numerical approximations concerning frequency resolution and direct space cutoff of the spin-spin correlations, these divergences are often suppressed and only manifest in finite peaks or kinks during the numerically-computed renormalization group flow. On the other hand, a featureless renormalization group flow without peaks or kinks is taken as an indication of a system that fails to develop conventional (dipolar) magnetic order signalled by magnetic Bragg peaks.

In most previous applications, the PFFRG has been used to study spin-isotropic Heisenberg models (see Refs.~\cite{Iida-2020,Kiese2020,thoenniss2020multiloop,Iqbal-2015,Iqbal-2016a,Iqbal-2016b,Iqbal16,Balz-2016,BuessenDiamond,Iqbal-2018a,Iqbal19,Niggemann-2019,MuellerBCC,Ghosh-2019,Chillal-2020,Ivica-2021,astrakhantsev2021pinwheel,ritter22,Hering-2022,keles22,schneider22,Kiese-2023} for a selection of recent applications of the PFFRG).
For our investigation of the Heisenberg-DM model, however, a recent extension needs to be implemented~\cite{Buessen19,buessen21}. The latter is capable of treating general two-body spin interactions including off-diagonal spin couplings $S_i^\alpha S_j^\beta$ with $\alpha,\beta\in\{x,y,z\}$, $\alpha\neq\beta$, in the absence of any continuous spin rotation symmetries. While this generalization increases the computational effort by a non-negligible factor of $32$, the numerical implementation of the flow equations still remains feasible.

Within PFFRG, the foremost magnetic properties of interest are probed via the static spin-spin correlations
\begin{equation}
    \bar{\chi}_{ij}^{\alpha\beta} = \int_0^{\infty} d\tau \langle  \hat{S}_{i}^\alpha(\tau) \hat{S}_{j}^\beta(0) \rangle, 
    \label{eq:chiij}
\end{equation}
where $\tau$ is the imaginary time. Fourier transforming the site dependence of $\bar{\chi}_{ij}^{\alpha\beta}$ yields the static spin susceptibility $\bar{\chi}^{\alpha\beta}({\bm q})$ as a function of momentum,
\begin{equation}
\label{eq:PFFRGsusceptibility}
    \bar{\chi}^{\alpha\beta}({\bm q})=\frac{1}{N}\sum_{ij}e^{i{\bm q}({\bm r}_i-{\bm r}_j)}\bar{\chi}_{ij}^{\alpha\beta},
\end{equation}
where ${\bm r}_i$ is the position of the pyrochlore site $i$. Here, the notation $\bar{\chi}$ is used to distinguish a static (zero frequency) response from an equal-time response $\chi$ as in Eq.~(\ref{eq:IrrepChi}). Note that in the general spin-anisotropic case, these susceptibilities are $3\times3$ tensors. However, as a result of the symmetries of our Heisenberg-DM model, the diagonal components $\bar{\chi}^{\alpha\alpha}({\bm q})$ with $\alpha \in\{x,y,z\}$ in the global coordinate frame are related by rotations in momentum space and similarly for the off-diagonal components. Below, we shall focus our discussions on the diagonal entries and, without loss of generality, consider the $\bar{\chi}^{zz}({\bm q})$ component. Since the magnetic orders relevant for this paper are all ${\bm q}= \bm{0}$ orders, it is impossible to distinguish them from their peak positions in $\bar{\chi}^{zz}({\bm q})$. However, in analogy to Eq.~(\ref{eq:IrrepChi}), a refined characterization is possible by computing the following susceptibilities
\begin{equation}
    \bar{\chi}_{\psi}=\frac{1}{N}\sum_{ij}\sum_{\alpha\beta}n_i^\alpha \bar{\chi}_{ij}^{\alpha\beta}n_j^\beta, 
    \label{eq:order-specific}
\end{equation}
where $n_i^\alpha$ is the $\alpha$-component of a normalized vector ${\bm n}_i$ which corresponds to the orientation of a spin on site $i$ in the specific spin configuration $\psi$. Later, the $\psi$ label will be assigned to $\psi_2$, $\psi_3$ or $T_{1\perp}$ orders to probe the system's tendency to order into these states. Note that the susceptibility $\chi_I$ in Eq.~(\ref{eq:IrrepChi}) becomes $\bar{\chi}_{\psi}$ in Eq.~(\ref{eq:order-specific}) if one switches from an equal-time formulation to a static one, which amounts to the replacement
\begin{equation}
\langle \bm{m}_{I}^{\boxtimes_1} \cdot \bm{m}_{I}^{\boxtimes_2}  \rangle \rightarrow \int_0^{\infty} d\tau \langle \bm{m}_{I}^{\boxtimes_1} (\tau) \cdot \bm{m}_{I}^{\boxtimes_2} (0) \rangle  
\end{equation}
in each contribution from two tetrahedra $\boxtimes_1$ and $\boxtimes_2$ in the first term of Eq.~(\ref{eq:IrrepChi}). Note that the second term of Eq.~\eqref{eq:IrrepChi} vanishes in the case of unbroken symmetries, i.e., it also does not contribute to order parameter susceptibilities calculated with PFFRG, since this method, by construction, can only treat symmetry-unbroken systems.

\subsection{High-temperature series expansion}
\label{subsec:HTSE}

Our primary goal of the high-temperature expansion (HTSE) is to study the selection within the $\Gamma_5$ manifold between $\psi_2$ and $\psi_3$ orders in the quantum model upon approaching the critical transition temperature $T_c$, coming from the paramagnetic side. For $\psi_2$ and $\psi_3$ long-range order, the spins orient along the local $x$ and $y$ axes, respectively. To probe this ordering, we add weak local fields $h_x$ and $h_y$ to our Hamiltonian of Eq.~\eqref{eq:hamiltonian}
along the local $x$ and $y$ axes (see Fig.~\ref{fig:gamma_5_local_basis}), described by the following perturbative field Hamiltonian term
\begin{equation}
    \mathcal{H}_{f}= -h_x \sum_{i} \hat{S}_i^x- h_y \sum_{i} \hat{S}_i^y .
\end{equation}
Here, $\hat{S}_i^x$ and $\hat{S}_i^y$ are the spin $S=\frac{1}{2}$ operators at site $i$ along the local $x$ (cubic $\langle 112\rangle$) and $y$ (cubic $\langle 110\rangle$) axes, respectively. The order parameter susceptibilities for the two cases ($\alpha=x,y$) are defined as
\begin{equation}\label{eq:ops}
    \chi_\alpha=\frac{-1}{\beta}\frac{\partial^2}{\partial h_\alpha^2}  \ln{Z(h_x,h_y)}\big |_{h_x=0,h_y=0}.
\end{equation}
High-temperature expansions are developed for the order parameter susceptibilities in powers of $\beta$ using the linked-cluster method \cite{oitmaa_book,Oitmaa-2013}. The series expansion coefficients for $\chi_x$ and $\chi_y$ are found to be identical term by term in powers of $\beta\equiv 1/T$, showing that the degeneracy within the $\Gamma_5$ manifold is not lifted at the level of the linear 
(response to ${\cal H}_f$) susceptibility. In fact, we prove in Appendix~\ref{appendix:OrderParameterSusceptibilityProof} that the $C_{3}$ symmetry of the model restricts the order parameter susceptibility for any long-range spin ordering within the $\Gamma_5$ manifold to be the same.


To go beyond the linear susceptibility, we compute high-temperature expansions for higher cumulants for $\alpha=x,y$, which we now define. Let,
\begin{equation}
    \hat{M}_\alpha \equiv \sum_{i} \hat{S}_i^\alpha.
\end{equation}
The cumulants $C_{n,\alpha}$ are defined as
\begin{equation}
    C_{2,\alpha}\equiv \langle \hat{M}_\alpha^2 \rangle,
\end{equation}
\begin{equation}
    C_{4,\alpha} \equiv \langle \hat{M}_\alpha^4 \rangle
    -3\langle \hat{M}_\alpha^2 \rangle^2,
\end{equation}
and
\begin{equation}\label{eq:sixth_order}
    C_{6,\alpha} \equiv \langle \hat{M}_\alpha^6 \rangle
    -15\langle \hat{M}_\alpha^4 \rangle \langle \hat{M}_\alpha^2 \rangle +30 \langle\hat{M}_\alpha^2 \rangle^3.
\end{equation}

Furthermore, we define $ C_n(\psi_2)\equiv C_{n,\alpha=x}$ and $C_n(\psi_3)\equiv C_{n,\alpha=y}$, in accordance with the spin orientations of the $\psi_2$ and $\psi_3$ states in the local coordinate frame shown in Fig.~\ref{fig:gamma_5_local_basis}.
These cumulants have the linked-cluster property, meaning that the cumulants evaluated for a disconnected cluster made of a disjoint union of two subclusters equals the sum of the cumulants for the two subclusters. This property ensures that their high-temperature expansions can be obtained by the linked-cluster method \cite{oitmaa_book,Oitmaa-2013}.
As found previously in a study of thermal ObD in 
Er$_2$Ti$_2$O$_7$~\cite{Oitmaa-2013}, we shall find below 
in Sec.~\ref{sec:high-T_results} that 
 the lowest order cumulants that can discriminate between the $\psi_2$ and $\psi_3$ orders are sixth order cumulants $C_{6,\alpha}$. 
All series expansions reported in this paper are calculated to order $\beta^8$. Results are discussed below in Sec.~\ref{sec:high-T_results}.


\section{\label{sec:results}Results}

\subsection{\label{subsec:mc_results}Classical Monte Carlo results}

\begin{figure}[ht!]
    \centering
    \begin{overpic}[width=0.9\columnwidth]{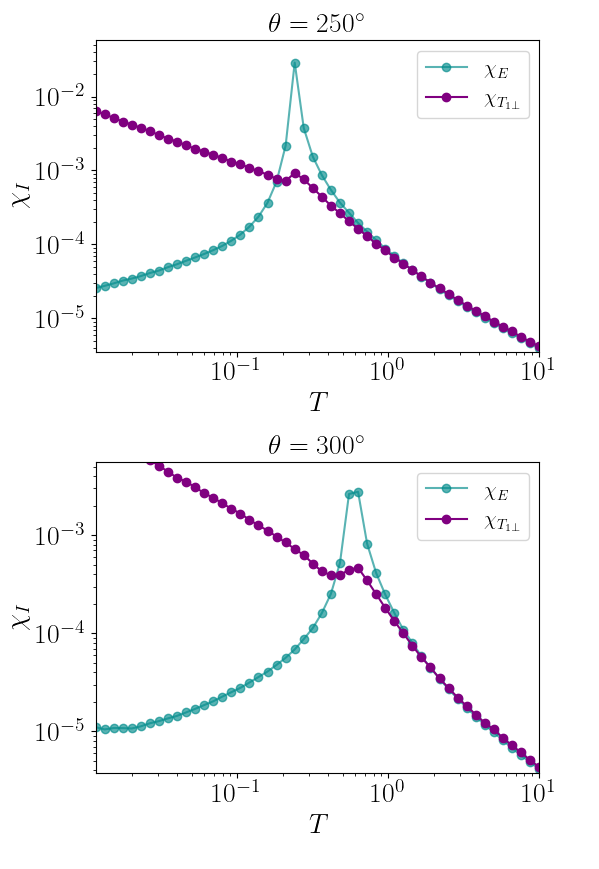}
        \put(7,48){(b)}
        \put(7,97){(a)}
    \end{overpic}
        \caption{Susceptibility $\chi_I$ [see Eq.~(\ref{eq:IrrepChi})] for the $E$ and $ T_{1\perp}$ irreps for $\theta=250\degree$ (a) and $\theta=300\degree$ (b). Both susceptibilities were obtained for a system with $L=10$. }
        \label{fig:chi_250_300}
\end{figure}

\begin{figure}[t]
    \centering
    \begin{overpic}[width=1.1\columnwidth]{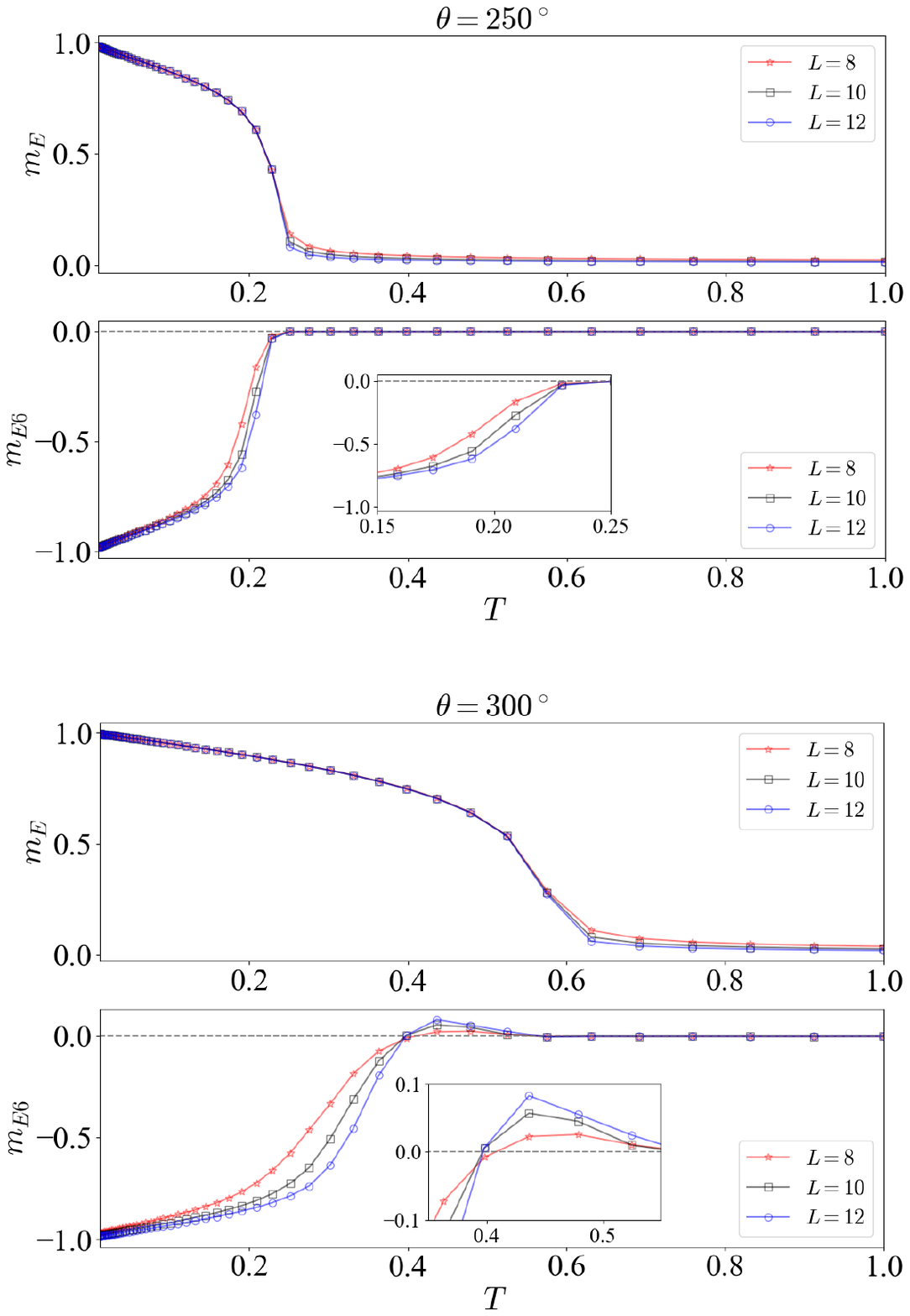}
     \put(3,96){(a)}
     \put(3,47){(b)}
     \put(10,10){}
    \end{overpic}
    \caption{Evolution of the $m_E$ and $m_{E6}$ order parameters [see Eq.~(\ref{eq:me6})] for (a) $\theta=250\degree$ and (b) $\theta=300\degree$ as a function of temperature for different systems sizes.}
        \label{fig:ME_ME6_250_300}
\end{figure}

Classical Monte Carlo simulations identify three distinct regions in parameter space consistent with the irrep analysis (see Sec.~\ref{subsection:state_class}) illustrated in Fig.~\ref{fig:Irreps}. In particular, we find AIAO order for $\theta\in(0\degree,135\degree)$ and colinear ferromagnetic order for $\theta\in[135\degree, 243\degree)$, as predicted by the irrep analysis. However, for $\theta\in(243\degree, 360\degree)$, where the irrep analysis suggests a degeneracy between the $E$ and $T_{1\perp}$ irreps (see Fig.~\ref{fig:Irreps}), our simulations find a transition into a $\Gamma_5$ phase which is indicated by a dominant susceptibility associated with the $m_E$ order parameter, as illustrated in Fig.~\ref{fig:chi_250_300}. The transition into this symmetry broken phase is further confirmed by the observation that $m_E$ saturates to one at low temperatures [see Figs.~\ref{fig:ME_ME6_250_300}(a) and (b)]. Although the $E$ irrep is ultimately selected as the low-temperature phase, the competition between the  $E$ and $T_{1\perp}$ irreps  caused by their energy degeneracy 
can be explicitly observed in the paramagnetic regime by considering the 
temperature evolution of the corresponding susceptibilities $\chi_E$ and $\chi_{T_{1\perp}}$, as illustrated in Figs.~\ref{fig:chi_250_300}(a) and (b) for $\theta=250\degree$ and $\theta=300\degree$, respectively. As mentioned in Sec.~\ref{subsection:state_class}, the
high-temperature limiting behavior of these correlation functions is consistent with a Ginzburg-Landau theory where the leading quadratic term evolves proportional to $(3T+a_I)$ \cite{Reimers-1991a,Enjalran}, only departing from each other significantly upon approaching the transition temperature $T_c$. A similar behavior for the $\chi_E$ and $\chi_{T_{1\perp}}$ susceptibilities is observed 
within the entire $\theta\in(243\degree, 360\degree)$ range (not shown).
The selection of the $E$ irrep in the low-temperature phase is a consequence of the entropic weight associated with the $\psi_3$ states and the pseudo-Goldstone modes resulting from this configuration. More precisely, and as noted in Ref.~\cite{Elhajal05}, both manifolds can be constructed by considering a $\psi_3$ state whose spin configuration is then continuously rotated along a local or a global axis, see Figs.~\ref{fig:phasediagram}(b)-\ref{fig:phasediagram}(d) and Eq.~\eqref{eq:psi3-t1p}. These two continuous degrees of freedom result in the observation of two pseudo-Goldstone modes associated with the $\psi_3$ states~\cite{Chern08}, which, at low temperatures, may favor these spin configurations.

\begin{figure}[ht]
\hspace*{-0.3cm} 
    \begin{overpic}[width=1.1\columnwidth]{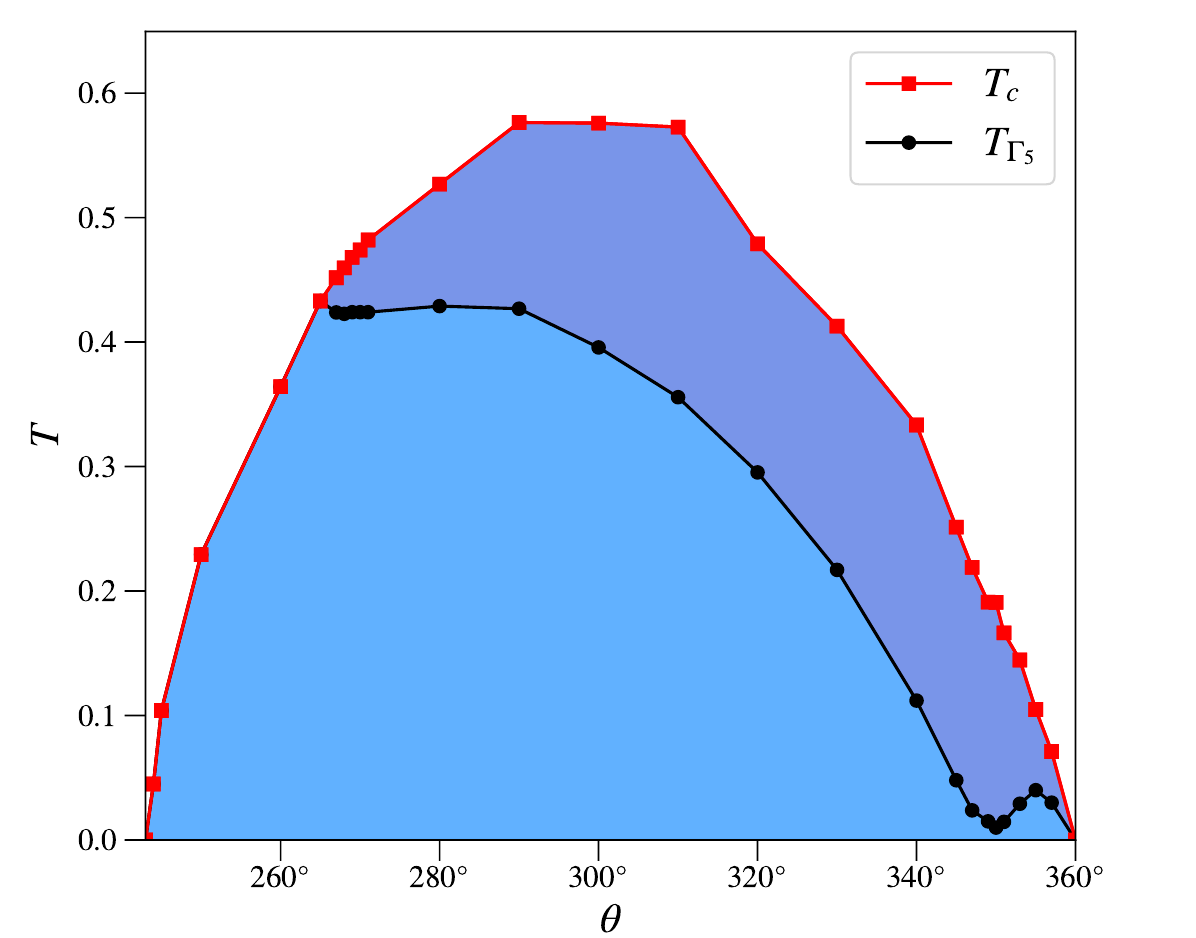}
     \put(50,57){$\psi_2$}
     \put(50,28){$\psi_3$}
     \end{overpic}
    \caption{Temperature-dependent phase diagram for the $\Gamma_5$ phase illustrating a selection between the $\psi_2$ and $\psi_3$ ordering as a function of $\theta$. This phase diagram was produced by considering system sizes $L=12$ to identify $T_{\Gamma_5}$ as the sign switching of the $m_{E6}$ order parameter obtained for both system sizes, as shown in the inset of Fig.~\ref{fig:ME_ME6_250_300}(b).}
        \label{fig:Phase_diagram_Gamma_5_T}
\end{figure}

Having identified $\Gamma_5$ as the low-temperature phase for $\theta\in(243\degree, 360\degree)$, the $m_{E6}$ order parameter is computed to further characterize the ordered configurations realized below $T_c$. Figures \ref{fig:ME_ME6_250_300}(a,b) show the temperature dependence of the $m_E$ and $m_{E6} $ order parameters below $T_c$ for $\theta=250\degree$ and $\theta=300\degree$, respectively. For $243\degree< \theta \lesssim 265\degree$, the $m_{E6}$ order parameter becomes rapidly negative immediately below $T_c$, see Fig.~\ref{fig:ME_ME6_250_300}(a). 
This indicates a $\psi_3$ selection from $T_c$ to $T=0^+$, following the  convention made in Sec.~\ref{subsec:class_mc} regarding  $\psi_2$ versus $\psi_3$ state labeling at  $0< T < T_c$. 
In contrast, for $265\degree\lesssim\theta<360\degree$,  $m_{E6}$ now being positive just below $T_c$ indicates, again in accord with the convention introduced in Sec.~\ref{subsec:class_mc}, a $\psi_2$ selection at $T\lesssim T_c$ followed by a transition from $\psi_2$ order to $\psi_3$ order at a temperature denoted $T_{\Gamma_5}$ ($T_{\Gamma_5} <T_c$), see  Fig.~\ref{fig:ME_ME6_250_300}(b).
Please refer to Appendix \ref{appendix:distribution} for  further discussion on the topic of the $\psi_2$ to $\psi_3$ transition at $0< T < T_c$, illustrated by the black line in  Fig.~\ref{fig:Phase_diagram_Gamma_5_T}.

\begin{figure}[ht]

    \hspace*{-10mm}   
    \begin{overpic}[width=1.1\columnwidth]{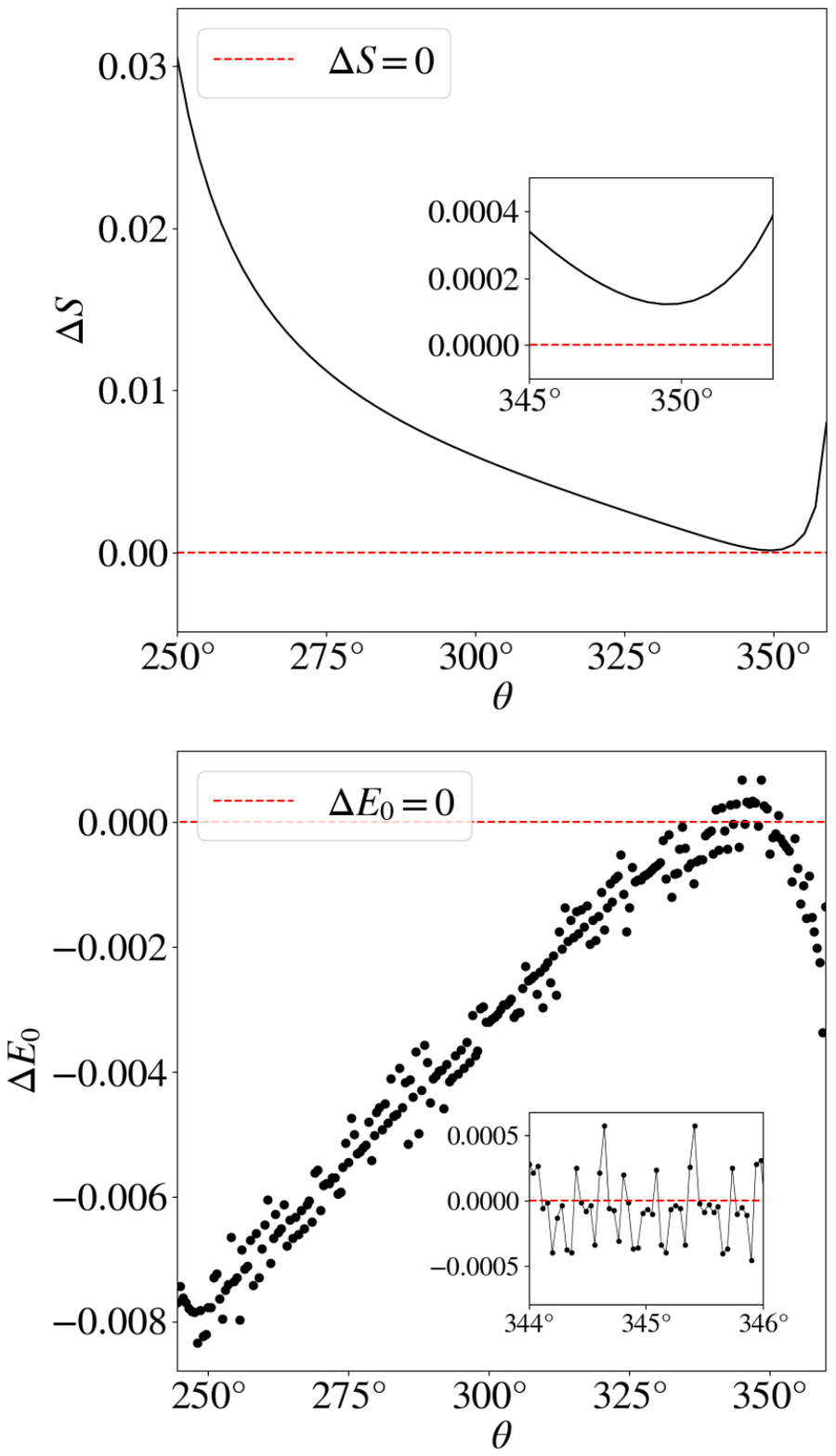}
      \put(10,94){(a)}
       \put(10,48){(b)}
    \end{overpic}
        
        \caption{(a) Entropic selection between the $\psi_3$ and $\psi_2$ orders as a function of $\theta$ obtained from a classical low-temperature expansion [see Eq.~(\ref{eq:entropy})], where $\Delta S>0$ signals $\psi_3$ selection. The inset illustrates the minimum in the entropy difference centered around $\theta\approx 350\degree$. (b) Zero-point energy difference [see Eq.~(\ref{eq:Zero_point_energy})] between the $\psi_3$ and $\psi_2$ configurations as a function of $\theta$ for $S=\frac{1}{2}$, see Eq.\eqref{eq:Zero_point_energy}, where $\Delta E_0<0$ signals $\psi_3$ selection and $\Delta E_0>0$ signals $\psi_2$ selection. The inset shows rapid oscillations in the zero-point energy as a function of $\theta$ for $\theta\in[344\degree,346\degree]$.}
        \label{fig:DS_DE}
\end{figure}

The selection of $\psi_3$ and $\psi_2$ orders at $T=0^+$ and $T_c$, respectively, is illustrated by the outermost and middle rings of Fig.~\ref{fig:phasediagram}(e), respectively. The phase diagram in the $\Gamma_5$/copl regime as a function of temperature and $\theta$ is depicted in Fig.~\ref{fig:Phase_diagram_Gamma_5_T}. Here, the critical temperature $T_c$ is obtained by identifying the temperature at which the specific heat peaks while the $\cramped{\psi_2 \rightarrow \psi_3}$ transition temperature $T_{\Gamma_5}$ is identified by the change of sign of the $m_{E6}$ parameter. The calculation of a more refined phase diagram would require a systematic finite-size scaling study for the identification of the critical temperature $T_c$ as well as the transition temperature $T_{\Gamma_5}$ (see the discussion in Sec.~\ref{subsec:class_mc} and Appendix~\ref{appendix:distribution} where we comment and study the dangerously irrelevant  $m_{E6}$ order parameter in a Ginzburg-Landau-Wilson theory~\cite{Zhitomirsky-2014,AndradePRL2018}), which is beyond the scope of this paper. Interestingly, we find a local minimum in $T_{\Gamma_5}$ for $\theta\approx 350\degree$. This reduction in $T_{\Gamma_5}$ is connected to the reduced entropy difference $\Delta S$ between the $\psi_2$ and $\psi_3$ states [see Eq.~\eqref{eq:entropy}].
We find that  $\Delta S>0$, hence $\psi_3$ order, for \textit{all} angles in $\theta\in(243\degree,360\degree)$ with a local $\Delta S$ minimum around $\theta \approx  350\degree$, as shown in Fig.~\ref{fig:DS_DE}(a). We note that a $\psi_2$ selection at intermediate temperatures followed by a $\psi_3$ selection at lower temperature for $J>0$ and $D<0$ had previously been reported for a finite range of $|D|/J$~\cite{chern10}.

The phase diagram in Fig.~\ref{fig:Phase_diagram_Gamma_5_T} also reveals the expected vanishing of $T_c$ in the Heisenberg limit $\theta=360\degree$ where the system realizes a low-temperature classical spin liquid~\cite{Moessner98,Moessner-1998b,Reimers-1992,Zinkin-1996,Henley-2010}. Interestingly, a rapid drop of $T_c$ is also observed at the lower boundary of the $\Gamma_5$/copl regime at $\theta\approx243\degree$ (i.e., at $D/J=2$). 
The critical ordering  temperature into the ferromagnetic phase 
grows rapidly for $\theta \lesssim 243\degree$ (not shown).
Like for the pure Heisenberg model, this behavior at $D/J=2$ can be explained by a large (extensive) ground state degeneracy which warrants further investigation.

\subsection{\label{subsec:lsw}Quantum $1/S$ spin waves results}

In this section, we begin extending our model from the classical to the quantum version, first by investigating the semiclassical case as realized in linear order in $1/S$. To study the quantum ObD selection between $\psi_2$, $\psi_3$ and $T_{1\perp}$ states, the zero-point energy $E_0(\psi)$ [see Eq.~\eqref{eq:Zero_point_energy}] associated with these spin configurations is computed. As reported by Ref.~\cite{Canals08}, for indirect DMI and antiferromagnetic Heisenberg couplings, the zero-point energy reveals a selection of the $\Gamma_5$ manifold over the $T_{1\perp}$ states where this selection is prevalent for $\theta\in(243\degree,360\degree)$ (not shown). Next, to investigate quantum ObD selection \emph{within} the $\Gamma_5$ manifold, we analyze the zero-point energy selection between the $\psi_2$ and $\psi_3$ configurations, illustrated in Fig.~\ref{fig:DS_DE}(b). This calculation reveals a $\psi_3$ selection for $\theta \in(243\degree,344\degree]$ and $\theta \in [352\degree,360\degree)$. Interestingly, for $\theta \in[344\degree,352\degree]$ the selection oscillates between $\psi_2$ and $\psi_3$. We have confirmed that these oscillations are not produced by numerical errors by implementing different ${\bm q}$-space integration schemes. In addition, as the zero-point energies are obtained from Eq.~\eqref{eq:Zero_point_energy}, as a consistency check, we verified the numerical correctness of the magnon frequencies obtained with the quantum ($1/S$) spin-wave theory by comparing with the frequencies obtained by a traveling wave solution to the classical torque equations in the limit of small oscillation~\cite{Shu-2019}, finding the two sets of results to be numerically identical to machine precision. It thus seems that these oscillations in $\Delta E_0$ are genuine to the problem and are not the consequence of computational accuracy. This phenomenon might be related to the observation of a thin sliver in the $\{J_{zz},J_{\pm},J_{\pm\pm},J_{z,\pm}\}$ \footnote{In terms of Heisenberg and DM couplings the local exchange interaction parameters~\cite{ross11,wong13} take the form $J_{zz}=-\frac{1}{3}(J+4D)$, $J_{\pm}=\frac{1}{6}(J-2D)$, $J_{\pm\pm}=\frac{1}{3}(J+D)$ and $J_{z\pm}=\frac{1}{3\sqrt{2}}(2J-D)$.} phase diagram where $\psi_2$ order is selected \cite{wong13}. The oscillations observed in $\Delta E_0$ might be caused by its dependence on the local exchange interaction parameters \cite{wong13} 
\begin{eqnarray}
    \Delta E_0\approx && J_{\pm}\Big[c_3 \left(\frac{J_{\pm\pm}}{J_\pm}\right)^3+c_2 \left(\frac{J_{\pm\pm}J_{z\pm}}{J_\pm^2}\right)^2\nonumber\\
    &&\quad + c_1 \left(\frac{J_{\pm\pm}J_{z\pm}^4}{J_\pm^5}\right)+c_0 \left(\frac{J_{z\pm}}{J_\pm}\right)^6 \Big] .
\end{eqnarray}
Here, the coefficients $c_i$ are dimensionless functions of $J_{zz}/J_\pm$\footnote{This expression was constructed by Ref.~\cite{wong13} through a power counting argument while only considering terms up to sixth order in the interaction parameters $\{J_{zz},J_{\pm},J_{\pm\pm},J_{z,\pm}\}$.} and the local interaction parameters are combinations of trigonometric functions resulting from the parametrization in Eq.~\eqref{eq:parametrization}~\footnote{Using the parametrization in Eq.~\eqref{eq:parametrization} the local interaction parameters exhibit the following $\theta$ dependence $J_{zz}=-\frac{1}{3}(\cos(\theta)+4\sin(\theta))$, $J_{\pm}=\frac{1}{6}(\cos(\theta)-2\sin(\theta))$, $J_{\pm\pm}=\frac{1}{3}(\cos(\theta)+\sin(\theta))$ and $J_{z\pm}=\frac{1}{3\sqrt{2}}(2\cos(\theta)-\sin(\theta))$.}. 
Although the above equation looks deceivingly simple, when the dependence on the parametric angle $\theta$ is introduced, we note that the undetermined $c_i$ coefficients are also non-trivial oscillating functions of $\theta$ which impact the oscillatory period of $\Delta E_0$. A similar non-monotonic (or rapidly varying) behavior in the zero-point energy $E_0$ can also be 
observed in Fig.~(2) of Ref.~\cite{Savary-2012},  which considers the selection of $\psi_2$ versus $\psi_3$ within the $\Gamma_5$ manifold for the 
model they consider, though this was not commented upon therein. This oscillatory behavior of $\Delta E_0(\theta)$ deserves further attention, which we leave for future work.
It is worth noting that the zero-point energy selection differs from the classical entropic selection in the region $\theta \in[344\degree,352\degree]$; contrast the results of Fig.~\ref{fig:DS_DE}(a) with those of Fig.~\ref{fig:DS_DE}(b).
Furthermore, at slightly nonzero temperature, the quantum ObD magnon free energy in Eq.~\eqref{eq:LSW_free_energy} selects the \textit{same} states as those selected by the zero-point energy for all theta value corresponding to the $\Gamma_5$ phase.

\subsection{\label{subsec:pffrg}PFFRG results}

\begin{figure}[t!]
    \centering
       \begin{subfigure}[b]{0.23\textwidth}
        \centering
        \begin{overpic}[width = 1\textwidth]{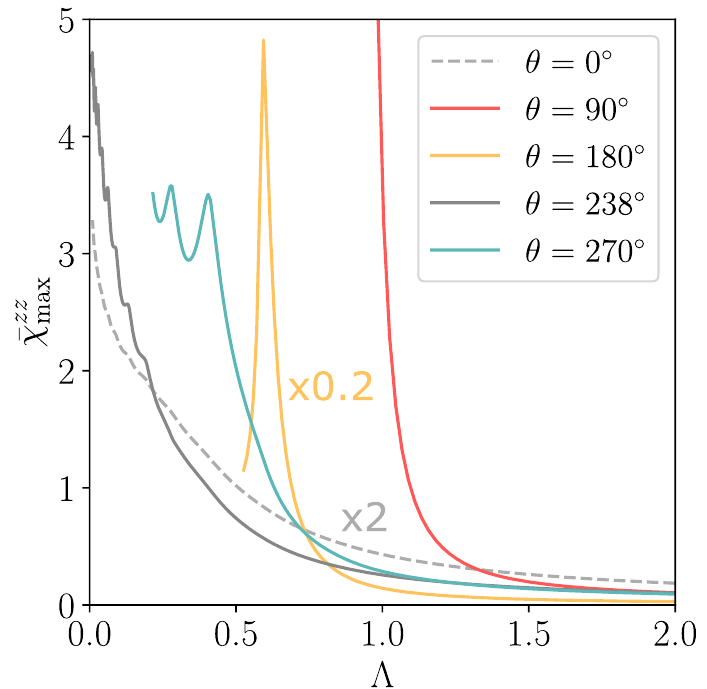}
        \put(-6,90){(a)}
        \end{overpic}
        \caption*{PFFRG flows of $\bar{\chi}^{zz}_{\text{max}}$ for different $\theta$}
        \label{fig:CharacteristicFlows}
    \end{subfigure}%
    ~\hspace{1mm}%
    ~
    \begin{subfigure}[b]{0.23\textwidth}
        \centering
        \begin{overpic}[width = 1\textwidth]{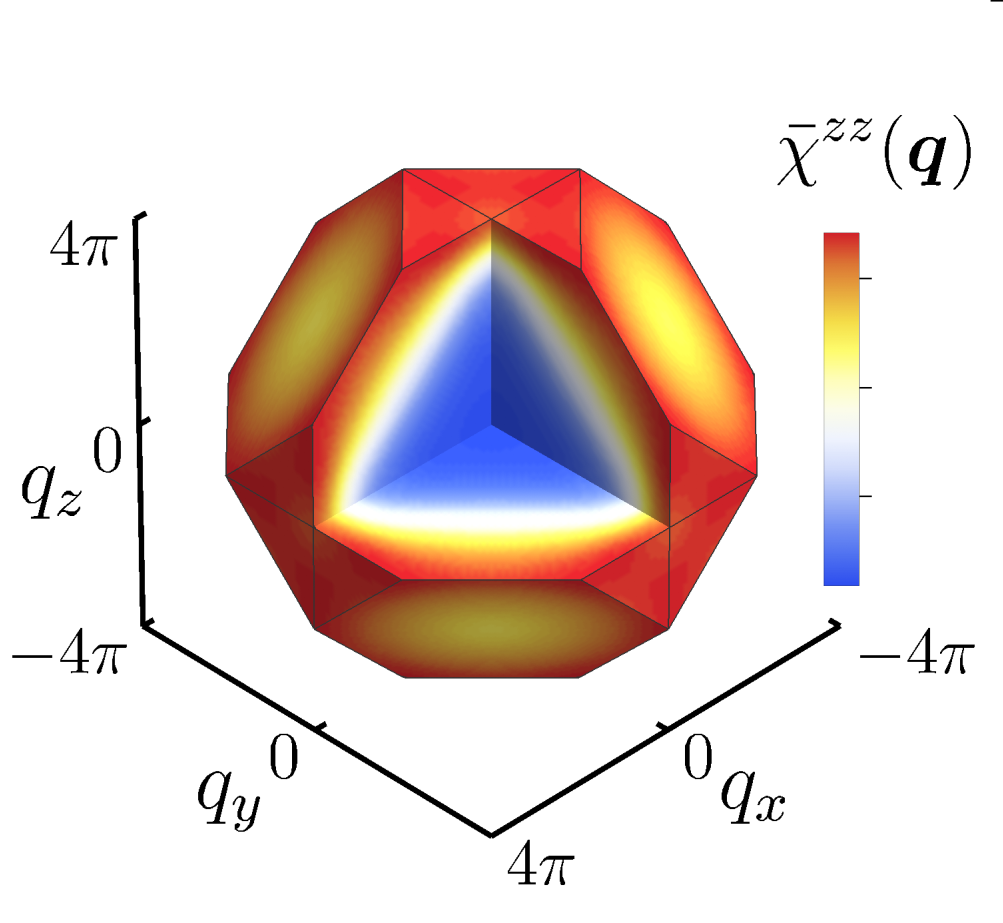}
            \put(-7,70){(b)}
            \put(88.5,40.8){\scriptsize{0.5}}
            \put(88.5,51.5){\scriptsize{1.0}}
            \put(88.5,62.4){\scriptsize{1.5}}
        \end{overpic}
        \caption*{$\theta=0^{\circ}$, antiferromagnetic interaction}
    \end{subfigure}
    \vskip -2pt 
    \begin{subfigure}[b]{0.23\textwidth}
        \centering
        \begin{overpic}[width = 1\textwidth]{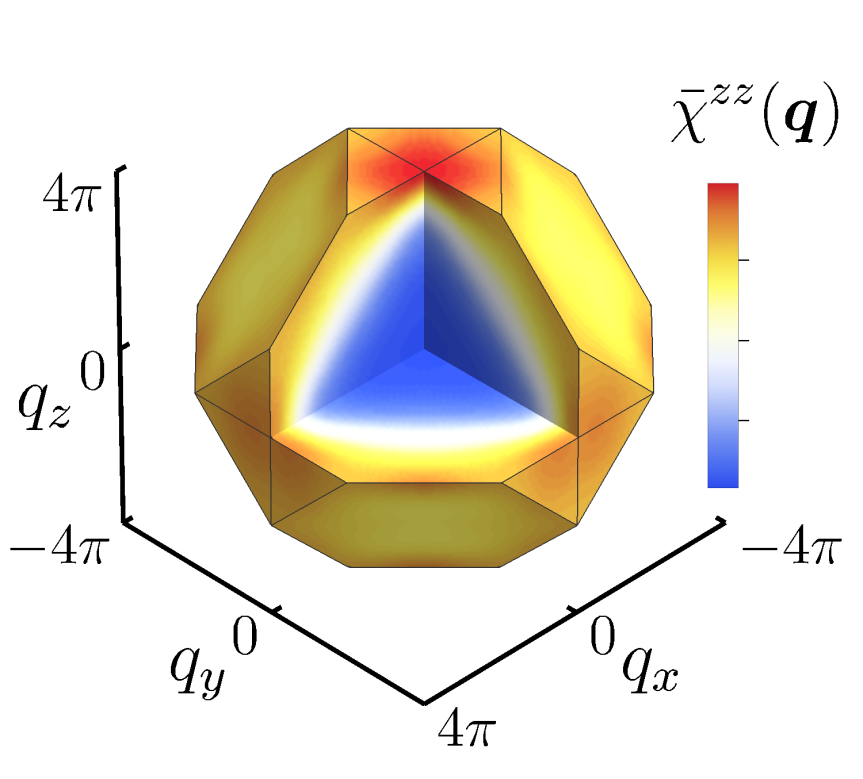}
            \put(-7,70){(c)}
            \put(88.5,40.0){\scriptsize{0.5}}
            \put(88.5,49.3){\scriptsize{1.0}}
            \put(88.5,58.5){\scriptsize{1.5}}
        \end{overpic}
        \caption*{$\theta=356^{\circ}$}
        \label{fig:DirectChi}
    \end{subfigure}%
    ~\hspace{1mm}%
    ~
        \begin{subfigure}[b]{0.23\textwidth}
        \centering
        \begin{overpic}[width = 1\textwidth]{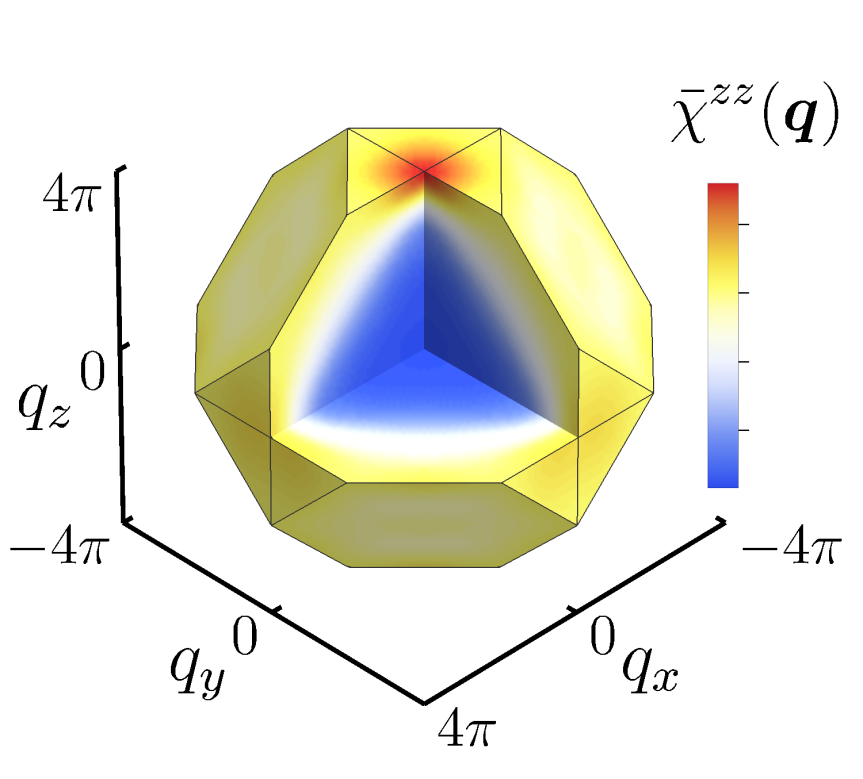}
            \put(-7,70){(d)}
            \put(88.5,38.6){\scriptsize{0.5}}
        \put(88.5,46.7){\scriptsize{1.0}}
        \put(88.5,54.9){\scriptsize{1.5}}
        \put(88.5,62.6){\scriptsize{2.0}}
        \end{overpic}
        \caption*{$\theta=4^{\circ}$}
        \label{fig:FMChi}
    \end{subfigure}
    \vskip -2pt 
    \begin{subfigure}[b]{0.23\textwidth}
        \centering
        \begin{overpic}[width = 1\textwidth]{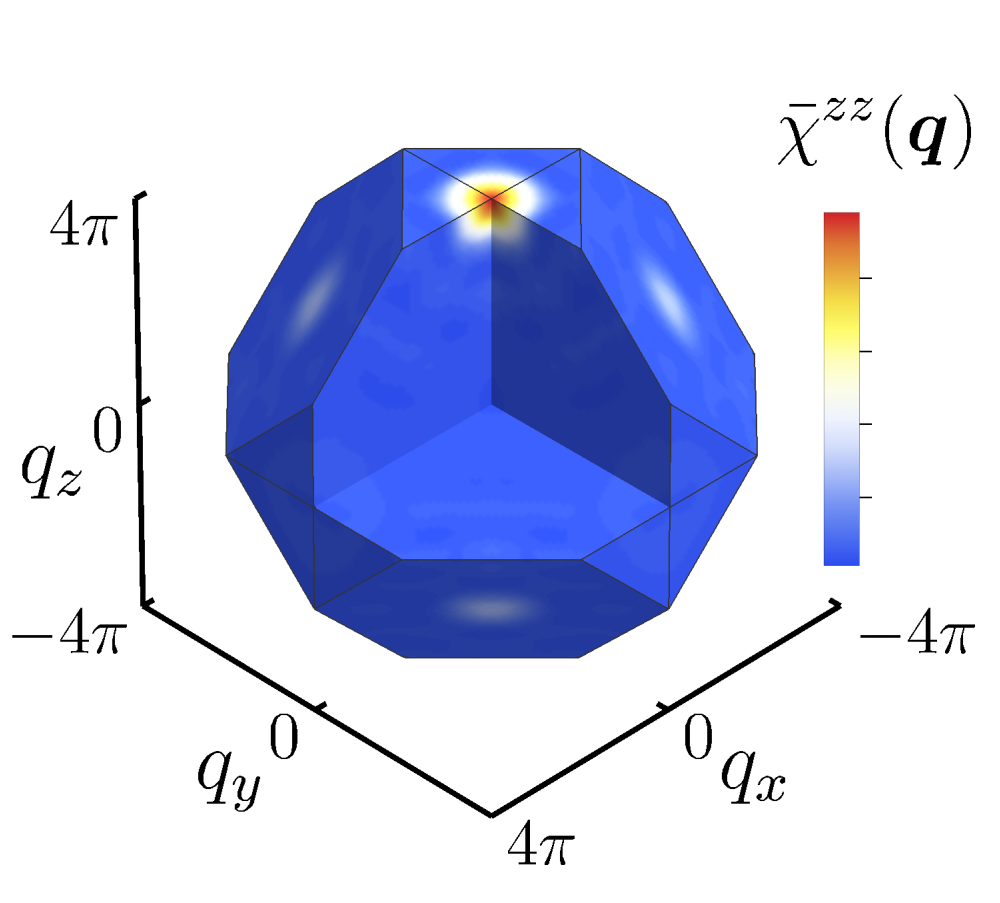}
            \put(-7,70){(e)}
            \put(89,38.7){\scriptsize{1}}
            \put(89,46.1){\scriptsize{2}}
            \put(89,53.3){\scriptsize{3}}
            \put(89,60.8){\scriptsize{4}}
        \end{overpic}
        \caption*{$\theta=90^{\circ}$, direct DMI\newline}
        \label{fig:DirectChi}
    \end{subfigure}%
    ~\hspace{1mm}%
    ~
        \begin{subfigure}[b]{0.23\textwidth}
        \centering
        \begin{overpic}[width = 1\textwidth]{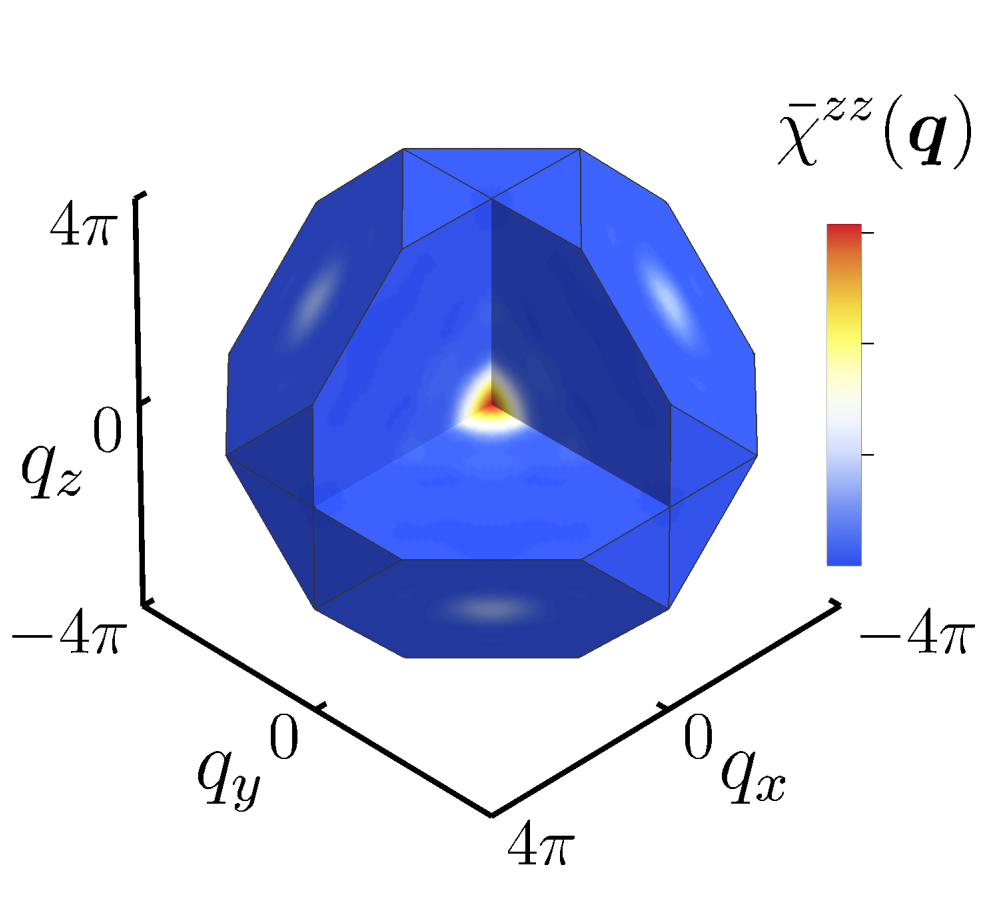}
            \put(-7,70){(f)}
            \put(89,43){\scriptsize{5}}
            \put(89,54.1){\scriptsize{10}}
            \put(89,65.1){\scriptsize{15}}
        \end{overpic}
        \caption*{$\theta=180^{\circ}$, ferromagnetic interaction}
        \label{fig:FMChi}
    \end{subfigure}
    \vskip -2pt 
    \begin{subfigure}[b]{0.23\textwidth}
        \centering
        \begin{overpic}[width = 1\textwidth]{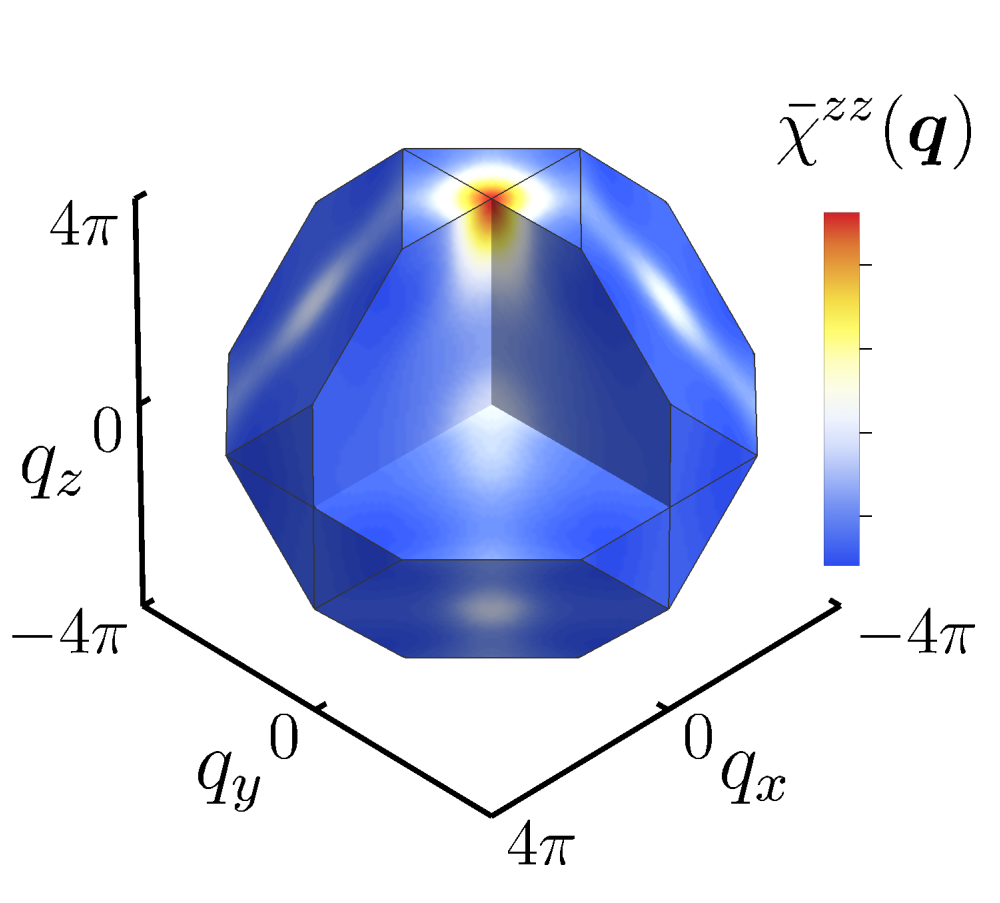}
            \put(-7,70){(g)}
            \put(89,37.0){\scriptsize{1}}
            \put(89,45.2){\scriptsize{2}}
            \put(89,53.5){\scriptsize{3}}
            \put(89,62.1){\scriptsize{4}}
        \end{overpic}
        \caption*{$\theta=238^{\circ}$, ferromagnetic and indirect DM interaction}
        \label{fig:DirectChi}
    \end{subfigure}%
    ~\hspace{1mm}%
    ~
    \begin{subfigure}[b]{0.23\textwidth}
        \centering
        \begin{overpic}[width = 1\textwidth]{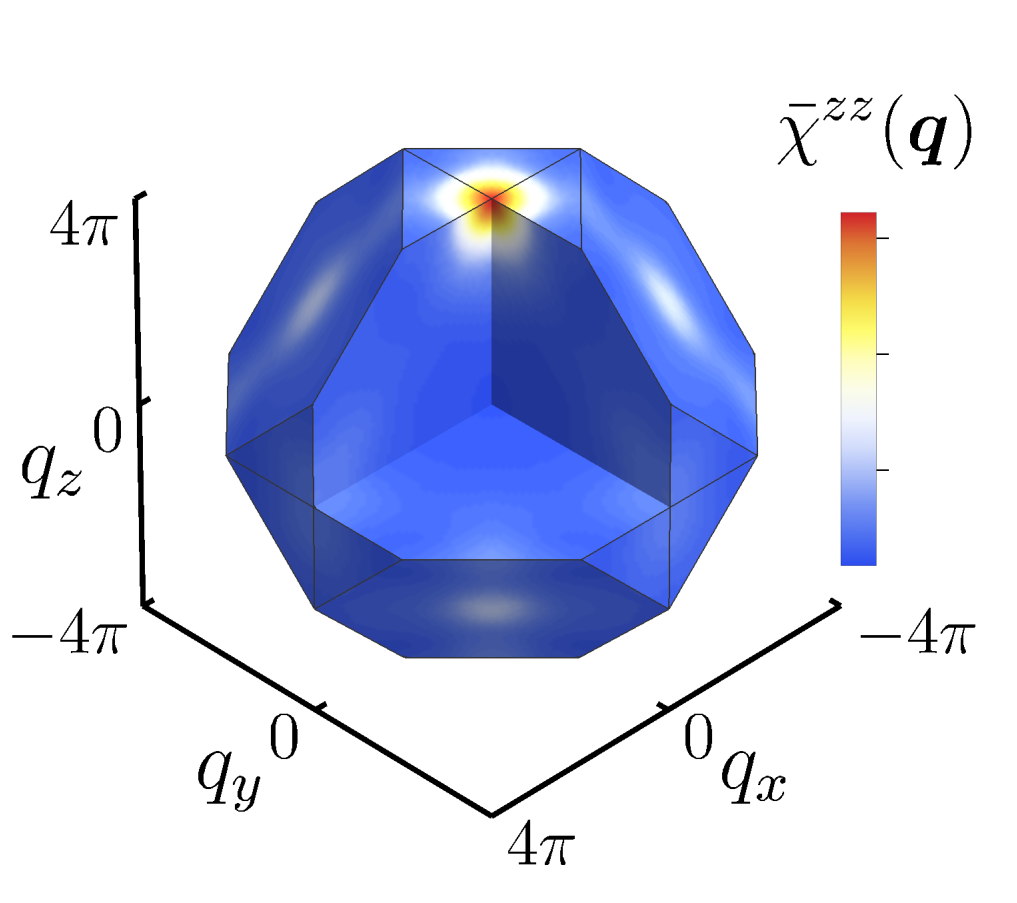}
            \put(-7,70){(h)}
            \put(88.7,41){\scriptsize{1}}
            \put(88.7,52.2){\scriptsize{2}}
            \put(88.7,63.3){\scriptsize{3}}
        \end{overpic}
        \caption*{$\theta=270^{\circ}$, indirect DMI\newline}
        \label{fig:IndirectChi}
    \end{subfigure}
    
    \caption{(a) Selected PFFRG flows of the maximum susceptibility $\bar{\chi}^{zz}_{\text{max}}$ and (b)-(h) examples for momentum resolved susceptibilities $\bar{\chi}^{zz}(\bm{q})$ over the extended Brillouin zone at the flow breakdown for magnetically ordered phases [i.e., for (e), (f), (h)] or in the low $\Lambda$ limit for nonmagnetic flows [i.e., for (b), (c), (d), (g)].}
    \label{fig:ExemplaryFlows}
\end{figure}
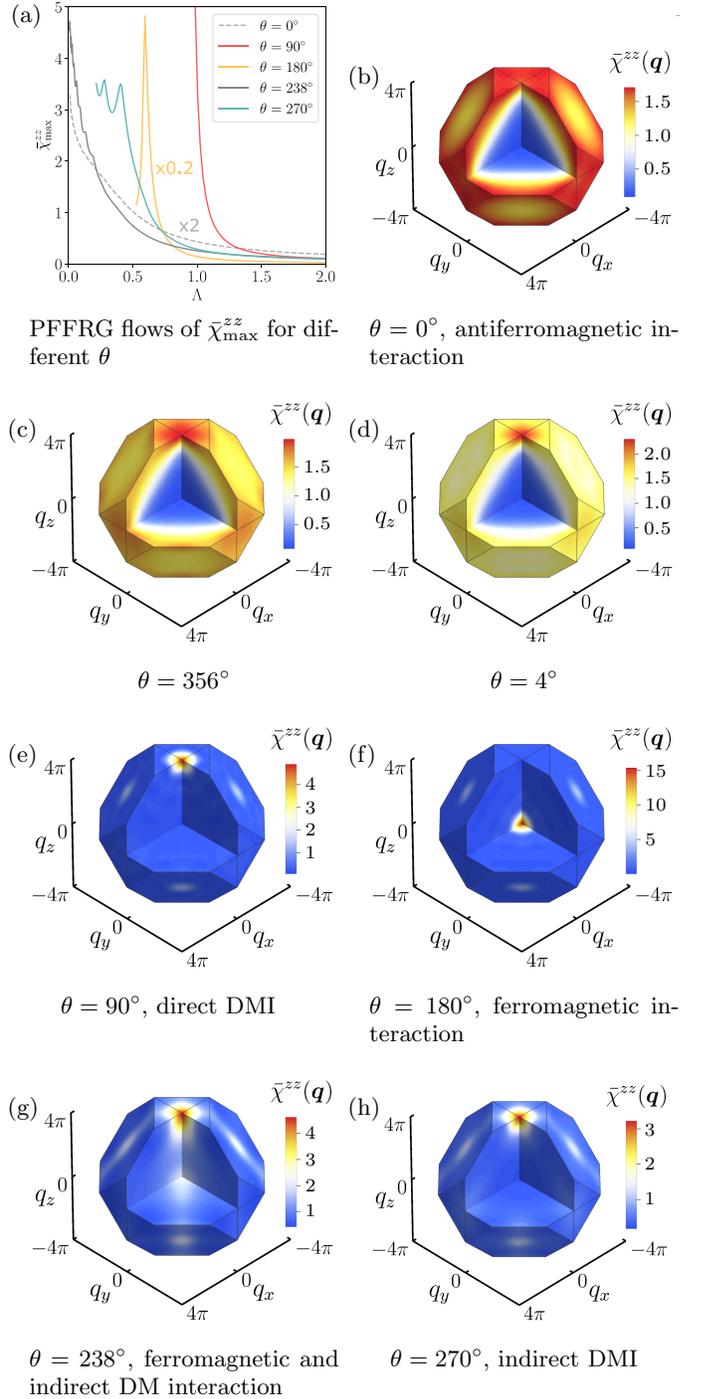

We now investigate the quantum $S=\frac{1}{2}$ model using the PFFRG approach. The crucial question in this extreme quantum limit is whether there are parameter regimes where quantum fluctuations are strong enough to destroy any type of magnetic long-range order at $T=0$. The PFFRG can address the question whether a spin system is magnetically ordered or disordered via the behavior of the renormalization group flow ($\Lambda$ flow) of the maximum of the static susceptibility, $\bar{\chi}^{zz, \Lambda}_{\text{max}}$, in momentum space,  
as discussed in Sec.~\ref{sec:PFFRG}. Particularly, peaks or kinks in the $\Lambda$ dependence of this susceptibility indicates a magnetically ordered system, while otherwise a nonmagnetic phase is signalled.
Figure~\ref{fig:ExemplaryFlows}(a) shows examples for different flow behaviors where the smooth flows at $\theta=0\degree$ and $\theta=238\degree$ indicate a nonmagnetic phase at $T=0$ while the pronounced kinks or divergences in $\bar{\chi}^{zz, \Lambda}_{\text{max}}$ at $\theta=90\degree,180\degree$ and $270\degree$ point towards magnetic long-range order. Based on this diagnosis, we can construct the quantum phase diagram in the full $\theta$ range as shown by the inner ring of Fig.~\ref{fig:phasediagram}(e). It needs to be emphasized, however, that close to quantum critical points, it is often difficult to distinguish between the two types of (smooth versus kinked) flow behaviors. This leads to regions of uncertainty marked by continuous color gradients between magnetic (colorful) phases and nonmagnetic (white) phases. Furthermore, we note that the small oscillations by which some curves in Fig.~\ref{fig:ExemplaryFlows}(a) are modulated are a numerical artifact from a combination of discretizing the frequency arguments of the vertex functions and a step-like regulator function defining the RG scale $\Lambda$ (within more involved numerical implementations, these oscillations can be suppressed~\cite{thoenniss2020multiloop,Kiese2020}).

To characterize the magnetic properties of the individual quantum phases, we plot $\bar{\chi}^{zz}({\bm q})$ over the extended Brillouin
zone in Figs.~\ref{fig:ExemplaryFlows}(b)-\ref{fig:ExemplaryFlows}(h), either at $\Lambda\rightarrow0$ for nonmagnetic or at the critical $\Lambda_c$ for magnetic phases. The AIAO and ferromagnetic ordered phases clearly exhibit the expected sharp peaks at the $\Gamma$-point and at ${\bm q}=(0,0,4\pi)$ (which indicates ${\bm q}={\bm 0}$ orders) as shown in Fig.~\ref{fig:ExemplaryFlows}(e) and Fig.~\ref{fig:ExemplaryFlows}(f), respectively. The order in the $\Gamma_5$/copl regime is, likewise, characterized by a sharp peak at ${\bm q}=(0,0,4\pi)$, see Fig.~\ref{fig:ExemplaryFlows}(h). We will study the quantum competition between the $\psi_2$, $\psi_3$ and $T_{1\perp}$ states in the $\Gamma_5$/copl regime further below.

The PFFRG indicates the existence of two nonmagnetic ground state phases which are signalled by an absence of any instability in the $\Lambda$ flow in the regimes $-9\degree \lesssim \theta \lesssim 8\degree$ and $237\degree \lesssim \theta \lesssim 241.5\degree$. The first originates from the pure Heisenberg limit $\theta=0$ and shows a broad distribution of signal in $\bar{\chi}^{zz}({\bm q})$ that spreads over large regions in momentum space, see Fig.~\ref{fig:ExemplaryFlows}(b). Based on a variety of powerful numerical methods (including PFFRG), the broadened remnants of pinch points and the possibility of a spontaneous breaking of point-group symmetries in this phase have recently been studied in Refs.~\cite{Iqbal19,hagymasi21,Astrakhantsev21,Hering-2022}. As shown in Figs.~\ref{fig:ExemplaryFlows}(c) and ~\ref{fig:ExemplaryFlows}(d), in parameter regimes where the DM coupling is non-vanishing but small enough to not induce magnetic long-range order, the signal still remains distributed over large parts of momentum space. However, the formation of broad peaks at ${\bm q}=(0,0,4\pi)$ already indicates the proximity to ${\bm q}={\bm 0}$ long-range ordered phases. Additional PFFRG data for this nonmagnetic phase in different planes or line cuts in momentum space are presented in Appendix~\ref{appendix:additional_PFFRG}. 

Most strikingly, the PFFRG shows indications for a thin sliver of a second nonmagnetic phase in the vicinity of the classical transition point between the ferromagnetic phase and the $\Gamma_5$ regime at $\theta\approx243\degree$ alluded to in Sec.~\ref{subsec:mc_results}. However, the sizable regions of uncertainty by which it is flanked complicate an unambiguous detection of this putative nonmagnetic phase. As a result of its location between the classical ferromagnetic and $\Gamma_5$/copl regimes, the momentum resolved susceptibility $\bar{\chi}^{zz}({\bm q})$ displays a smeared signal at and between the corresponding wave vectors [$\Gamma$ point and ${\bm q}=(0,0,4\pi)$], see Fig.~\ref{fig:ExemplaryFlows}(g). We propose a possible connection between this nonmagnetic phase in the quantum $S=\frac{1}{2}$ case and the observation of vanishing $T_c$ in the classical model at $\theta\approx243\degree$, see Fig.~\ref{fig:Phase_diagram_Gamma_5_T}, but leave a more detailed investigation for future work.

\begin{figure}
\hspace*{-0.3cm} 
        \centering
\includegraphics[width = 0.5\textwidth]{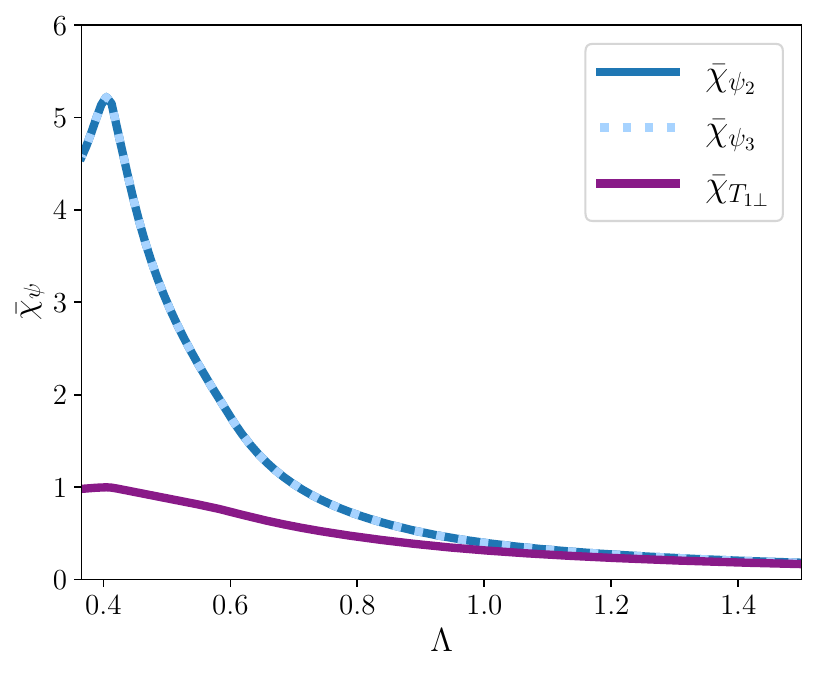}
\caption{PFFRG flows of the order parameter susceptibility ${\bar{\chi}}_{\psi}$ for $\psi_2$, $\psi_3$ and $T_{1\perp}$ orders in the pure indirect DMI limit at $\theta=270^{\circ}$. Note that the curves for ${\bar{\chi}}_{\psi_2}$ and ${\bar{\chi}}_{\psi_3}$ lie on top of each other, see text for details. The kink at $\Lambda\approx0.4$ signals a flow breakdown associated with magnetic long-range order.}
\label{fig:specificOrderFlow}
\end{figure}

We finally discuss the magnetic order in the $\Gamma_5$/copl regime which cannot be fully characterized by the peak position in $\bar{\chi}^{zz}({\bm q})$ alone. A more detailed investigation is possible with the order parameter susceptibilities $\bar{\chi}_{\psi}$ [Eq.~(\ref{eq:order-specific})] which we calculate for the $\psi_2$, $\psi_3$ and $T_{1\perp}$ states, see Fig.~\ref{fig:specificOrderFlow} for a representative flow behavior at $\theta=270\degree$. 
Our results show that $\bar{\chi}_{T_{1\perp}}$ is significantly smaller than $\bar{\chi}_{\psi_2}$ and $\bar{\chi}_{\psi_3}$, indicating that quantum fluctuations select the $\Gamma_5$ manifold over the $T_{1\perp}$ states, in agreement with the low-temperature selection found in the classical model and presented in Fig.~\ref{fig:chi_250_300}. 

As discussed in Sec.~\ref{sec:PFFRG}, despite the formulation of the PFFRG at $T=0$, the renormalization group parameter $\Lambda$ shares similarities with the physical temperature $T$. Hence, the type of selection which we detect at the critical $\Lambda_c$ should be interpreted as the (combined quantum and thermal) selection at $T_c$ rather than strictly at $T=0$. The order parameter susceptibilities, for the system approaching its instability point towards either $\psi_2$ or $\psi_3$ states upon lowering the cutoff scale $\Lambda$, i.e., $\bar{\chi}_{\psi_2}$ and $\bar{\chi}_{\psi_3}$ from Eq.~\eqref{eq:order-specific}, are found to be precisely identical in Fig.~\ref{fig:specificOrderFlow}.
While the $\psi_{2}$ and $\psi_{3}$ orders are not related by symmetry, the equivalence $\bar{\chi}_{\psi_2} =\bar{\chi}_{\psi_3}$ can nonetheless be derived from the $C_{3}$ symmetry of the model, as shown in Appendix \ref{appendix:OrderParameterSusceptibilityProof}~\footnote{That the PFFRG computation gives numerically identical results for 
$\bar{\chi}_{\psi_2}$ and $\bar{\chi}_{\psi_3}$ 
down to the lowest $\Lambda$ value is a useful test for the correctness of the implementation of the RG flow equations.}. This property is also observed in high-temperature series expansion [see discussion below Eq.~(\ref{eq:ops})] when temperature approaches the critical temperature $T_c$ from above~\cite{Oitmaa-2013}.
A distinction between $\psi_2$ and $\psi_3$ requires the sixth-order cumulant~\cite{Oitmaa-2013} in Eq.~(\ref{eq:sixth_order}), which corresponds to vertex functions that are not incorporated in existing PFFRG codes and whose calculation would exceed currently available computational resources. To conclude, in its current algorithmic implementation, PFFRG's inability to further characterize the nature of the $\Gamma_5$-ordered phase for the $S=\frac{1}{2}$ version of model~\eqref{eq:hamiltonian} in
the range $\theta \gtrsim 243\degree$ necessitates the usage of HTSE, which we present in the next section.

\subsection{High-temperature series expansion results}
\label{sec:high-T_results}

In the previous sections, we investigated the zero-temperature regime in the quantum $S=\frac{1}{2}$ model and the nature of the transition at $T_c$ in the classical variant. To tie things together, we need to study the transition at $T_c$ in the quantum model. Because of its underlying sign problem, this frustrated $S=\frac{1}{2}$ model~\eqref{eq:hamiltonian} is not amenable to large scale quantum Monte Carlo simulations. As an alternative to study the $\theta$-dependent selection at $T_c$ in the most interesting $\theta \in [250^\circ,360^\circ]$ region, we use HTSE. We analyze the linear order-parameter susceptibility series by d-log Pad\'e approximants to obtain estimates of the critical temperature~\cite{oitmaa_book}.

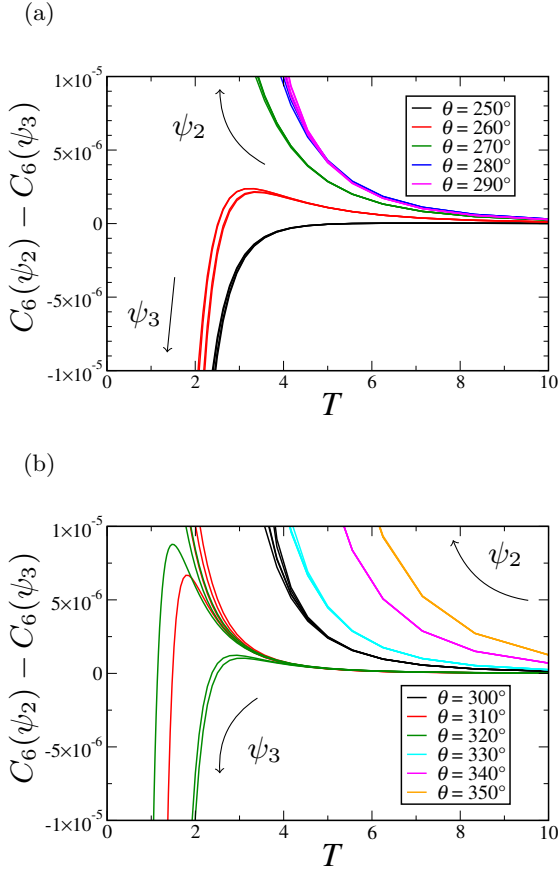
\begin{figure}[ht]
\centering
\begin{tikzpicture}
    \draw (0, 6) node[inner sep=0] {\includegraphics[width=0.8\columnwidth]{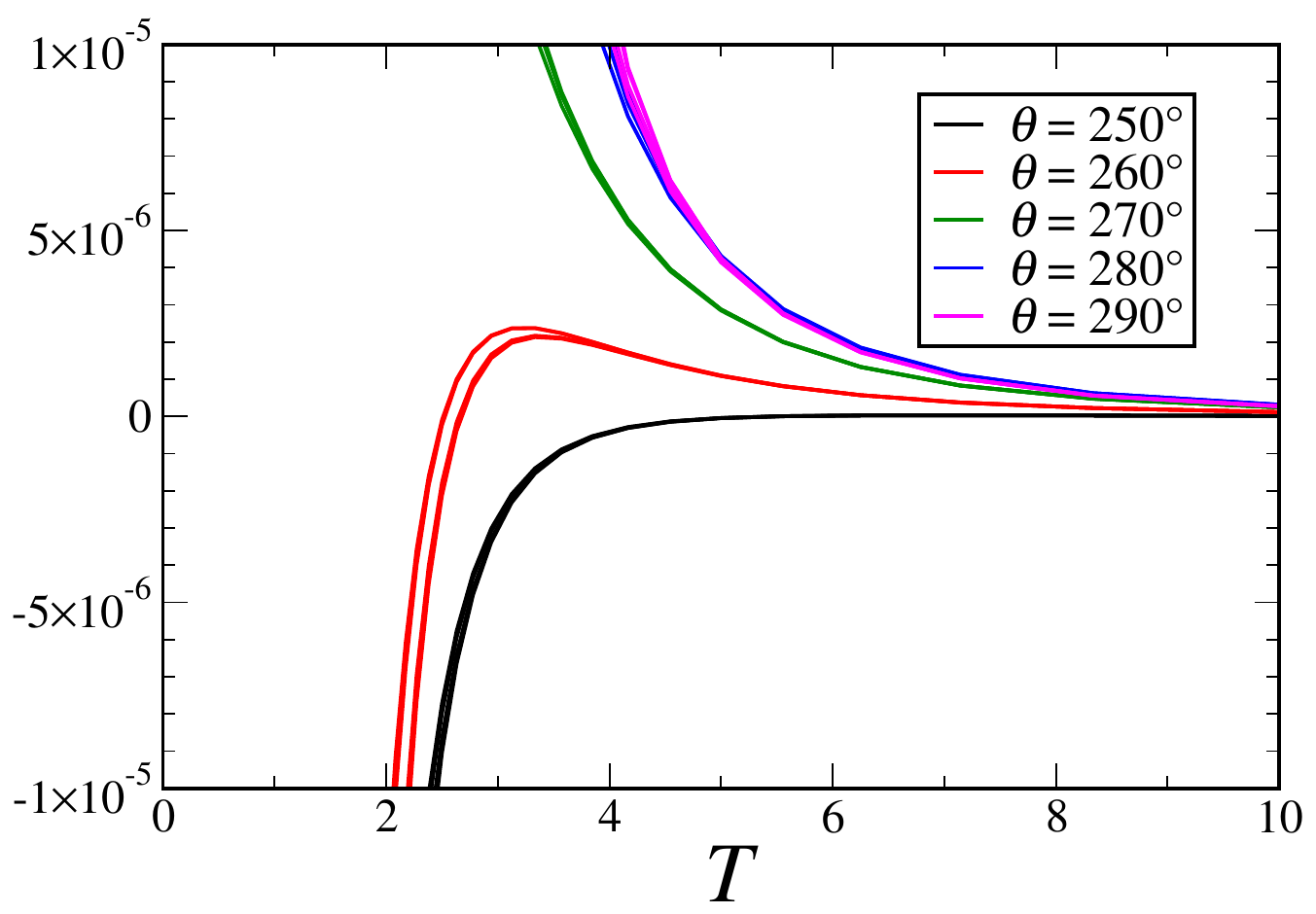}};
     \draw (-3.7, 6.3) node [rotate=90,scale=1.3]{$C_6(\psi_2)-C_6(\psi_3)$};
   
    \draw (0, 0) node[inner sep=0] {\includegraphics[width=0.8\columnwidth]{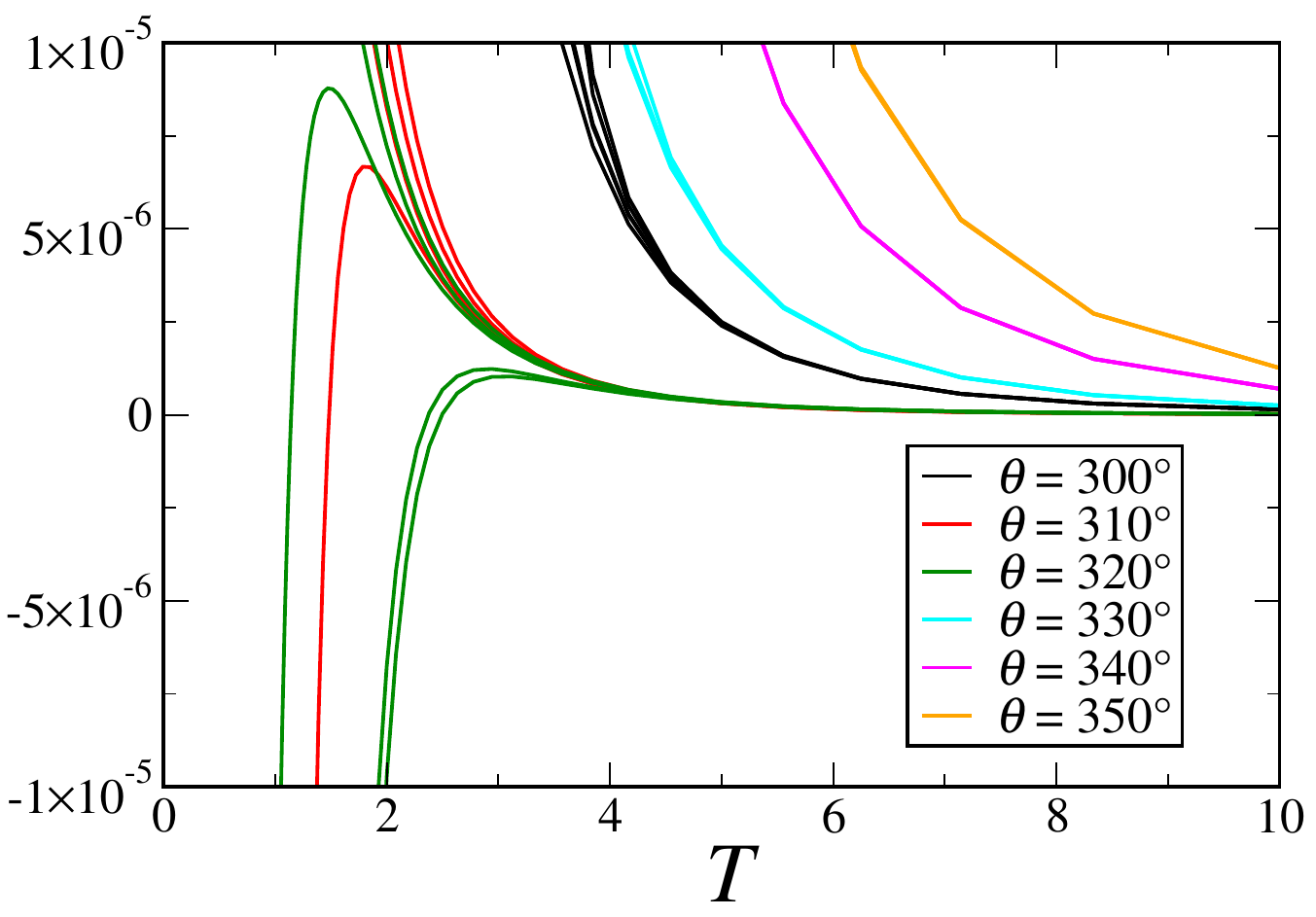}};
    \draw (-3.7, 0.3) node [rotate=90,scale=1.3]{$C_6(\psi_2)-C_6(\psi_3)$};
    \draw (-3.5,9) node [scale=1]{(a)};
    \draw (-3.5,3) node [scale=1]{(b)};

     \coordinate (p5) at (3,1.2);
    \coordinate (p6) at (2,2);
    \draw[->] [black,bend left] (p5) to (p6);

     \coordinate (p7) at (-0.6,-0.1);
    \coordinate (p8) at (-1.1,-1.1);
    \draw[->] [black,bend right] (p7) to (p8);

    \draw (2.7, 1.8) node [scale=1.3]{$\psi_2$};
    \draw (-0.5, -0.8) node [scale=1.3]{$\psi_3$};

    \coordinate (p1) at (-0.5,7.);
    \coordinate (p2) at (-1.1,8);
    \draw[->] [black,bend left] (p1) to (p2);

     \coordinate (p3) at (-1.7,5.5);
    \coordinate (p4) at (-1.8,4.5);
    \draw[->] [black] (p3) to (p4);

    \draw (-1.5, 7.5) node [scale=1.3]{$\psi_2$};
    \draw (-2.1, 5.0) node [scale=1.3]{$\psi_3$};
\end{tikzpicture}
        \caption{Pad\'e approximants for difference series between sixth order cumulants of $\psi_2$ and $\psi_3$ orders. A number of approximants are shown for each value of $\theta$. In (a), we see that for $\theta= 250\degree$ and $260\degree$, $\psi_3$ order is selected whereas for $\theta=270\degree$ through $290\degree$, $\psi_2$ order is selected. In (b), we show the approximants for angles between $300\degree$ and $350\degree$. While $\psi_2$ order seems to be clearly selected for $\theta=300\degree$ and for $330\degree$ and higher, for $\theta=310\degree$ and $320\degree$, some approximants indicate a selection of $\psi_2$ whereas others indicate selection of $\psi_3$ order. This may reflect enhanced competition between the two orders at high temperatures in this small $\theta\in[310\degree,320\degree]$ angular window.
        }
        \label{HTE:X6-Y6}
\end{figure}

We also analyze the difference between the sixth order $\psi_2$ and $\psi_3$ cumulants by Pad\'e approximants to study the selection of order within the $\Gamma_5$ manifold~\cite{Oitmaa-2013}. At high temperatures $T\approx10$, the approximants change slowly with temperature and the difference is of order $10^{-6}$,  reflecting the leading order difference between the two series. Upon decreasing $T$, the difference can become non-monotonic reflecting the competition between different order coefficients. We follow the approximants down in temperature until they begin to grow rapidly in magnitude. At this point,  the sign of the approximant is taken as an indication of the selected order. 
Several Pad\'e approximants~\cite{oitmaa_book,Oitmaa-2013} for this difference series as a function of temperature for angular parameters between $250\degree$ and $350\degree$ are shown in Fig.~\ref{HTE:X6-Y6}.
Our main conclusions based on the analysis are fourfold.

Firstly, the series analysis of the linear susceptibility defined in Eq.~\eqref{eq:ops}  using d-log Pad\'e approximants~\cite{oitmaa_book,Oitmaa-2013} shows best convergence at $\theta=290\degree$, where $T_c$ is estimated to be in the range $0.45-0.47$ and the order parameter susceptibility exponent $\gamma$ is roughly consistent with the three-dimensional XY universality class value of approximately $1.32$~\cite{PhysRevB.63.214503}.

Secondly, moving progressively away from this parameter angle of $\theta=290\degree$, the analysis of the linear susceptibility series becomes ill-behaved in both estimating $T_c$ and $\gamma$, which is an indication that the critical temperature of the quantum $S=\frac{1}{2}$ model goes down rather rapidly as one moves away from this angle. A similar peak in $T_c$ is also seen in the classical model at $\theta=290\degree$ (see Fig.~\ref{fig:Phase_diagram_Gamma_5_T}) and hence the drop in $T_c$ found in these high-$T$ expansion calculations is not solely due to quantum effects.

Thirdly, from Fig.~\ref{HTE:X6-Y6}, we see that $\psi_3$ order is favored for parameter angles $\theta=250\degree$ and $\theta=260\degree$, but $\psi_2$ is favored at most other angles as the temperature is lowered towards $T_c$, coming from the paramagnetic high-temperature side. Note that in Fig.~\ref{HTE:X6-Y6}, we restrict attention to temperatures well above $T_c$ where the 
$C_6(\psi_2)-C_6(\psi_3)$ difference series begins to grow rapidly in magnitude in either positive or negative direction.

Fourthly, in Fig.~\ref{HTE:X6-Y6}(b), we also see that within the parameter range where $\psi_2$ is favored, there is a small window $\theta\in [310\degree,320\degree]$ where some Pad\'e approximants favor $\psi_3$ order as $T\rightarrow T_c^+$ over $\psi_2$. This may be an indication of enhanced competition between the two orders in the middle of the region where $\psi_2$ selection occurs upon approaching $T_c$ from above.
Note that $\theta\approx 310\degree$ appears to correspond to the angle where the $T_c$ for the transition from the paramagnetic phase into $\psi_2$ displays a sudden fast drop in the classical phase diagram of Fig.~\ref{fig:Phase_diagram_Gamma_5_T}.

The HTSE results are broadly consistent with the study of the classical model and the phase diagram shown in Fig.~\ref{fig:Phase_diagram_Gamma_5_T}. The paramagnetic transition temperature is largest in the classical model in a small angular range near $\theta=300\degree$. The quantum and classical models also show similar behaviors in that $\psi_3$ order is favored at the paramagnetic phase boundary only at the smallest angles within the $\Gamma_5$/copl regime [i.e. $\theta=250 \degree$ and $\theta=260\degree$ within HTSE, see Fig.~\ref{HTE:X6-Y6}(a)], whereas $\psi_2$ order is favored over the rest of this regime. In the quantum model studied via HTSE, the enhanced uncertainty in the selection of order at $\theta\in [310\degree,320\degree]$ [see Fig.~\ref{HTE:X6-Y6}(b)] could simply be due to the rapid decrease in $T_c$. The precise angles at which $T_c$ is maximum or whether there is a local minimum in the ratio of outer 
(paramagnetic) $T_c$ and inner $T_{\Gamma_5}$ transition temperatures indicating an enhanced competition between the $\psi_2$ and $\psi_3$ phases (see Fig.~\ref{fig:Phase_diagram_Gamma_5_T}) may also differ between the classical and quantum models. Such quantitative differences may in fact be expected as the two models are not quantitatively the same. 

Given the uncertainties of the series expansion study and allowing for some quantitative differences between the classical and quantum models, these HTSE results affirm that the basic physics of order and selection in the classical and quantum models at the paramagnetic transition at $T_c$ is largely the same.

\section{\label{sec:discussion}Summary and Outlook}

Our comprehensive Monte Carlo analysis of the classical model shows that within the parameter regime $\theta\in(243\degree,360\degree)$ where the $\Gamma_{5}$ and $T_{1\perp}$ states are energetically degenerate, thermal fluctuations select the $\Gamma_{5}$ phase at low temperatures, as revealed from the corresponding susceptibility data. A subsequent analysis of the $m_{E6}$ order parameter helps to decipher the precise nature of ordered configurations stabilized below $T_{c}$, revealing a cascade of phase transitions. Indeed, while for $\theta\in(243\degree,265\degree]$ there occurs a single step $\psi_{3}$ selection from $T_{c}$ down to $T=0^{+}$, for $\theta\in(265\degree,360\degree)$ there is a two-step transition whereby there takes place a $\psi_{2}$ selection at $T_{c}$ followed by a lower temperature $\psi_{3}$ selection at $T_{\Gamma_{5}}$. Interestingly, at $\theta=350\degree$, we observe a local minimum in $T_{\Gamma_{5}}$ which can be rationalized within a low-temperature  expansion wherein the behavior of the entropic weight difference $\Delta S$ between $\psi_{3}$ and $\psi_{2}$ shows an enhanced competition at $\theta \approx 350\degree$. A suppression of $T_c$ is observed at the lower boundary of the $\Gamma_5$/copl regime at $\theta\approx243\degree$ (i.e., at $D/J=2$). Similar to the pure Heisenberg antiferromagnetic model, this behavior at $D/J=2$ can be rationalized on the basis of  an extensively degenerate ground state manifold at this $D/J$ value. We defer a more detailed investigation of this interesting point to a forthcoming publication~\cite{Gomez-DJ_2}.

At $T=0$, an analysis of the ObD due to quantum fluctuations treated in the semiclassical limit $(1/S\ll1)$ via linear spin-wave theory finds the $\psi_{3}$ state being selected within large portions of the $\Gamma_5$/copl regime, for $\theta\in(243\degree,344\degree]$ and $\theta\in[352\degree,360\degree)$, akin to the selection effect from thermal fluctuations for classical spins, i.e., $1/S=0$.
Interestingly, for $\theta\in(344\degree,352\degree)$, we find that the energetic selection oscillates between $\psi_{2}$ and $\psi_{3}$. It would be interesting to investigate whether a treatment of anharmonic terms in spin-wave expansion could resolve the subtle selection effects at work within this region.

For the extreme quantum value of spin $S=\frac{1}{2}$, a HTSE analysis provides insights into the selection effects at $T_{c}$, which show a qualitatively similar behavior compared to that found in the classical version of the model. In particular, $\psi_{3}$ order is favored only at the smallest angles $\theta$ within the $\Gamma_5$/copl regime, while in the rest of this regime,  $\psi_{2}$ order is selected at $T_c$. Furthermore, similar to what is found for classical spins, there exists a sliver in the fourth quadrant $\theta\in[310\degree,320\degree]$, i.e., in the middle of the region where $\psi_{2}$ is selected at $T_{c}$, where an enhanced competition between the $\psi_{2}$ and $\psi_{3}$ ordering tendencies is observed for $S=\frac{1}{2}$. These results indicate that selection effects for classical and $S=\frac{1}{2}$ spins are broadly similar at $T_c$.

A salient finding obtained from the $S=\frac{1}{2}$ PFFRG analysis is that, upon inclusion of direct and indirect DMI in the Heisenberg antiferromagnetic model, the quantum paramagnetic ground state persists over an appreciable parameter range. Specifically, no indications for magnetic instabilities are found for $-9\degree \lesssim \theta \lesssim 8\degree$ ($-0.16 \lesssim D/J \lesssim 0.14$). 
Guided by the PFFRG results of Ref.~\cite{Iqbal19} for the $S=1$ pyrochlore Heisenberg antiferromagnetic, where a disordered phase is found, we speculate that it is likely that the nonmagnetic phase survives for $S=1$, albeit with a reduced extent in parameter space and possibly of a different nature. 
This regime could be host to novel types of nonmagnetic phases, and it would be important to investigate further their microscopic nature. At the pure Heisenberg point ($\theta=0\degree$), for $S=\frac{1}{2}$, various numerical studies provide compelling indications for a ground state with broken lattice symmetries~\cite{hagymasi21,Astrakhantsev21,Hering-2022}, a valence bond crystal being a promising candidate. Most recently, this possibility was further confirmed by the identification of energetically favorable hard-hexagon crystal tilings corresponding to translational symmetry broken valence bond crystal states with large 48-site unit cells~\cite{schaefer22}. In the presence of finite DMI beyond the gap size of such states, however, the ground state scenario can be expected to change considerably, since the system is no longer SU(2) spin-symmetric which disfavors singlet formation, possibly driving the formation of another nonmagnetic ground state phase. In particular, it would be worthwhile exploring the scenario where an exotic chiral spin liquid is potentially induced by a DMI similar to what has been proposed for the kagom\'e lattice~\cite{Messio-2017}. The question of the microscopic identification of the nature of the nonmagnetic phase can be addressed within the PFFRG framework itself by combining it with a self-consistent Fock-like mean-field scheme to calculate low-energy effective theories for emergent spinon excitations in $S=\frac{1}{2}$ systems~\cite{Hering-2019psg,Hering-2022}. One could also perform Gutzwiller projected variational wave function studies employing Monte Carlo methods~\cite{Hering-2022,Iqbal-2013} based on a projective symmetry group classification of ans\"atze with different low-energy gauge groups~\cite{Liu-2021,Schneider-2022} aimed at identifying the variational ground state, and determine its spectrum of excitations~\cite{Ferrari-2022}. 

Our PFFRG analysis also hints at the existence of a quantum paramagnetic state in the regime where ferromagnetic Heisenberg interactions compete with indirect DMI, i.e., for $237\degree \lesssim \theta \lesssim 241.5\degree$ ($1.54 \lesssim D/J \lesssim 1.84$). As this region is proximate to the phase boundary with FM order, it opens up the exciting possibility of realizing spin nematic orders~\cite{Andreev-1984} on the 3D pyrochlore lattice, similar to what has been reported on square~\cite{Shannon-2006,Iqbal-2016a} and kagome lattices~\cite{Wietek-2020}.

Taken together, besides the rich magnetic behavior of the Heisenberg and Dzyaloshiskii-Moriya model on the pyrochlore lattice that we have uncovered here, this system continues to host a multitude of fascinating open aspects which are worth addressing in future works.

\section*{Acknowledgements}
We acknowledge stimulating and useful discussions with Eric Andrade, Alex Hickey, Subhankar Khatua and Jeffrey Rau.
V.N. would like to thank the HPC Service of ZEDAT and Tron cluster service at the Department of Physics, Freie Universität Berlin, for computing time.
The work in Waterloo was supported by the NSERC of Canada and the Canada Research Chair (Tier 1, MJPG) program. 
Simulations done at Waterloo were performed thanks to the computational resources of Compute Canada.
J.R. thanks the Indian Institute of Technology Madras, Chennai for a Visiting Faculty Fellow position under the IoE program which facilitated the completion of the research work and writing of the manuscript.
Y.I. acknowledges financial support by the Science and Engineering Research Board (SERB), Department of Science and Technology (DST), India through the Startup Research Grant No.~SRG/2019/000056, MATRICS Grant No.~MTR/2019/001042, and the Indo-French Centre for the Promotion of Advanced Research (CEFIPRA) Project No. 64T3-1. This research was supported in part by the National Science Foundation under Grant No.~NSF~PHY-1748958, ICTP through the Associates Programme and from the Simons Foundation through Grant No. 284558FY19, IIT Madras through the Institute of Eminence (IoE) program for establishing the QuCenDiEM CoE (Project No. SP22231244CPETWOQCDHOC), the International Centre for Theoretical Sciences (ICTS), Bengaluru, India during a visit for participating in the program “Frustrated Metals and Insulators” (Code: ICTS/frumi2022/9). Y.I. thanks IIT Madras for IoE travel grant which facilitated progress on the research work and this manuscript. Y.I. acknowledges the use of the computing resources at HPCE, IIT Madras.\\

\appendix

\section{Irreducible representations in the global spin basis}
\label{appendix:configurations}

The spin modes defined by the different irreps may be specified by identifying the corresponding spin configuration. To this end, we first define the local basis at the $0$th sublattice in
the global coordinate frame as
\begin{eqnarray}
\bm x_0=\frac{1}{\sqrt{6}}\begin{pmatrix}-1\\ -1\\ 2\end{pmatrix},\quad
\bm y_0=\frac{1}{\sqrt{2}}\begin{pmatrix}1\\ -1\\ 0\end{pmatrix},\quad
\bm z_0=\frac{1}{\sqrt{3}}\begin{pmatrix}1\\ 1\\ 1\end{pmatrix},\nonumber
\end{eqnarray}
where the local basis for the other sublattices can be obtained by the application of a two-fold rotation symmetry $C_2$. With these local bases in hand, the irrep mode $\bm{m}_I^\boxtimes$ defined for a single tetrahedron $\boxtimes$ for the different irreps, labeled $I$, are given by 
\begin{eqnarray}
    m_{A_2}^\boxtimes&=&\frac{1}{2\sqrt{3}}\sum_{\mu} (\bm z_\mu \cdot\bm S_\mu),\label{eq:irreps_appendix_A1}\\
    \bm m_{E}^\boxtimes&=&\frac{1}{2}\sum_{\mu}\begin{pmatrix} \frac{1}{\sqrt{6}} \bm x_\mu \cdot\bm S_\mu \\ \frac{1}{\sqrt{2}}\bm y_\mu\cdot \bm S_\mu\end{pmatrix},\\
    \bm m_{T_{1\parallel}}^\boxtimes&=&\frac{1}{2}\sum_{\mu}\begin{pmatrix}S_\mu^x \\  S_\mu^y \\ S_\mu^z\end{pmatrix},\\
    \bm m_{T_{1\perp}}^\boxtimes&=&\frac{1}{2\sqrt{2}}\sum_{\mu}\begin{pmatrix} z_{\mu}^x  \bm v_\mu^{yz}\cdot \bm S_\mu  \\ z_{\mu}^y \bm v_\mu^{xz}\cdot \bm S_\mu^y \\ z_{\mu}^z \bm v_\mu^{xy} \cdot\bm S_\mu^z \end{pmatrix},\\
    \bm m_{T_{2}}^\boxtimes&=&\frac{1}{2\sqrt{2}}\sum_{\mu}\begin{pmatrix} (\bm z_\mu \times \bm S_\mu)^x \\  (\bm z_\mu \times \bm S_\mu)^y \\ (\bm z_\mu \times \bm S_\mu)^z \end{pmatrix},\label{eq:irreps_appendix_A5}
\end{eqnarray}
where $z^\alpha_\mu$ ($S^\alpha_\mu$) is the $\alpha$ component of the \textit{local}-$\bm z$ direction (spin $\bm{S}$) on sublattice $\mu$ expressed in the global Cartesian frame, and $\bm v^{\alpha\beta}_\mu$ are the normalized bond vectors attached to a sublattice site $\mu$ belonging to an up-tetrahedron and which lie on the cubic $\alpha$-$\beta$ plane defined as

\begin{eqnarray}
\bm v^{xy}_{0}&=&-\bm v^{xy}_{3}=\frac{1}{\sqrt{2}}\begin{pmatrix}1\\ 1\\ 0\end{pmatrix},\qquad
 \bm v^{xy}_{2}=-\bm v^{xy}_{1}=\frac{1}{\sqrt{2}}\begin{pmatrix}-1\\ 1\\ 0\end{pmatrix},\nonumber\\
\bm v^{xz}_{0}&=&-\bm v^{xz}_{2}=\frac{1}{\sqrt{2}}\begin{pmatrix}1\\ 0\\ 1\end{pmatrix},\qquad
\bm v^{xz}_{3}=-\bm v^{xz}_{1}=\frac{1}{\sqrt{2}}\begin{pmatrix}1\\ 0\\ -1\end{pmatrix},\nonumber\\
\bm v^{yz}_{0}&=&-\bm v^{yz}_{1}=\frac{1}{\sqrt{2}}\begin{pmatrix}0\\ 1\\ 1\end{pmatrix},\qquad
\bm v^{yz}_{3}=-\bm v^{yz}_{2}=\frac{1}{\sqrt{2}}\begin{pmatrix}0\\ -1\\ 1\end{pmatrix}\nonumber.
\end{eqnarray}

\section{Irreducible representations parameters}
\label{appendix:Irrep_Energy_Parameters}
As discussed in Sec.~\ref{subsection:state_class}, the hierarchy of the interaction energy parameters (IEP), and more specifically the irrep corresponding to the minimum IEP, defines the classical ground state spin configuration. In terms of the Heisenberg and DM interaction parameters, $J$ and $D$, the IEP are given by
\begin{eqnarray}
a_{A_2}&=&-J-4D,\\
a_{E}&=&-J+2D,\\
a_{T_2}&=&-J-2D,\\
a_{T_{1\parallel}}&=& 3J,\\
a_{T_{1\perp}}&=&-J+2D.
\end{eqnarray}
In particular, note that the $a_{E}$ coincides with the $a_{T_{1\perp}}$ IEP for all values of $J$ and $D$, as shown in Fig.~\ref{fig:Irreps}. With the above equations, the boundary between different low-temperature phases can be identified by studying the ratios $D/J$ for which the IEP are degenerate. For example, for the phase boundaries in the second and third quadrant of Fig.~\ref{fig:phasediagram}(e), setting $a_{A_2}=a_{T_{1\parallel}}$ results in 
\begin{equation}
    -J-4D=3J\rightarrow \frac{D}{J}=-1,
\end{equation}
which corresponds to an angle $\theta=135\degree$, signaling the transition between AIAO order and the colinear ferromagnetic order. Likewise, setting $a_{E}=a_{T_{1\parallel}}$ yields
\begin{equation}
    -J-2D=3J\rightarrow \frac{D}{J}=2,
\end{equation}
which corresponds to an angle $\theta\approx243\degree$ and signals  the transition between colinear ferromagnetic order and the orders in the $\Gamma_5$/copl manifold.

Finally, and for completeness, we provide the IEP in terms of the $\theta$-parametrization given in Eq.~\eqref{eq:parametrization}:

\begin{eqnarray}
a_{A_2}&=&-\cos(\theta)-4\sin(\theta),\\
a_{E}&=&-\cos(\theta)+2\sin(\theta),\\
a_{T_2}&=&-\cos(\theta)-2\sin(\theta),\\
a_{T_{\parallel}}&=&3\cos(\theta),\\
a_{T_{1\perp}}&=&-\cos(\theta)+2\sin(\theta).
\end{eqnarray}

\section{Curie-Weiss behaviors of $\Gamma_5$ and $T_{1\perp}$ susceptibilities}
\label{appendix:HTSE_proof}

To further illustrate how the energy degeneracy between the $E$ and the $T_{1\perp}$ states results in an equality of the associated irrep susceptibilities $\chi_{I}$ in their respective Curie-Weiss behavior and their value in a high-temperature expansion to order $\beta^2$, we compute these in a HTSE to order $\beta^2$. To this end, we first relabel the different terms in the Hamiltonian of Eq.~\eqref{eq:hamiltonian} as 
\begin{eqnarray}
    \mathcal{H}&=&\underbrace{J\sum_{\langle i ,j \rangle } \hat{\bm S}_i\cdot \hat{\bm S}_j}_{\mathbb{J}}+\underbrace{ \sum_{\langle ij \rangle}{\bm D}_{ij} \cdot ( {\hat{\bm S}}_{i} \times { \hat{\bm S}}_{j} )}_{\mathbb{D}}+\underbrace{\sum_{\boxtimes} h^\boxtimes_I  m^\boxtimes_I}_{\mathbb{M}}\nonumber\\
    &\equiv&\mathbb{J}+\mathbb{D}+\mathbb{M},\label{eq:HTSE_Hamiltoniand}
\end{eqnarray}
where $ m^\boxtimes_I$ is one of the irrep modes defined in Eqs.~\eqref{eq:irreps_appendix_A1}-\eqref{eq:irreps_appendix_A5}, 
and  $h^\boxtimes_I$ is the corresponding conjugate field defined on that tetrahedron \footnote{For example, for the $\psi_2$ states this would correspond to local fields along the local $x$ direction.}, for more details we refer the reader to Sec.~\ref{subsec:HTSE}. 
For a given irrep mode $m^\boxtimes_I$, the HTSE for its associated susceptibility is computed and gives
\begin{eqnarray}
\chi_{m^\boxtimes_I}&=&\frac{1}{2^N}\left[\frac{\beta}{2}\frac{\partial^2}{\partial h^2}\rm{Tr}(\mathbb{M}^2)-\frac{\beta^2}{3}\frac{\partial^2}{\partial h^2 }\rm{Tr}\left((\mathbb{J}+\mathbb{D})\mathbb{M}^2 \right)\right]\bigg\rvert_{h=0}\nonumber\\
&& +\mathcal{O}(\beta^3),\label{eq:HTSE_susceptibility}
\end{eqnarray}
where $N$ is the number of spins. With the above equation, the susceptibility of the $x$-$y$ $T_{1\perp}$ state, i.e. the $T_{1\perp}$ state where all the spins lie on the cubic $x$-$y$ plane while pointing along the nearest-neighbor bonds [illustrated in Fig.~\ref{fig:phasediagram}(b)], denoted as $T_{1\perp}^{xy}$, is found to be
\begin{eqnarray}
\chi_{T_{1\perp}^{xy}}&=&\frac{N_\boxtimes}{2^N}\left[\beta+\frac{\beta^2}{3}(J+2D)\right]\nonumber +\mathcal{O}(\beta^3)\label{eq:HTSE_susceptibility_M_t1_xy},
\end{eqnarray}
where $N_\boxtimes$ is the number of tetrahedra. The susceptibilities for the other $T_{1\perp}$ states defined for the other cubic planes are found to be equivalent. The same calculation for the $\psi_3^{xy}$ state, i.e., the $\psi_3$ state where all spins lie in the $x$-$y$ plane, reveals that the susceptibility associated with this configuration is, to order $O(\beta^2)$, identical to the one found for the $T_{1\perp}^{xy}$ states:
\begin{equation}
    \chi_{\psi_3^{xy}}=\chi_{T_{1\perp}^{xy}}.
\end{equation}
Finally, since the $\psi_2$ states are described by the same $E$ irrep as the $\psi_3$ states the associated susceptibility is again the same. In other words, to order $\beta^2$, the HTSE of the susceptibilities for orders defined by the $\Gamma_5$ and $T_{1\perp}$ states are equal.

\section{Distribution of the $\phi$ angle in the $\Gamma_5$ phase}
\label{appendix:distribution}

\begin{figure}[ht]
    \begin{overpic}[width=1.0\columnwidth]{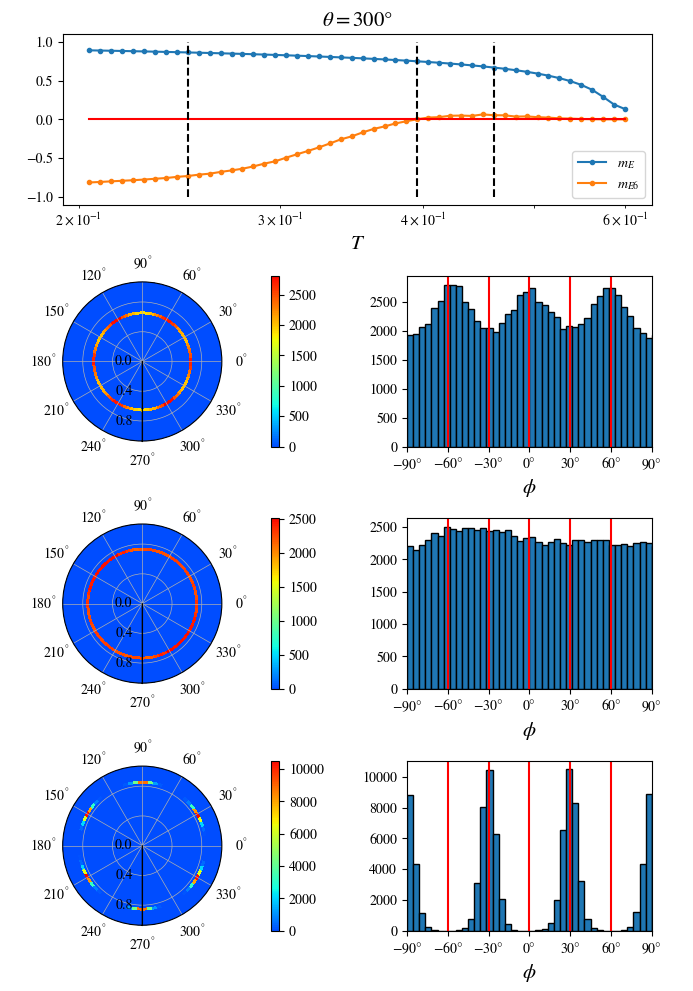}
     \put(2,98){(a)}
     \put(2,72){(b)}
      \put(36,72){(c)}
     \put(2,49){(d)}
     \put(36,49){(e)}
     \put(2,24){(f)}
     \put(36,24){(g)}
     \put(46,90){$T_0$}
     \put(38,90){$T_1$}
     \put(15,90){$T_2$}
     \end{overpic}
    \caption{Polar  and $1D$ histograms of the $\phi$ angle as defined in Eq.~(\ref{eq:phidef}) for the temperatures marked by the dashed black lines in panel (a). Panel (a) illustrates the $m_E$ and $m_{E6}$ parameters measured for $\theta=300\degree$. Panels (b), (d) and (f) show polar histograms of $\bm m_{E,\rm av}$. Panels (c), (e) and (g) show $1D$ histograms of the angle $\phi$ where the vertical red lines mark multiples of $\pi/6$. Here, the second row [panels (b) and (c)] corresponds to data taken at $T_0$, the third row [panels (d) and (e)] corresponds to data taken at $T_1$, and the fourth row [panels (f) and (g)] corresponds to data taken at $T_2$. 
    All temperatures are in units of     $\sqrt{J^2+D^2}/k_{\text{B}}$, with $T_0= 0.461$, $T_1= 0.395$, $T_2= 0.249$.
     }
        \label{fig:Histograms}
\end{figure}
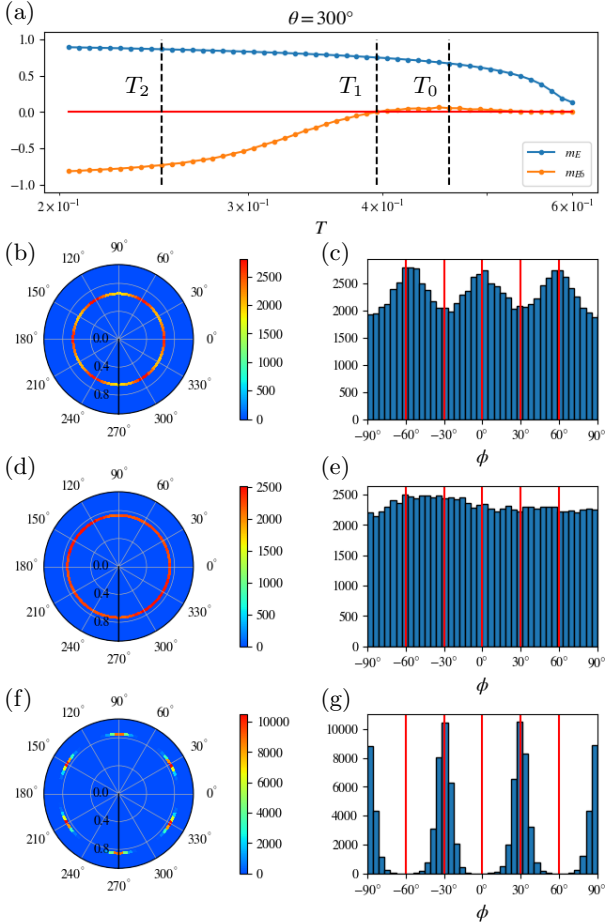

We discussed in Sec.~\ref{subsec:mc_results}, the nature of the  classical ordered state of the Heisenberg-DM model in the  $\theta \in (243 \degree, 360\degree)$ $\Gamma_5$/copl region where the system orders in the $\Gamma_5$ phase at $T_c$; see Figs. \ref{fig:phasediagram}(e), \ref{fig:chi_250_300},
\ref{fig:ME_ME6_250_300}, \ref{fig:Phase_diagram_Gamma_5_T} and \ref{fig:DS_DE}(a).   Two seemingly clear and robust key results were obtained in regard to the state selection below $T_c$: (i) at $T=0^+$, thermal fluctuations stabilize a $\psi_3$ state  throughout the $\theta \in (243 \degree, 360\degree)$ range because of its higher entropy; (ii) the transition at $T_c$ is into a $\psi_3$ state for   $\theta \in (243 \degree, 265\degree)$  while it is into a $\psi_2$ state for 
$\theta \gtrsim 265\degree$ (see Figs.~\ref{fig:ME_ME6_250_300} and \ref{fig:Phase_diagram_Gamma_5_T}). The state selection at $T_c$ for the quantum $S=\frac{1}{2}$ model was discussed in Sec.~\ref{sec:high-T_results} (see  Fig.~\ref{HTE:X6-Y6}). The results for the classical system for $\theta \gtrsim 265\degree$ with the noted 
distinction between the $\psi_2$ state selection at $T=T_c$ versus the  $\psi_3$ selection at $T=0^+$ motivates a brief discussion of how the system evolves from $\psi_2$ to $\psi_3$ as the temperature is reduced from $T\lesssim T_c$ to $T=0^+$.

According to a Ginzburg-Landau theory argument~\cite{Javanparast15}, the two simplest scenarios for the $\psi_2 \rightarrow \psi_3$ transition are (i) a first order transition or (ii) two consecutive Ising transitions at  $T_{\rm I}^+$ and $T_{\rm I}^-$, with  
$\cramped{T_{\rm I}^- < T_{\rm I}^+ < T_c}$. In the first case, the order parameter $m_{E6}$ should jump discontinuously from a positive value to a negative value at $T_{\Gamma_5}$ for a thermodynamically large system and the internal energy should exhibit a discontinuity. In the second case, the system would be found to have a distribution of the spins (local $xy$) in-plane projection to be peaked at $\phi=n\pi/3$ in the $\psi_2$ phase, with this distribution continuously starting to shift to an angle $\phi(T)$,  $n\pi/3 <\phi(T) < (2n+1)\pi/6$, at $T=T_{\rm I}^+$.  
For the system size considered, we were unable to detect any sign of  peaks in the specific heat (not shown) that could indicate either a single first order transition or two second order transitions.
The same difficulty in detecting the $\psi_2$ to $\psi_3$ transition in a classical Monte Carlo study of a model of the disordered Er$_{2-x}$Y$_x$Ti$_2$O$_7$ rare-earth pyrochlore compound was reported in Ref.~\cite{AndradePRL2018}.
Similarly to the approach taken in our paper, and with the results presented in Sec.~\ref{subsec:mc_results}, the authors of Ref.~\cite{AndradePRL2018} characterized the ordered phases below the critical temperature $T_c(x)$ using the sign of an order parameter akin to our $m_{E6}$ order parameter in Eq.~(\ref{eq:me6}).
In the same vein, Chern in his study of the Heisenberg-DM pyrochlore model~\cite{chern10} also characterized the nature of the ordered phase below $T_c$ through the sign of a quantity $\zeta_6$, again analogous to our $m_{E6}$ order parameter.

As it stands, because of computational limitations, a precise quantitative  determination of the phase and phase transition occurring within the $\Gamma_5$ phase for $T<T_c$ in the pyrochlore Heisenberg-DM model is still wanting. 
We attempted to identify in our Monte Carlo simulations which of the above two scenarios (one first order transition or two Ising transitions) is operating by focusing on $\theta=300\degree$ and considering the histograms of distribution of the local $xy$ in-plane orientation $\phi$  of the spins. These results are illustrated in  Fig.~\ref{fig:Histograms}. Panel (a) shows the evolution of $m_E$ and $m_{E6}$ for $\theta=300\degree$. Three representative temperatures are considered: $T_0<T_c$ and in the state labeled $\psi_2$ according to our convention in Sec.~\ref{subsec:class_mc}; $T_1<T_0$ 
where $m_{E6}$ changes sign and signals a transition from
$\psi_2$ to $\psi_3$, and $T_2$, deep at $T_2 \ll T_c$ in the $\psi_3$ regime. Panels (b), (d), and (f) show the distribution of $\phi$, $P(\phi)$, in a polar plot. The average radius of the circles corresponds to $m_E$, increasing as $T$ decreases from
$T_0$ to $T_2$. In panel (b) for $T=T_0$, the azimuthal distribution is slightly peaked at angles $\phi=n\pi/3$, also illustrated in a histogram form in panel (c), and which indicates a $\psi_2$ state.
For $T=T_1$, where the system transits from $\psi_2$ to $\psi_3$ according to our sign-of-$m_{E6}$ convention,
panel (d) shows that the azimuthal distribution is essentially unmodulated, as confirmed by the rather flat (though noisy) histogram in panel (e).
Finally, at $T_2\ll T_0$, the azimuthal $P(\phi)$  distribution in panel (f) is highly modulated and strongly peaked at $\phi=(2n+1)\pi/6$, as also seen in the histogram of panel (g), and indicating a well ordered $\psi_3$ state.
Note that these histograms were constructed using 100 independent Monte Carlo simulations of the model at $\theta=300\degree$.

Considering the histograms $P(\phi)$ for $\theta=300\degree$, we found that the peaks of $P(\phi)$ evolve \emph{very rapidly} from
being located at $\phi=n\pi/3$ values (indicating a $\psi_2$ state) at $T_1 < T_0 < T_c$, to being located at 
$\phi=(2n+1)\pi/6$ for $T<T_1$, indicating a $\psi_3$ state. 
Moreover, we found no clear signature that the histograms at $T_1<T<T_c$ display clear peaks at intermediate angles $\phi$ between $n\pi/3$ and $(2n+1)\pi/6$, that might suggest two Ising transitions~\cite{Javanparast15}. 
These results could tentatively be suggestive of a first order 
transition between $\psi_2$ to $\psi_3$, but not readily detectable in the Monte Carlo thermodynamic data. 
Clearly, a more systematic study 
considering much larger system sizes is  needed in order to resolve the question of the $\psi_2$ to $\psi_3$ transition within the regime of $\Gamma_5$ ordering in the Heisenberg-DM model.

\section{Static susceptibility in the nonmagnetic phase around $\theta=0$ from PFFRG}
\label{appendix:additional_PFFRG}

\begin{figure}[ht!]
        \centering
        \hspace*{-0.25cm} 
\includegraphics[width = 0.5\textwidth]{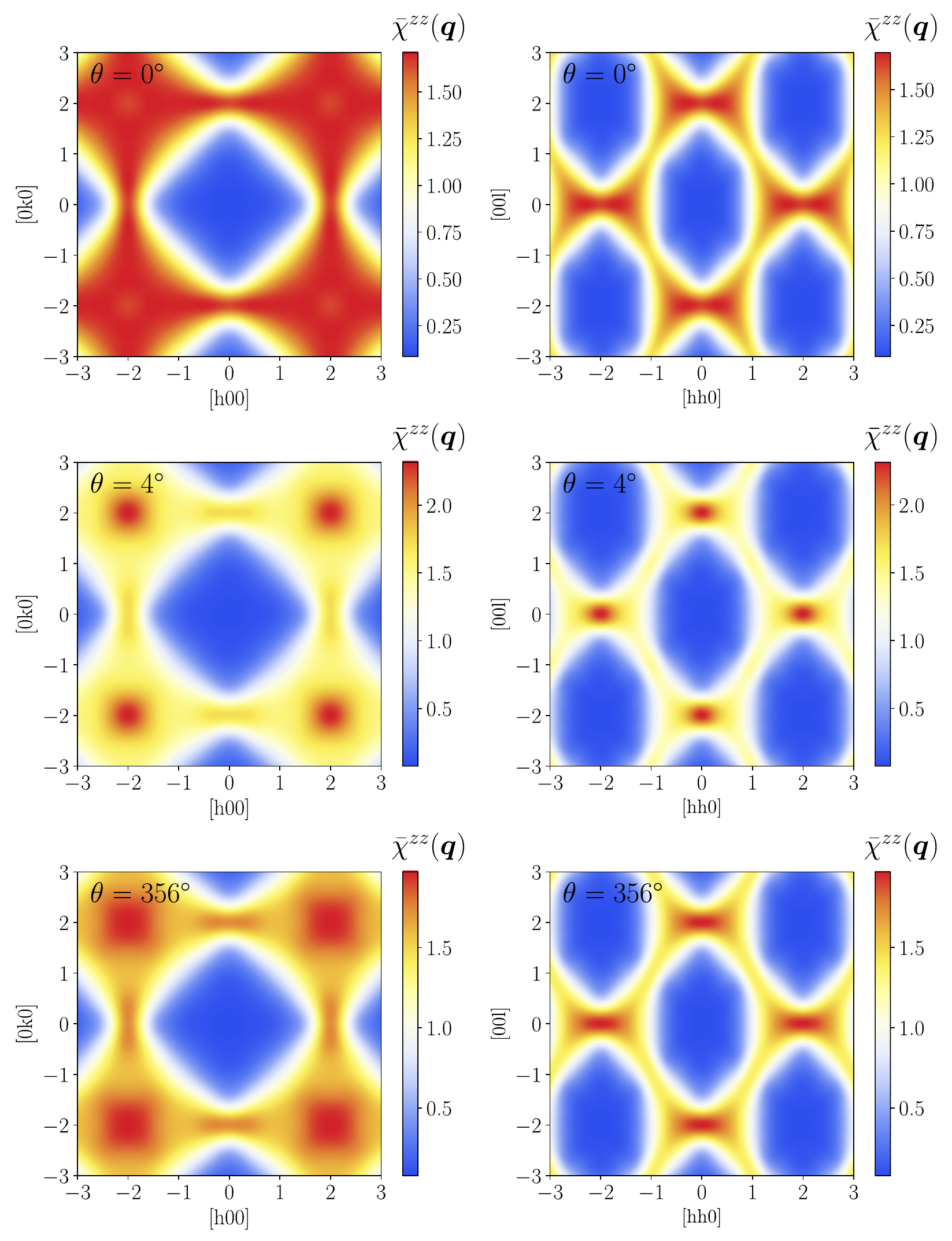}
    \caption{Static $zz$ susceptibility, as defined in Eq. \eqref{eq:PFFRGsusceptibility}, of the pure and DMI perturbed Heisenberg antiferromagnet, obtained from PFFRG in the low cutoff limit. The $[hk0]$ and $[hhl]$ planes are shown in the left and right columns, respectively.}
    \label{fig:PFFRGStaticSusceptibility}
\end{figure}

In Fig.~\ref{fig:PFFRGStaticSusceptibility}, we show  the effects of a small DMI on the ground state static susceptibility $\bar{\chi}^{zz}({\bm q})$ for the cutoff-free system ($\Lambda\rightarrow0$) in the $[hk0]$ and $[hhl]$ planes. For the Heisenberg model ($\theta=0\degree$), broadened pinch points are visible at $hkl=002$. For $\theta=\pm4\degree$, still in the nonmagnetic regime, the weights partially redistribute towards the $hkl=002$ positions. This redistribution is more pronounced for a direct ($D>0$) DMI.

\begin{figure}[ht!]
        \centering
        \includegraphics[width = 0.45\textwidth]{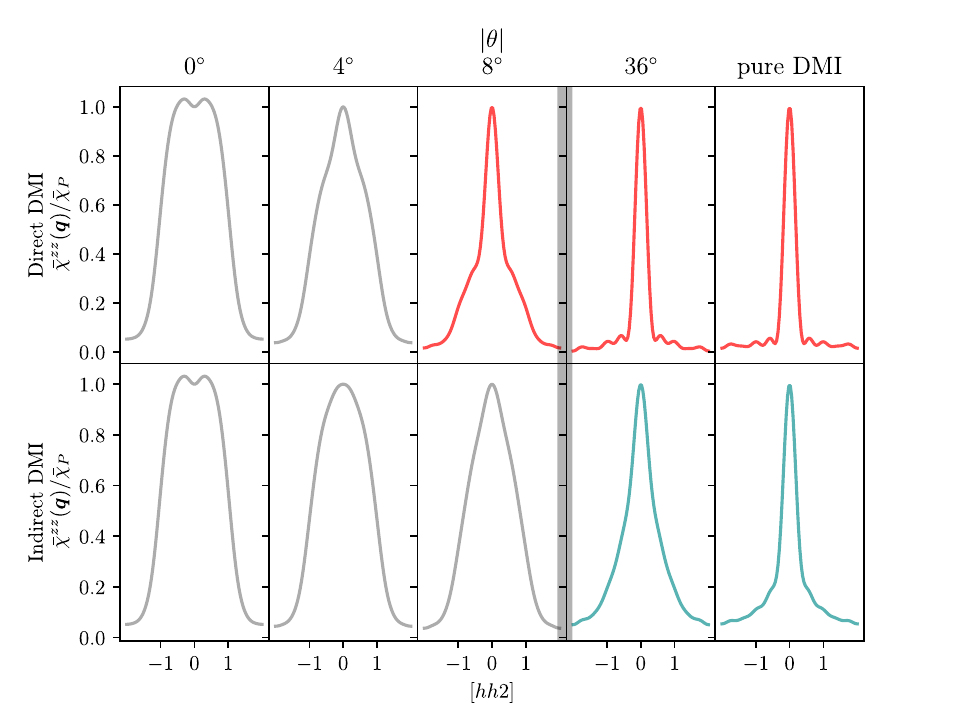}
        \caption{Normalized line shape of the static $zz$ susceptibility along $[hh2]$, i.e. across the pinch point/peak $\bar{\chi}_{P}$ at $hkl=002$. From left to right, starting from a pure antiferromagnetic Heisenberg model, the evolution of the line shape with increasing absolute DMI is shown. The transitions between paramagnetic and ordered phases are highlighted by bold vertical lines.}
        \label{fig:peakCrossSection}
\end{figure}

An alternative perspective on the effect of a finite DMI on the pinch point shape is illustrated in Fig. \ref{fig:peakCrossSection}, which shows the PFFRG static susceptibility along $[hh2]$ for different parameter angles $\theta$.

\section{Proof of equal $\Gamma_5$ order parameter susceptibilities}
\label{appendix:OrderParameterSusceptibilityProof}

Both HTSE and PFFRG find equal order parameter susceptibilities $\chi_{\psi_2}$ and $\chi_{\psi_3}$ for the $\psi_2$ and $\psi_3$ orders, respectively. In this appendix, we provide an explanation for this equivalence by demonstrating that all order parameter susceptibilities associated with states of the $\Gamma_{5}$ manifold are equal due to the $C_{3}$ lattice symmetry.
To this end, we consider spin correlations between sites $i$ and $j$ expressed as $3\times3$ matrices 
\begin{equation}
\label{eq:SpinCorrelationMatrix}
    \bm{\chi}_{ij} = \frac{\langle \bm{S}_{i} \bm{S}^{T}_{j} \rangle -\langle \bm{S}_{i} \rangle \langle \bm{S}^{T}_{j} \rangle}{T}.
\end{equation}
Here, the matrix structure results from multiplications of column and row vectors and the entries correspond to the different spin components. Furthermore, here, we assume that the components of ${\bm S}_i$ are given in the {\it local} coordinate frame, similar to the discussion in Sec.~\ref{subsec:HTSE}. The susceptibility matrix is obtained via Fourier transform
\begin{equation}
\label{eq:susceptibilityMatrix}
    \bm{\chi}({\bm q})=\frac{1}{N}\sum_{ij}e^{i{\bm q}\cdot ({\bm r}_i-{\bm r}_j)}\bm{\chi}_{ij},
\end{equation}
in which we sum over all pyrochlore lattice sites with real space positions $\bm{r}_{i}$.
For the computation of order parameter susceptibilities such as $\chi_{\psi_2}$ or $\chi_{\psi_3}$ whose corresponding spin configurations $\psi_2$, $\psi_3$ are {\it ferromagnetic} in the local coordinate frame, we set $\bm{q}=\bm{0}$ in Eq.~(\ref{eq:susceptibilityMatrix}) and define $\bm{\chi}$ as
\begin{equation}
\label{eq:q0susceptibilityMatrix}
    \bm{\chi}=\bm{\chi}({\bm q}={\bm 0})=\frac{1}{N}\sum_{ij} \bm{\chi}_{ij}.
\end{equation}
Since for $\psi_2$ orders ($\psi_3$ orders) the spins point along the local $x$ axis ($y$ axis), with these conventions, the two relevant order parameter susceptibilities $\chi_{\psi_2}$ and $\chi_{\psi_3}$ are simply given by matrix entries of ${\bm \chi}$, that is, $\chi_{\psi_2}=\bm{\chi}_{xx}$ and $\chi_{\psi_3}=\bm{\chi}_{yy}$. Since the correlation function in Eq.~(\ref{eq:SpinCorrelationMatrix}) obeys ${\bm \chi}_{ij}={\bm \chi}_{ji}^T$, and considering Eq.~(\ref{eq:q0susceptibilityMatrix}), it follows that ${\bm \chi}={\bm \chi}^T$, or, more concretely, ${\bm \chi}$ has the general form
\begin{align}
\label{eq:localRestrictedchi}
    \bm{\chi} = \begin{pmatrix}
A & D & E\\
D & B & F\\
E & F & C
\end{pmatrix},
\end{align}
with $A,B,C,D,E,F \in \mathbb{R}$. Applying a lattice rotation symmetry transformation onto ${\bm \chi}_{ij}$ changes the site indices ($ij\rightarrow kl$) and rotates the spins (via a rotation matrix ${\bm R}$) such that, overall, ${\bm\chi}_{ij}$ transforms as ${\bm \chi}_{ij}\rightarrow{\bm R}^T{\bm\chi}_{kl}{\bm R}$. Since the transformation of the site indices just reshuffles the order of the terms in the sum of Eq.~(\ref{eq:q0susceptibilityMatrix}), the symmetry transformation acts on ${\bm \chi}$ solely as ${\bm \chi}\rightarrow\bm{R}^{T} \bm{\chi} \bm{R}$. 
Now, assuming that this transformation corresponds to a symmetry of the Hamiltonian in Eq.~(\ref{eq:hamiltonian}) that is {\it not} spontaneously broken means that ${\bm \chi}$ has to fulfill
\begin{equation}\label{eq:chi_condition}
    \bm{\chi} = \bm{R}^{T} \bm{\chi} \bm{R}.
\end{equation}
If one specifically considers the system's $C_3$ lattice symmetry, the matrix ${\bm R}$ corresponds to a $120\degree$ rotation about the local $z$ axis and it follows from Eqs.~(\ref{eq:localRestrictedchi}) and (\ref{eq:chi_condition}) that $\bm{\chi}$ is restricted to be of the form
\begin{align}
\label{eq:localRestrictedchi2}
    \bm{\chi} = \begin{pmatrix}
A & 0 & 0\\
0 & A & 0\\
0 & 0 & C
\end{pmatrix}.
\end{align}
This form now makes the equivalence $\chi_{\psi_2}=\chi_{\psi_3}=\bm{\chi}_{xx}=\bm{\chi}_{yy}=A$ obvious.

\newpage

\bibliography{references}

\end{document}